\documentclass[twocolumn,graphics,10pt,floatfix,a4paper,superscriptaddress,longbibliography, aps]{revtex4-1}
\usepackage{color}
\usepackage[subpreambles=true]{standalone}
\usepackage[utf8]{inputenc}
\usepackage{import}
\usepackage{stackengine, graphicx}
\graphicspath{{../figs/}{figs/}}

\usepackage{pgfplots}
\usepackage{amsmath} 
\usepackage{amssymb}
\usepackage{mathtools}
\usepackage{enumitem}
\usepackage{booktabs}

\usepackage{multirow}

\usepackage{pifont}

\usepackage{flowchart}
\usepackage{tikz}
\usetikzlibrary{shapes,arrows,external}
\usetikzlibrary{decorations.pathreplacing,angles,quotes, calligraphy,matrix,positioning,calc}

\usepackage{floatrow}
\usepackage[caption=false, labelformat=simple]{subfig}

\usepackage{algorithm}
\usepackage{algpseudocode}
\usepackage{url}
\usepackage{flushend}

\setlist{nosep}

\usepackage{hyperref}

\def\equationautorefname~#1\null{Eq.(#1)\null}

\hypersetup{
pdfnewwindow=true, colorlinks=true,
linkcolor=blue, anchorcolor=blue,
citecolor=blue, filecolor=blue,
menucolor=blue, urlcolor=blue}

\usepackage{titlesec}

\newcommand{\be}{\begin{equation}}
\newcommand{\ee}{\end{equation}}

\newcommand{\bea}{\begin{eqnarray}}
\newcommand{\eea}{\end{eqnarray}}

\newcommand{\ba}{\begin{align}}
\newcommand{\ea}{\end{align}}

\newcommand{\bd}{\begin{description}}
\newcommand{\ed}{\end{description}}

\newcommand{\bi}{\begin{itemize}}
\newcommand{\ei}{\end{itemize}}

\newcommand{\benum}{\begin{enumerate}}
\newcommand{\eenum}{\end{enumerate}}

\newcommand{\cmark}{\ding{51}}%
\newcommand{\xmark}{\ding{55}}

\newcommand{\smax}{A_{max}}

\newcommand{\signal}{A}

\newcommand{\validmin}[1][]{V^{L}_{#1}}
\newcommand{\validmax}[1][]{V^{H}_{#1}}
\newcommand{\validrange}[1][]{[\validmin[#1], \validmax[#1]]}

\newcommand{\safemin}[1][]{V^{\text{safe min}}_{#1}}
\newcommand{\safemax}[1][]{V^{\text{safe max}}_{#1}}

\newcommand{\safetyrange}[1][]{[\safemin[#1], \safemax[#1]]}

\newcommand{\V}[1][]{V_{#1}}

\newcommand{\vL}[1][]{V^{L}_{#1}}
\newcommand{\vH}[1][]{V^{H}_{#1}}
\newcommand{\vT}[1][]{V^{T}_{#1}}

\begin{document}

\title{Autonomous tuning and charge state detection of gate defined quantum dots}
%
\author{J. Darulov\'a }
\address{Theoretische  Physik,  ETH  Zurich,  8093  Zurich,  Switzerland}

\author{S.J. Pauka}
\address{ARC Centre of Excellence for Engineered Quantum Systems,School of Physics, The University of Sydney, Sydney, NSW 2006, Australia}

\author{N. Wiebe}
\address{Pacific Northwest National Laboratory,  Richland,  Washington 99354,  USA}
\address{Department of Physics, University of Washington, Seattle, Washington 98195, United States}


\author{K. W. Chan}
\address{ Microsoft  Quantum,  The  University  of Sydney, Sydney, NSW 2006, Australia}

\author{G. C. Gardener}
\address{Birck Nanotechnology Center, Purdue University, West Lafayette, Indiana 47907, USA}
\address{Microsoft Quantum Purdue, Purdue University, West Lafayette, Indiana 47907, USA}

\author{M. J. Manfra}
\address{ Department of Physics and Astronomy, Purdue University, West Lafayette, Indiana 47907, USA}
\address{Birck Nanotechnology Center, Purdue University, West Lafayette, Indiana 47907, USA}
\address{Microsoft Quantum Purdue, Purdue University, West Lafayette, Indiana 47907, USA}
\address{School of Materials Engineering and School of Electrical and Computer Engineering, Purdue University, West Lafayette, Indiana 47907, USA}

\author{M.C. Cassidy}
\address{ Microsoft  Quantum,  The  University  of Sydney, Sydney, NSW 2006, Australia}

\author{M. Troyer}
\address{Microsoft Quantum, Redmond, Washington 98052, USA}

\date{\today}

\begin{abstract}

Defining quantum dots in semiconductor based heterostructures is an essential step in initializing solid-state qubits. With growing device complexity and increasing number of functional devices required for measurements, a manual approach to finding suitable gate voltages to confine electrons electrostatically is impractical. 
Here, we implement a two-stage device characterization and dot-tuning process which first determines whether devices are functional and then attempts to tune the functional devices to the single or double quantum dot regime.
We show that automating well established manual tuning procedures and replacing the experimenter's decisions by supervised machine learning is sufficient to tune double quantum dots in multiple devices without pre-measured input or manual intervention.
The quality of measurement results and charge states are assessed by four binary classifiers trained with experimental data, reflecting real device behaviour. We compare and optimize eight models and different data preprocessing techniques for each of the classifiers to achieve reliable autonomous tuning, an essential step towards scalable quantum systems in quantum dot based qubit architectures.

\end{abstract}

\maketitle

\section{Introduction}
\label{sec:intro}

Quantum computers are expected to solve a range of problems intractable for classical computers, such as factoring of large integers \citep{Shor:1997:PAP:264393.264406}, simulating quantum systems \citep{Reiher201619152, Wecker2015, Rungger2019, Zhang2017} and efficiently sampling correlated probability distributions \citep{Chiesa:2019uc, PhysRevLett.119.170501}. Useful quantum computers with millions of high quality qubits are still some way off, but early prototypes with 50-100 qubits, referred to as Noisy Intermediate-Scale Quantum (NISQ) systems, are already available \citep{Preskill2018quantumcomputingin, Arute:2019fg, Chiesa:2019uc}.
For these NISQ systems, accuracy and reliability of quantum gate operations, and hence qubit quality, are essential. Even before qubits can be used to execute quantum algorithms, large scale material and design studies are necessary to optimize fabrication and device yield, involving both characterization and initial qubit tuning.

Electrostatically defined quantum dots are the building blocks for a range of solid state qubit architectures based on charge \citep{PhysRevLett.105.246804, PhysRevLett.95.090502, Yang:2019wu}, spin \citep{PhysRevA.57.120, RevModPhys.79.1217, Petta2180, Veldhorst:tr} and topologically protected states \citep{Kitaev:2001gb, Karzig:2017if, Alicea:2011fe}.
Fabrication variances as well as defects within the material lead to a disordered background potential landscape, which must be compensated for by different gate voltages when defining quantum dots. As this disorder is random, each gate within each device has a unique optimal tuning voltage that must be set and updated over time.
This tuning step is an essential, yet repetitive task well suited to automation. 
The increasing complexity as the number of qubits per chip grows has motivated several approaches to partially automate key components of this process. Neural networks have been successfully used to detect charge states on nearly noiseless data \citep{Kalantre_dnn,zwolak_qlite, Zwolak:2019tx} and procedures to automate fine tuning of the inter-dot tunnel coupling \citep{Botzem, vanDiepen_automation, Teske:760139} as well as fitting of charge transitions in charge stability diagrams \citep{Mills:2019fy} have been implemented. Together with studies improving measurement efficiency \citep{Lennon:2019uq} and the tune-up automation of a known device \citep{Baart_automation}, these results are important stepping stones for automated tuning of electrostatically defined quantum dots. They have been, however, demonstrated only on single, pre-tuned devices requiring  significant manual input and ideal measurement results. Autonomous tuning of unknown devices, a key milestone required for applying these techniques to practical tuning situations, has yet to be established.


Here we make significant progress towards this goal by demonstrating autonomous device characterization and tuning on multiple devices without pre-measured input and manual intervention.
We implement a so called minimal viable product, a standard software development technique in which a product is developed with sufficient features to satisfy early users and initiate a feedback loop to guide future improvements.
Based on the QCoDeS Python control software \citep{qcodes}, we developed a software package tuning semiconductor qubit devices into double dot regime. We only require the devices' gate layout, bonding scheme, line mappings, safe gate voltage ranges and the setup specific noise floor as input.
Using gate defined quantum dots in GaAs as a proxy qubit device we demonstrate that our software is capable of tuning different devices at different locations within the same wafer over two cooldowns.

\begin{figure}[!t]
\resizebox{ \columnwidth}{!}{\input{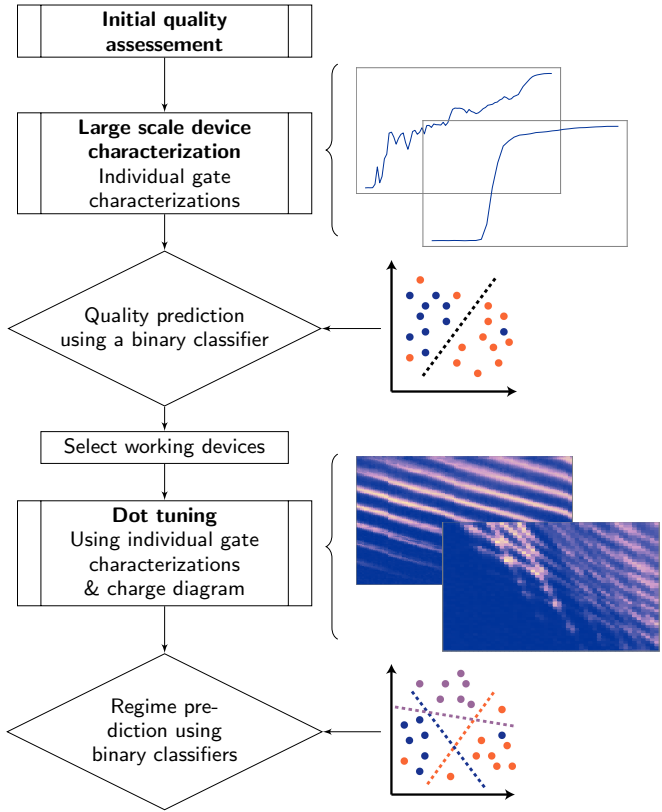}}
 \caption{The workflow of characterizing and tuning an unknown device. In an initial basic quality assessment we determine if any current is flowing through the device. We then individually characterize all gates of the device through 1D measurements. The quality of the gate response is assessed using a binary classifier and all gate responses need to be classified as good for a device to be selected for tuning. With the initial gate quality assessment and device characterization  we are able to distinguish two types of failure: no current through the device, or unresponsive gates. The tuning algorithm is a sequence of 1D measurements reducing the relevant voltage space and one or more 2D measurements establishing a charge stability diagram. Charge state and quality are established by a combination of three binary classifiers. }
    \label{fig:workflow}
\end{figure}

\section{Tuning approach}\label{sec:approach}
Quantum dots are systems confining electrons or holes in regions small enough to make their quantum mechanical energy levels observable. To form gate-defined quantum dots in GaAs, lithographically fabricated gate electrodes on the surface of the GaAs/AlGaAs heterostructure are used to deplete a two-dimensional electron gas (2DEG) beneath them, forming isolated puddles of charge with a physical dimension on the order of the Fermi wavelength \citep{vanderWiel:2002gr}.

Manual tuning of gate defined quantum dots involves a series of 1D and 2D measurements, measuring current through the device as a function of one or two gate voltages respectively, and utilizes informed guesses and the intuition of the experimenter.
These measurements narrow down the large parameter space to find a voltage combination defining the desired quantum dot structure. Material defects, impurities, fabrication variations, capacitive couplings between gates and hysteresis make these voltages unique to each device and difficult to find. Automation of these procedures will remove a manual, time consuming task and enable many devices to be tuned in parallel, which is an important element for a large scale quantum processor. To fully benefit from such a progress, automated procedures need to be fully independent and not rely on pre-measured input.

Given the success across a variety fo fields, machine learning techniques are promising tools to use.
Machine learning is commonly considered for two types of problems. First, it is used to solve problems too complex for predefined, structured programs such as face recognition,  speech-to-text processing and spam filtering. Our lack of understanding of good algorithms to perform this task makes them natural candidates for machine learning. Second, it can be used to improve solutions even where good classical algorithms are already known, as for example the simulation of quantum many body systems \citep{Carleo602, Carleo:2018tj, Torlai:2018wn} as well as designing methods for encoding and decoding quantum information within error correcting codes \citep{nautrup2018optimizing, sweke2018reinforcement}.

As quantum dot tuning is routinely performed by scientists and quantum dots have been studied extensively in the past \citep{vanderWiel:2002gr}, we choose an approach which automates existing procedures with simple machine learning tools to achieve autonomous quantum dot tuning.
Specifically, we implement a deterministic tuning sequence and replace the scientists' knowledge by binary classifiers trained on experimental data. We use approximately 10000 hand-labelled datasets to ensure that  features difficult to simulate, such as realistic noise and poor or intermediate dot regimes, are learned. By automating 1D and 2D measurements and their respective data analysis (see \autoref{sec:pinchoff} and \autoref{sec:charge_diagram}), and using four binary classifiers available in Python's scikit-learn, \autoref{sec:classification}, we are able to characterize and tune several devices with no pre-measured input.
Figure \ref{fig:workflow} illustrates this strategy: After an initial quality assessment identifying devices featuring a current above the setup specific noise floor, we use 1D measurements to characterize all gates of all devices available, as described in \autoref{sec:device_characterization}. Using a binary classifier we predict the quality of each measurement and establish a list of working gates of each device. If all gates respond well, the device is selected for tuning. The tuning algorithm uses a sequence of 1D measurements to narrow down and set voltages and a 2D measurement classified by three binary classifiers to assess the dot regime,  see \autoref{sec:tuning_algorithm}.
We show that a predefined sequence of fast 1D and a few 2D measurements together with binary classifiers is sufficient to reach the desired regime and that no complex machine learning algorithms are required to perform this well studied task.

\section{Devices}\label{sec:devices}

\begin{figure}[!t]
\begin{tabular}{cc}
       \sidesubfloat[]{%
       \vspace{-2in}
       \hspace{-0.1in}%
       \includegraphics[width=0.43\columnwidth]{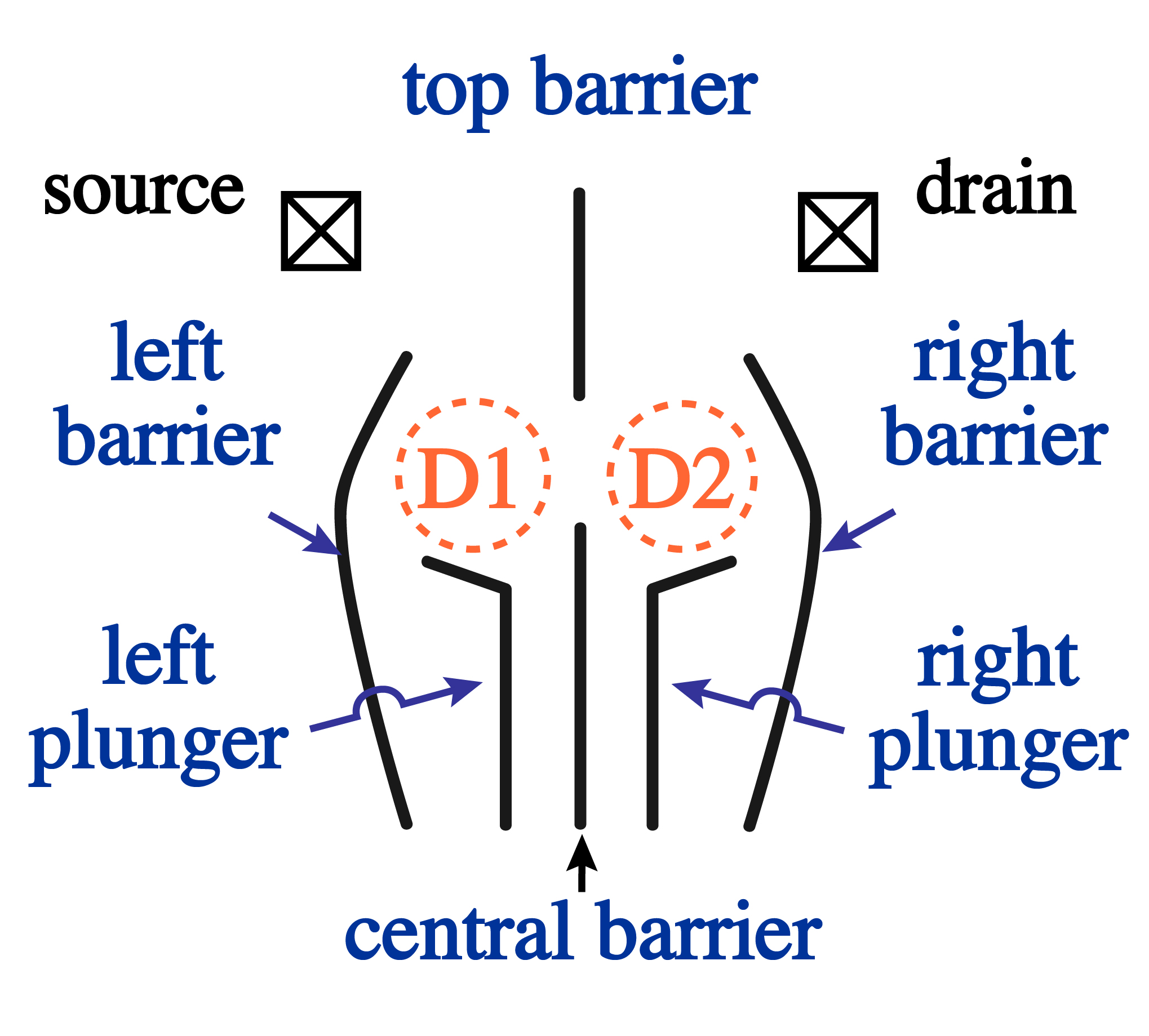} \label{fig:device_long_names}
      }
     \sidesubfloat[]{%
     \hspace{-0.05in}%
     \includegraphics[width=0.43\columnwidth]{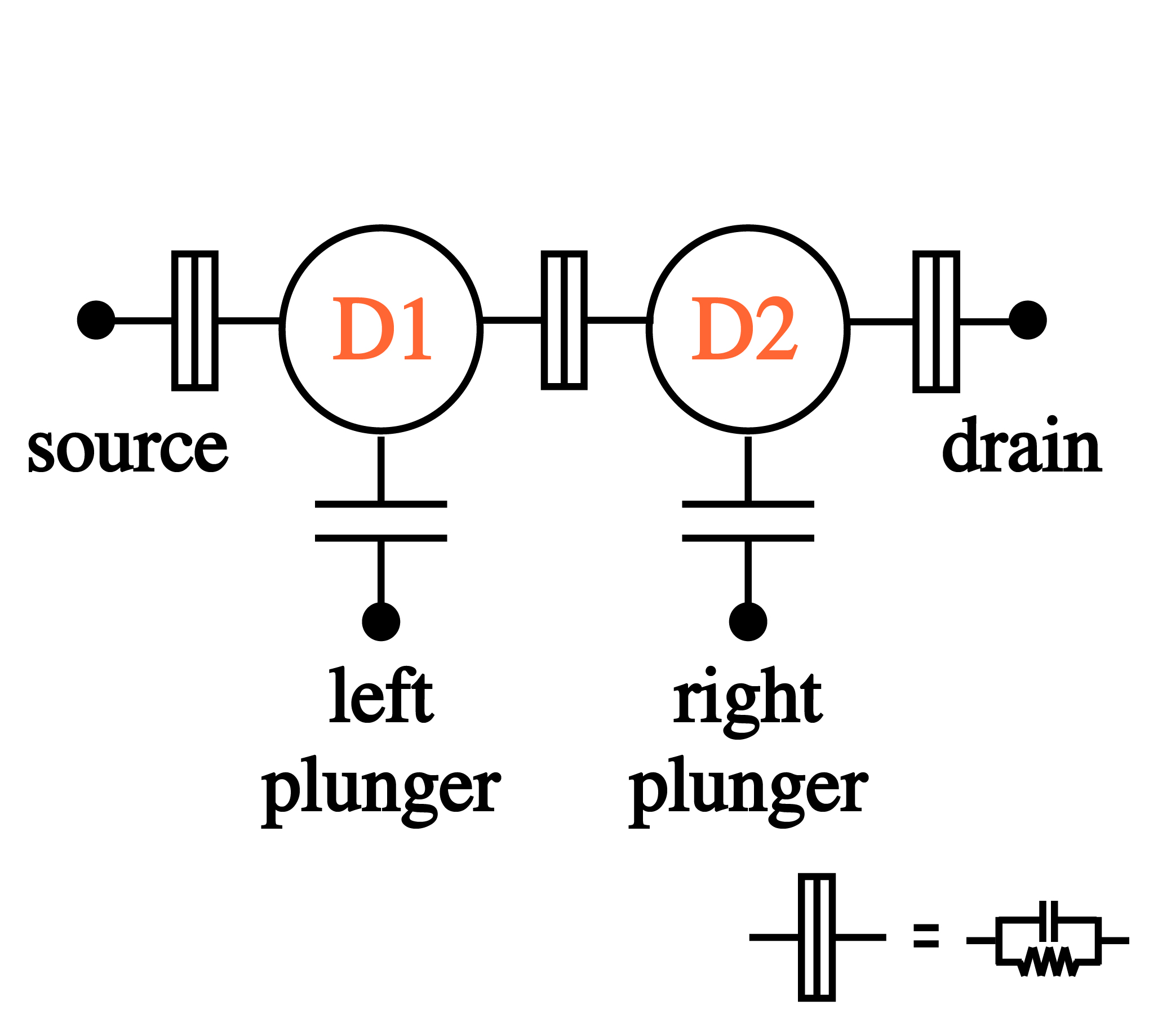}\label{fig:capa_model}
     } \\[5pt] 
          \sidesubfloat[]{%
    \includegraphics[width=0.4\columnwidth]{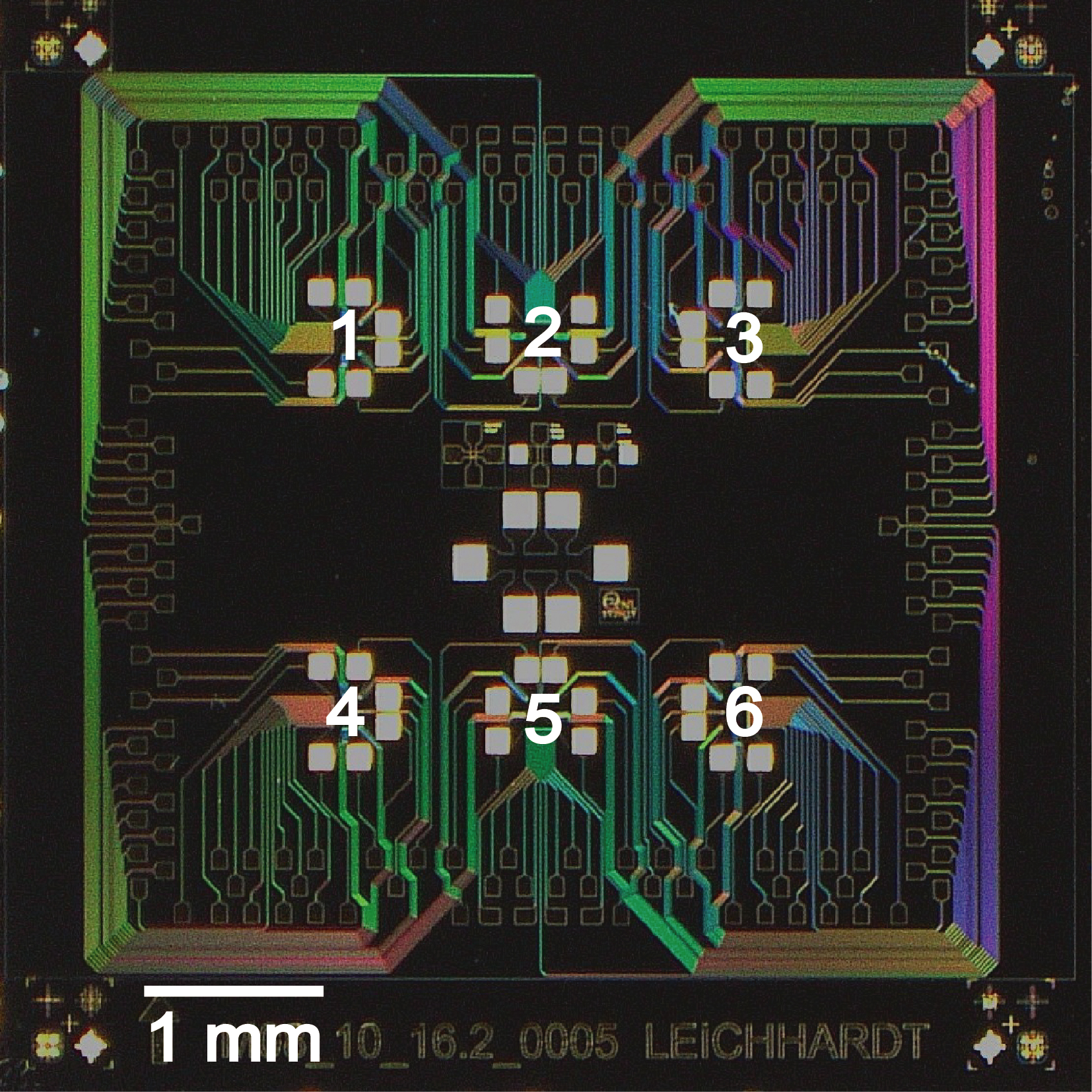} \label{fig:qdp_chip}
     }
       \sidesubfloat[]{%
       \vspace{-2in}
       \includegraphics[width=0.4\columnwidth]{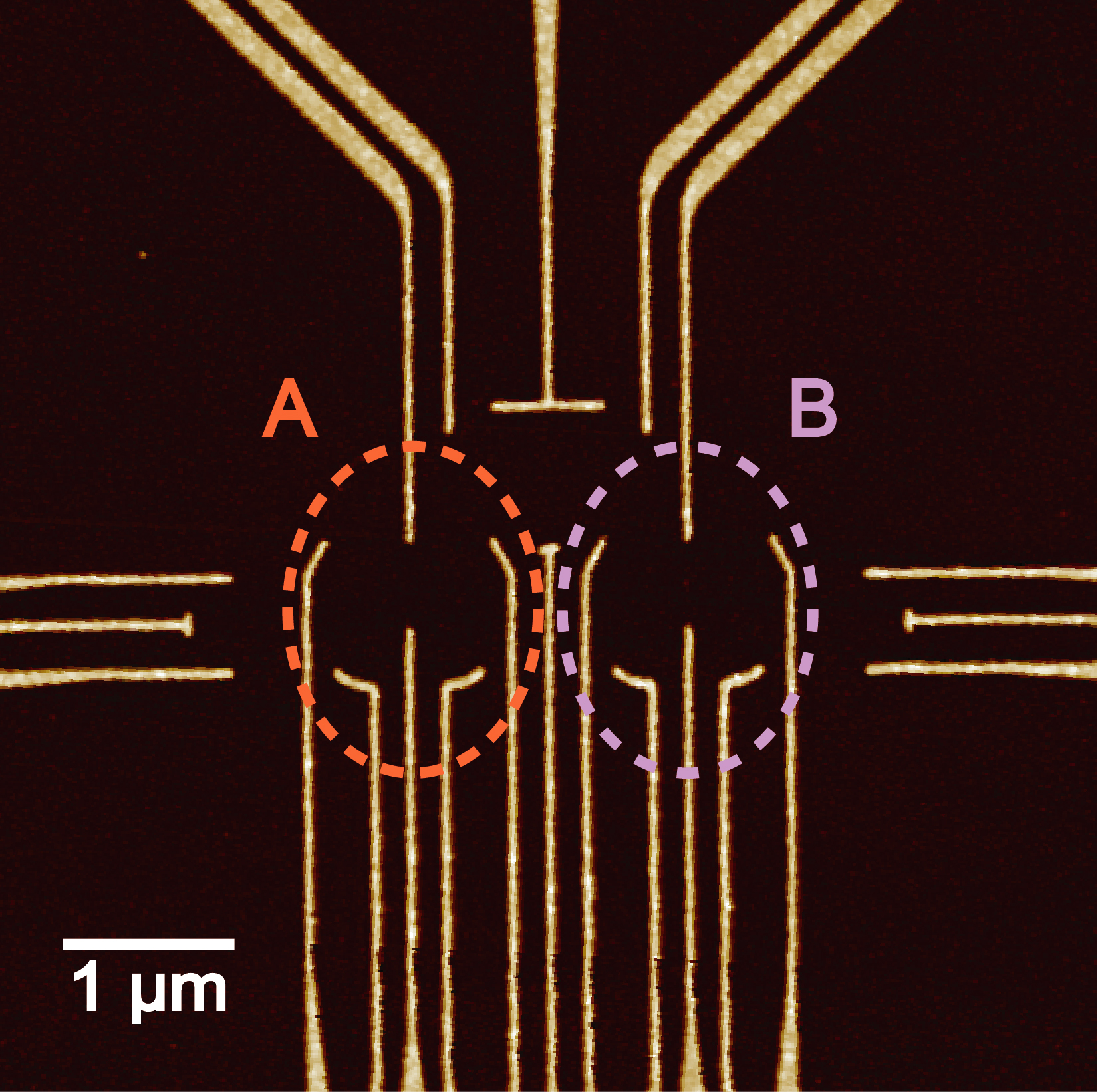} \label{fig:device}
      }
      \end{tabular}
     \caption{(a) The gate arrangement of a device. Six metallic gates are used to deplete the 2DEG underneath. Barrier gates create potential wells confining electrons within the 2DEG structure while plunger gates vary the electrochemical potential of the dots. Current through the device is measured between source and drain.%
    (b) Simplified capacitance model representing a double quantum dot. Two dots are coupled to source, drain and two gates. Barriers are characterized by tunnel resistors and a capacitor.%
    (c) Optical micrograph of a chip containing six device pairs, a prototype for future scalable quantum processor. It consists of 132 DC lines connected to metallic gates, for each of which a valid voltage needs to be determined. %
    (d) A false colour scanning electron micrograph showing the active region of one of the device pairs. It can host up to five dots, two double dots (A and B) used as qubits and one single middle dot for coupling \citep{Croot:2018iq}. We tune one double dot at a time, encircled in red.%
    }
\end{figure}

We use double quantum dots formed in a GaAs/AlGaAs two dimensional electron gas (2DEG) depleted by the gate structure illustrated in \autoref{fig:device_long_names} as example system to demonstrate the performance of our autonomous tuning package.
This system of gates and dots can be approximated classically by a network of tunnelling resistances and capacitors, called the constant interaction model \citep{vanderWiel:2002gr}, illustrated in \autoref{fig:capa_model}.

The 2DEG is located 91 nm  below  the  surface  of the  GaAs/AlGaAs  heterostructure of  density  $1.5 \times 10^{11}\text{cm}^{-2}$  and  mobility  $2.4 \times 10^{6} \text{cm}^{2}\: /\text{Vs}$. An aluminium oxide layer  deposited  using  atomic  layer  deposition separates TiAu gates from the heterostructure and enables  positive  voltages  to  be  applied  without  gate-leakage. Our test chip holding six pairs of double quantum dots on a 5mm x 5mm chip is displayed in \autoref{fig:qdp_chip} and a false colour scanning electron micrograph of one device pair in  \autoref{fig:device}. It allows us to study multiple double-quantum dots formed in a  single fabrication run, and reduce variation caused fabrication. We bonded four of the six available device pairs: number 1, 3, 4 and 6, located in the corners of the chip.

The device layout, illustrated in \autoref{fig:device_long_names}, consists of two types of gates: barriers and plungers. The left barrier (LB), right barrier (RB) and central barrier (CB) create potential wells, defining tunnel resistances between quantum dots and reservoirs. Left plunger (LP) and right plunger (RP) tune the electrochemical potential and are used to change the number of electrons in a quantum dot. These six gates constitute what is called a \textit{device}.
All voltages applied to these gates need to lie within a known safety range $\safetyrange[i]$, protecting against dialectric breakdown.

The purpose of tuning is to determine gate voltages leading to two well defined states with either a single or two separate quantum dots formed, referred to as single and double dot regimes.

Devices A and B of each device pair on the chip are tuned separately. Experiments were performed in a dilution refrigerator with a base temperature of 20 mK and at zero magnetic field. All measurements are direct transport measurements using standard lock-in techniques.

%
\section{Methods}\label{sec:method}
%
 The device characterization and dot tuning algorithms are implemented by a modular software consisting of two main modules, each representing an essential manual tuning step: A 1D, single gate characterization and a 2D measurement generating a charge stability diagram. These modules take and analyse data for tuning and preprocess it for machine learning classification. Outcomes are classified using binary classifiers, assessing quality, and, in case of charge stability diagrams, charge regime. The classification outcomes determine next actions to take, i.e. whether a device will be tuned after characterization or which gate voltage to adjust during tuning.
 
 At the beginning of each tuning we determine the saturation current $\smax$ when all $\V[\text{i}] = 0$, which is used to normalize data before classification. If $\smax$ is above the setup specific noise level, the device is declared not working and not considered for characterization or tuning.

\begin{figure}[!t]
    \begin{tabular}{l}
    \smallskip
     \sidesubfloat[]{%
     \includegraphics[width=0.9\columnwidth]{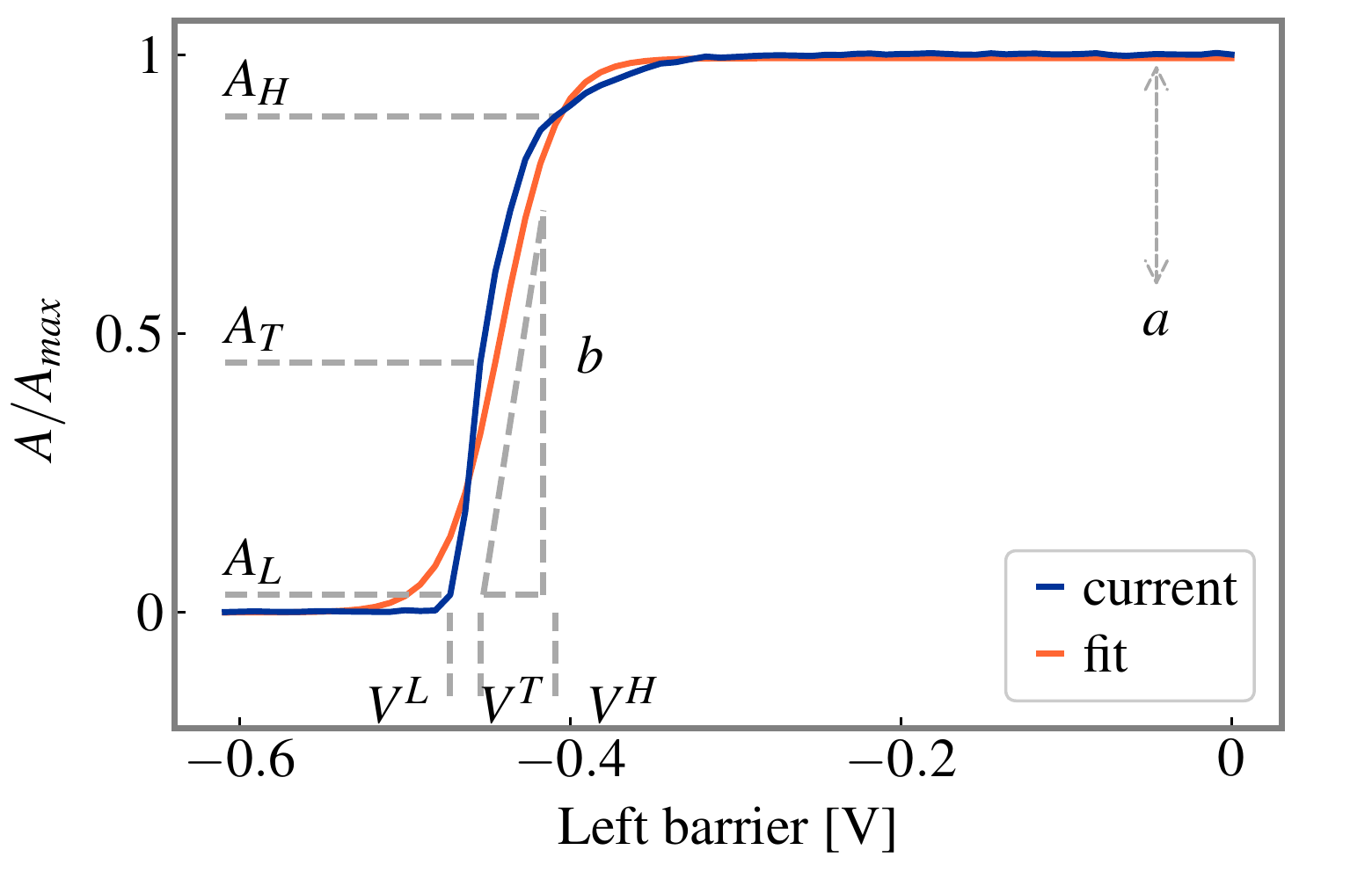} \label{fig:pinchfit}
     } \\
     \sidesubfloat[]{%
     \includegraphics[width=0.92\columnwidth]{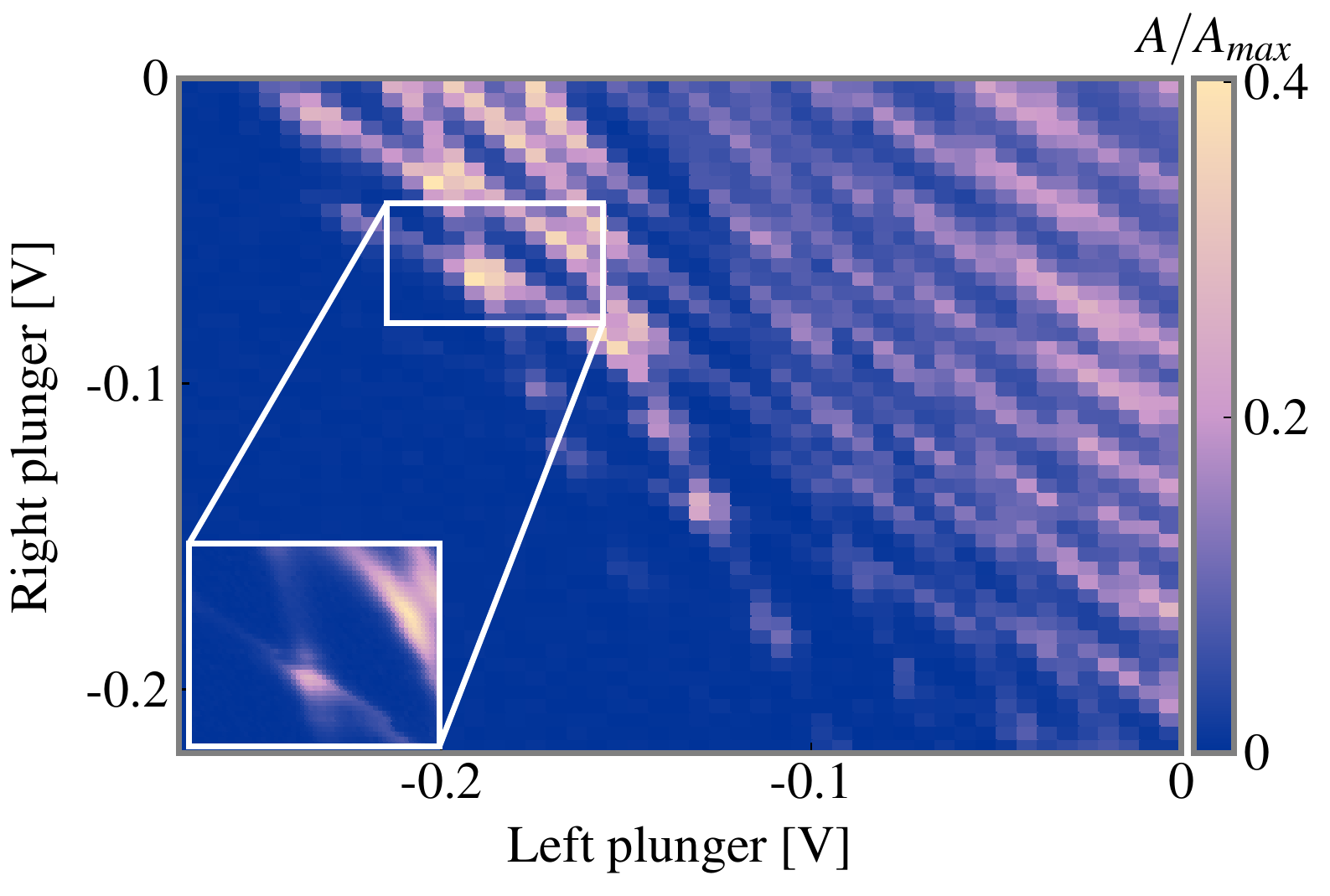} \label{fig:charge_diagram}
     }
    \end{tabular}
     \caption{%
     (a) Data analysis of an individual gate characterization. A gate is stepped from high to low safety limits until the signal falls below noise level. The fitting procedure extracts relevant features for classification, such as amplitude, slope, offset as well as relevant voltages for subsequent tuning $\vL[i], \vT[i], \vH[i]$. These voltages are used to determine a suitable voltage to set or range to sweep.
    (b) Charge stability diagram of a good double dot regime. Two gates, typically plungers, are stepped within their valid ranges $[\vL[i], \vH[i]]$ to measure a 2D map. The results is segmented into  $0.05 \: V \times 0.05 \: V$ regions which are classified individually to determine their quality and charge state.}
\end{figure}
 
\subsection{Individual gate characterization}\label{sec:pinchoff}
%
The individual gate characterization module determines the voltage range in which a gate partially depletes the underlying 2DEG for a given device configuration. This is the interesting voltage range for further tuning. The voltage $v$ of the gate in question is varied while keeping all other gates constant. The measured I-V curve, also called pinch-off curve, is normalized by $\smax$ and smoothed using Gaussian kernel convolution. The curve is then fitted to
\be
f(x, a, b, c) = a  (1 + \tanh(bx +  c)),
\ee
extracting amplitude $a$, slope $b$ and offset $c$. Here $x$ is the normalized gate voltage 
 \be
 x = \frac{v - v_{min}}{v_{max} - v_{min}},
 \ee
 where $v_{min}$ and $v_{max}$ the highest and lowest voltage setpoints respectively.
Similar to conventional semiconductor transistors, we are interested in gate voltages separating cut-off, transition and saturation regions $\vL[i], \vH[i]$, see \autoref{fig:pinchfit}. 
Using methods described in \citep{OrtizConde:2002jg}, we define $\vT[i]$ as the point of highest variance of the measured current, the lower bound of the transition region $\vL$ as the voltage axis intercept of the tangent at $\vT[i]$ and the upper bound $\vH[i]$ as the minimum of the second order derivative of the current.
Further tuning will continue by considering the range $\validrange[i]$.

Following standard machine learning procedure we select a set of features representing the measurement, also known as feature selection. Similar to a transistor, we are looking for a noiseless, sharp transition between fully open and fully closed regimes. The features extracted during fitting are amplitude, slope, offset, low current, high current, residuals, offset and transition current. Voltages, i.e transition voltage, are not considered at this stage as we do not optimize for transition locations in voltage space.
By training and testing classifiers with all possible subsets of features we find that the following ones are most relevant:
\bi
\item amplitude $a$ 
\item slope $b$
\item residuals
\item pinch-off current $A_{L}$.
\ei
Exposing a classifier to more than these parameters results in lower accuracy due to overfitting. We can, however, replace amplitude by high and low current or low current by offset.
Including both results in a decrease in accuracy due to overfitting.

\subsection{Charge stability diagram}\label{sec:charge_diagram}

Charge stability diagrams are measured by varying left and right plunger voltages over the voltage range $\validrange[i]$ to change electrochemical potentials of nearby dots while measuring the current though the device. The purpose is to find regions in voltage space where single and double dots are formed. While gates could be swept in any order, we focus on stepping over the left plunger's range on the $x$ axis and the the right plunger's range on the $y$ axis.
The voltage ranges are sampled over 50 equidistant points. The number of setpoints is increased if the step size is larger than a threshold $\delta V$, which is smaller than both a device specific safe voltage step and a desired voltage resolution. As a compromise between measurement time and voltage resolution we choose $\delta \V < 0.005 \: V$. The resulting diagram is transformed using using skimage's resize method \citep{scikit-image}.
%
Based on previously measured data we know that averages of currents in good quantum dot regimes are within $[0.004 \smax, 0.1 \smax]$. We compare the averages $\bar{A}$ for each of the boundaries to these limits and update the the plunger's current voltage ranges:

\bi
\setlength\itemsep{0.3em}
\item if $\bar{A}_{\text{left vertical}} > 0.1 \smax$: decrease $\vL[\text{LP}], \vH[\text{LP}]$ 
\item if $\bar{A}_{\text{bottom horizontal}} > 0.1 \smax$ : decrease $\vL[\text{RP}], \vH[\text{RP}]$
\item if $\bar{A}_{\text{right vertical}} < 0.004 \smax$: increase $\vL[\text{LP}], \vH[\text{LP}]$
\item if $\bar{A}_{\text{top horizontal}} < 0.004 \smax$: increase $\vL[\text{RP}], \vH[\text{RP}]$.
\ei

The measurement result is segmented into  $0.05\: V \times 0.05\: V$ regions, each of which is classified individually. We use three binary classifiers to assess quality and regime, each trained to predict single dot quality, double dot quality and dot regime respectively, see \autoref{sec:classification}.
The module returns the success of measuring a charge diagram with respect to signal strength as well as quality and regime predictions of each segment.
Our attempts in defining suitable features to represent charge stability diagrams failed due to noise and large variability of the data. We thus use the full current map for classification.

\subsection{Classification}\label{sec:classification}
The task of supervised machine learning is to approximate an unknown target function $f$
\be
Y = f(X),
\ee
mapping input variables $X$ to output variables $Y$. In the present dot tuning algorithm, $X$ is either post-processed data or a smaller feature vector extracted from it and $f$ the mapping onto either quality (good/bad) or regime (single/double dot), summarized in \autoref{tab:clf_input_output}.
After labelling existing data, i.e attaching a label $Y$ to each $X$, we can train a machine learning model to learn an approximation $m$ of the unknown mapping function $f$ and use it to predict labels of new measurements.
In order to find an approximation, assumptions about its form need to be made. 
Different classifiers implement different types of functions $m(X, h)$ with hyper parameters $h$. The task is to find an accurate model $m$ and good hyper parameters $h$. A good model will not only accurately fit known $X$ and $Y$ relations, but also generalize well to new data.
The more complex a model is, the more flexible it is to learn general concepts of $X$ and the more data it needs for training to avoid overfitting, a situation when noise and random fluctuations are learned as concepts. 
 
\begin{table}[!t]
    \setlength{\tabcolsep}{7pt}
    \centering
    \begin{tabular}{ccc}
    \hline
    \hline
       classifier &  $X$ & $Y$ \\
    \hline
        pinch-off & features & good/bad\\ 
        single dot & current data  & good/bad\\ 
        double dot & current data  & good/bad\\ 
        dot regime & current data  & single/double\\ 
    \hline
    \hline
    \end{tabular}
    \caption{Summary of input $X$ and output $Y$ for each classifier used. Features extracted during individual gate characterizations form the 1D input vector to the classifier. The output predicts the quality of the gate's response, also called pinch-off. Single, double and dot regime classifiers use the normalized current and its Fourier Transform (FT) or both combined to predict qualities and regime respectively. The outcomes of these classifiers guide the dot tuning algorithm in \autoref{sec:tuning_algorithm}.}
    \label{tab:clf_input_output}
\end{table}

\begin{figure}[!t]
\includegraphics[width=\columnwidth]{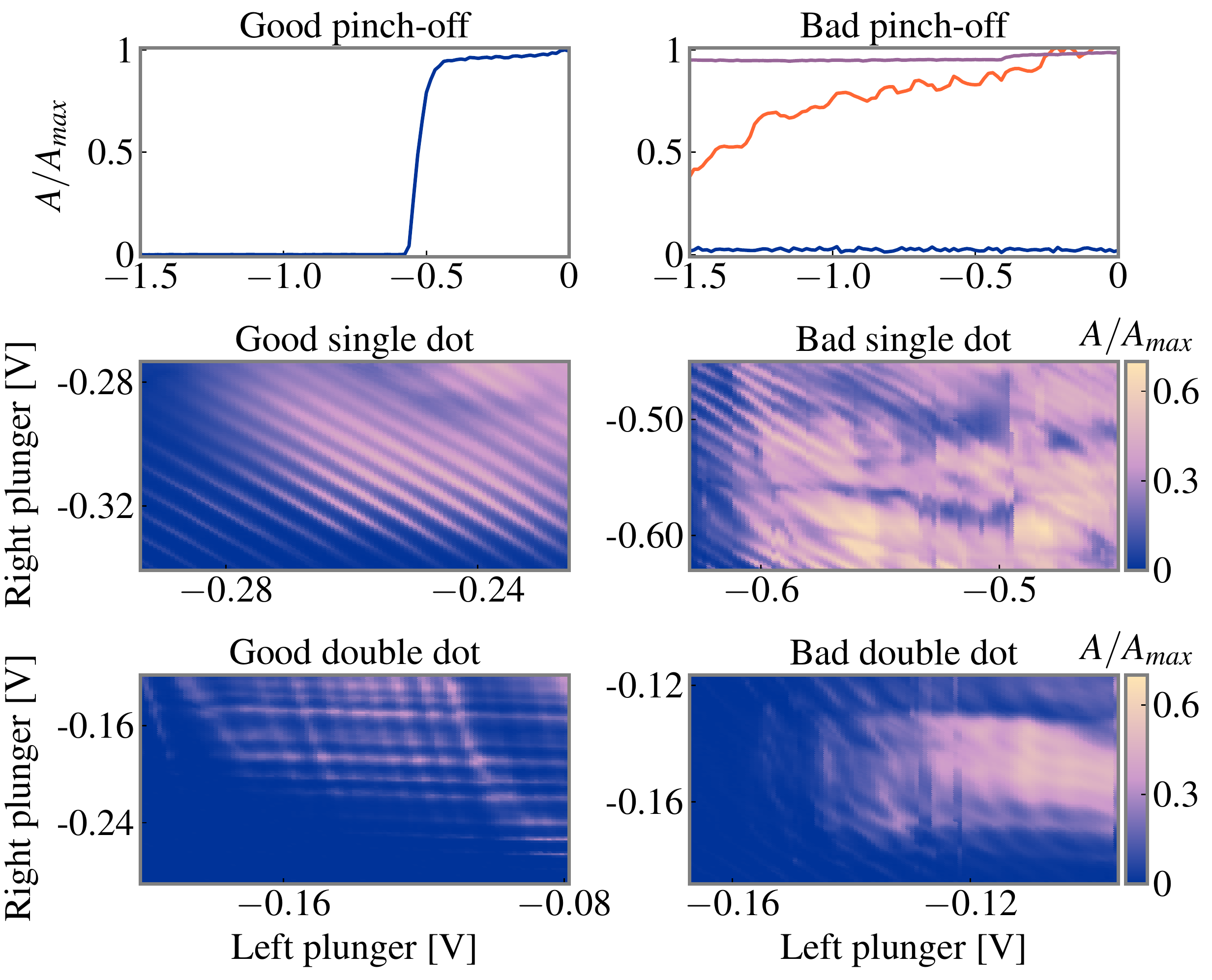}
 \caption{Examples of data used for classification, normalized current $A/A_{max}$ as a function of one or two gate voltages. Left column shows measurement results labelled as good, right column shows examples of results labelled as bad.
 Top, middle and bottom row correspond to individual gate characterizations, single dot and double dot measurements respectively. All data used to train classifiers originates from devices other than those characterized and tuned here. The examples above stem from the device presented in Ref. \citep{Croot:2018iq}.}
    \label{fig:data_examples}
\end{figure}

The large variety of noise and intermediate regimes of quantum dot measurements makes it difficult to generate synthetic data reflecting real device behaviour. For this reason, we use experimental data for training. We collected and labelled 4000 pinch-off curves, 4000 single dot and 2500 double dot current maps. Examples of good and bad results for each category are shown in \autoref{fig:data_examples}. Data used for classifier training originate from different devices of similar design as characterized and tuned here.

As the size of our training set is not large enough for algorithms containing many free parameters, such as deep neural nets, we investigate the potential of simpler models that are easy to train and optimize.
We compare classifiers readily available in Python's scikit-learn package: Logistic Regression, Support Vector Machines (SVM), Multilayer Perceptron (MLP), Gaussian Process, Decision Tree, Random Forest, Quadratic Discriminant Analysis and $k$-Nearest Neighbors ($k$-NN).
To estimate the potential of a model and to exclude overfitting, we average performances over $n$ train and test splits. At each iteration, we first randomly select a subset with equal populations of all available data, i.e equal numbers of good and poor results, and then split into training and test datasets. We use a percentage ratio of 80/20 between test and training data and we choose $n = 20$ for 1D data and $n = 10$ for 2D data based on performance fluctuation studies detailed in \autoref{ax:data_fluctuation}.

Classification performances are evaluated using the accuracy score (ACC), defined as the ratio of correctly classified samples over the total number of samples. In machine learning terms the correctly labelled samples are the sum of true positive (TP) and true negative (TN) predictions, while the total number of samples is the sum of all positive (P) and all negative (N) samples:
\be
\text{ACC} = \frac{\text{TP} + \text{TN}}{\text{P} + \text{N}}.
\ee

We compare classifier performances trained and tested on differently preprocessed data. In addition to the normalized current we use the current map's Fourier transform  as well as principle components determined by performing a Principle Component Analysis (PCA) implemented in sklearn \citep{scikit-learn}. All transformations are applied to training, testing and prediction data before classification. Principle components are extracted and hence defined from training data, which is expressed as amplitudes of each of these components. Test and prediction data is projected onto components and the amplitudes used for prediction.  
We compare performances using just the normalized current, its Fourier transform or both, with and without PCA. Results are summarized in \autoref{sec:results} and detailed in \autoref{ax:classifiers}.

\section{Device characterization}\label{sec:device_characterization}
%
The objective of device characterization is to choose devices suitable for tuning among a large number of unknown devices. It establishes a list of working gates and their pinch-off voltages $\vL[i], \vT[i], \vH[i]$, providing a quality measure and enabling early device comparison.
We implement a predefined sequence of measurements by using the individual gate characterization module introduced in \autoref{sec:pinchoff}, sweeping each gate individually and preprocess the data for machine learning classification.

The algorithm initializes all gates to their upper safety limit $\safemax$ to measure the saturation current $\smax$.
For a gate to create an entirely opaque barrier, a negative voltage on a second gate opposite needs to be set. We use the top barrier to deplete the 2DEG in the upper half of the device and to facilitate a complete depletion: It is set to its negative safety limit $\safemin$, which in ideal devices will not result in a reduced current through the device. A signal drop drop below $0.8 \:\smax$ without additional gate voltages set on the lower barrier gates indicates the presence of offset charges on one or several of the remaining gates. In this case the top barrier voltage is increased in steps of $0.2 \:V$ until the signal is above the desired threshold $0.8 \:\smax$. Safety ranges of all gates are shifted towards more positive values by $0.5 \:V$.

\begin{figure}[!t]
\includegraphics[width=\columnwidth]{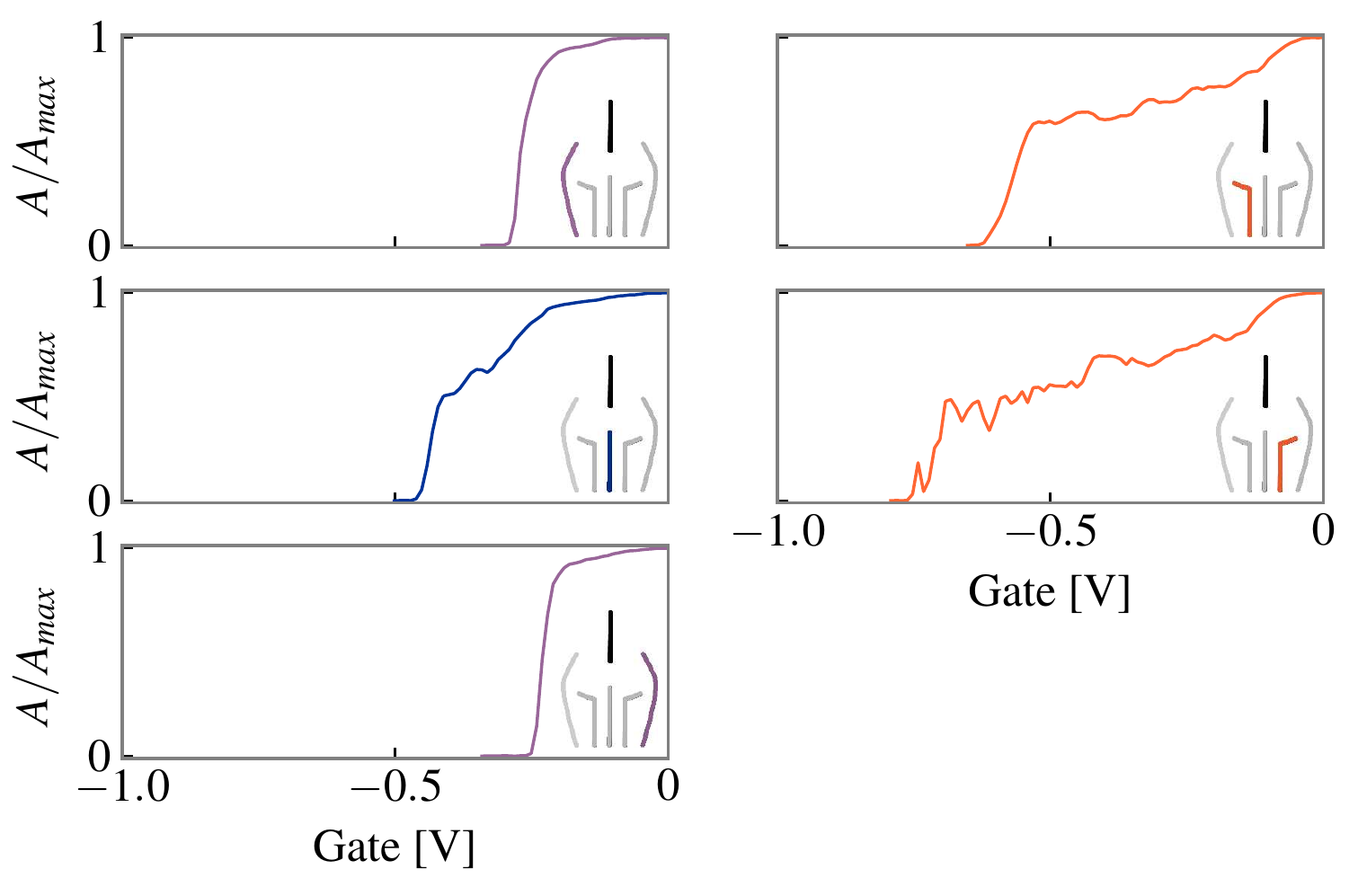}
 \caption{Example of device characterization. Normalized current is plotted as a function of gate voltage. The top barrier is set to its negative safety limit $\safetyrange[\text{TB}] = -3 \:V$ and all remaining gates are characterized individually. In this case all gate responses are identified as good by the binary classifier, selecting the device for further tuning.}
    \label{fig:characterization}
\end{figure}

The left, central and right barrier as well as left and right plungers are characterised individually. An example of this process, performed on device 3.A is shown in \autoref{fig:characterization}. The quality of measurement results is assessed by a binary classifier, differentiating between good and bad gate responses. If all responses are classified as good, featuring a high amplitude, sharp transition and zero pinch-off current, a device is declared as working and considered for double dot tuning.

Devices classified as working are characterized further by establishing a valid range for the top barrier, defined as the voltage range in which the remaining barriers are able to deplete the 2DEG. This is determined by decreasing the top barrier's voltage starting from zero and in steps of $0.2\:V$, characterizing left, right and central barrier at each iteration. If a barrier responds well, i.e. is able to deplete the 2DEG, it is not considered in subsequent iterations. The top barrier voltage at which the last barrier is able to deplete the 2DEG defines the top barrier's lower valid range value $\validmin[\text{TB}]$. To establish the upper valid range voltage $\validmax[\text{TB}]$, the last barrier to pinch off is set to its lower safety limit $\safemin$ and the top barrier is characterized. The pinch-off voltage $\vL$ defines $\validmax[\text{TB}]$. This step is specific to our device design and variances may need to be considered for other types of devices.
%
\section{Tuning algorithm}\label{sec:tuning_algorithm}
%
The purpose of the tuning algorithm is to tune fully working devices into either a single or double dot regime. We illustrate the algorithm for the case of a double dot regime. It uses the same individual gate characterizations as in the previous section but for a different top barrier. It determines suitable gate voltages of barriers and narrows down valid plunger ranges. The procedure used to determine the device's state and the next actions to take are illustrated in \autoref{fig:tuning_algorithm}.

\begin{figure}[!t]
\noindent\stackinset{r}{-10pt}{t}{-25pt}{ 
\includegraphics[width=0.42\linewidth]{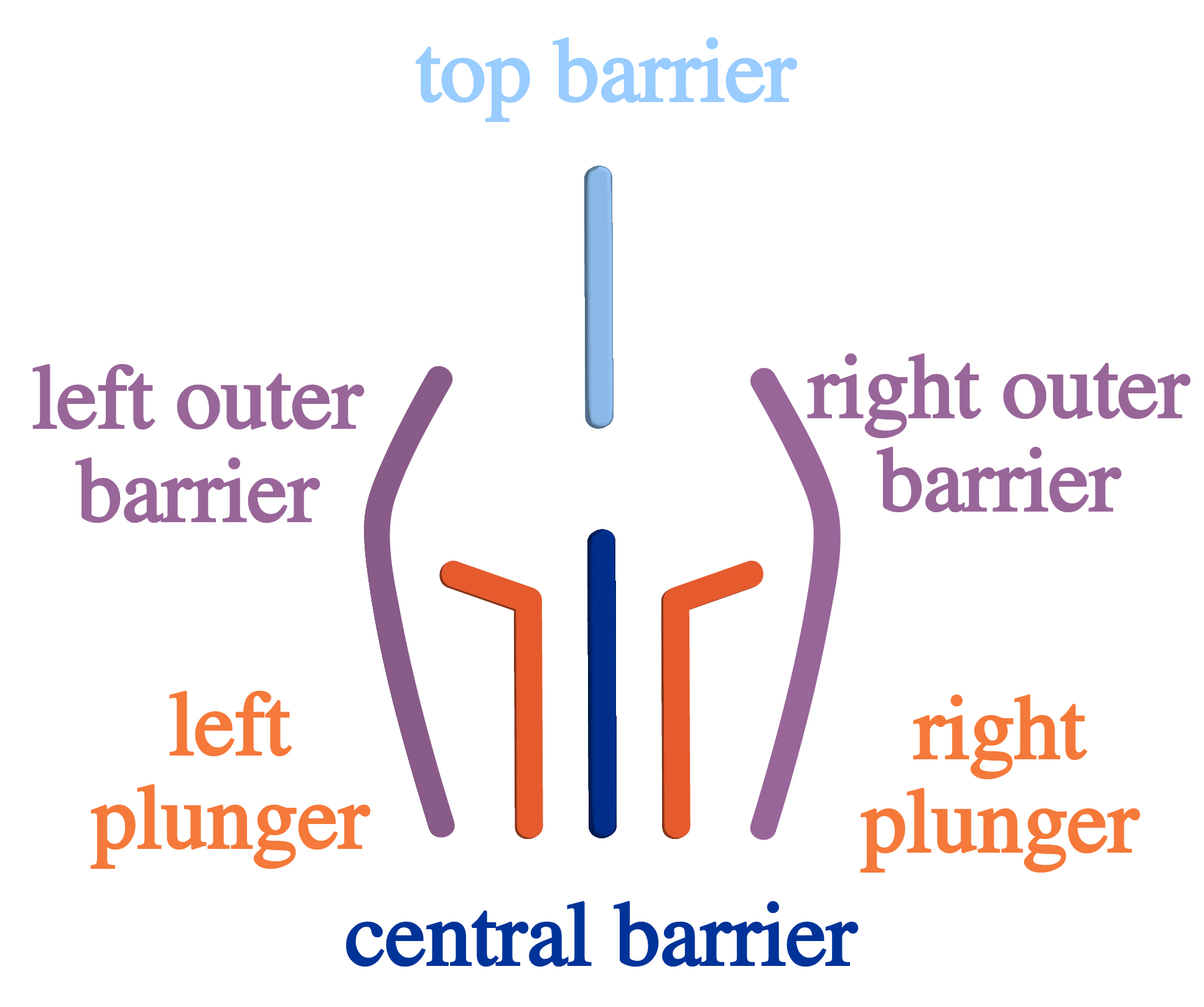}}
 {\hspace*{-20pt} 
%
%
%
%
%
%
%
%
%


\definecolor{myred}{HTML}{FF6633}
\definecolor{myblue}{HTML}{003399}
\definecolor{mygreen}{HTML}{996699}
\definecolor{mygray}{HTML}{99CCFF}

\definecolor{myred}{RGB}{255, 102, 51}
\definecolor{myblue}{RGB}{0, 51, 153}
\definecolor{mygreen}{RGB}{153, 102, 153}
\definecolor{mygray}{RGB}{153, 204, 255}

\begin{tikzpicture}[>=latex',font={\sf \small},
			    auto,
			    node distance = 2cm,
			    inner sep=2pt,
			    text centered,] 
			    
\def\smbwd{2cm}

\tikzstyle{init_return} = [draw,
			terminal,
    			text width=5em,
			inner sep=0.1cm,
						]
			
\tikzstyle{my_decision} = [draw,
			diamond,
    		text width=8em, 
			aspect=1.6,
			inner xsep=-5mm,
			minimum height=-1cm,
			]

\tikzstyle{1Dstage} = [predproc, draw,
			     inner sep=0.1cm,
			      text centered,
			       text width=6em,
			       minimum height=0.1cm,
			     minimum width=1cm,
			     ]

\tikzstyle{2Dstage} = [predproc, draw,
			     inner sep=0.1cm,
			      text centered,
			      text width=6em,
			      minimum height=0.1cm,
			     minimum width=3cm,
			     ]
 
\tikzstyle{action} = [draw,
			     predproc,
    			     text width=6em, 
			     text centered,
			    minimum height=0.1cm,
			     minimum width=1cm,
			      inner sep=0.1cm,
			     ]

\tikzstyle{action_simple} = [draw,
			     process,
    			     text width=6em,
			     minimum height=0.1cm,
			     minimum width=2cm,
			     inner sep=0.1cm,
			     ]

\tikzstyle{line} = [draw, -latex']
\tikzstyle{line_dashed} = [draw, -latex', dashed]

    \node [init_return] (init) {Initialize $\vec{V} = \vec{V}_{max}$};

    \node[action_simple, below of=init, node distance = 1cm](norm_constant){ Measure $A_{max}$};
    
      \path [line] (init) -- (norm_constant);
     
  
   \node[action_simple, fill=mygray!20, below of=norm_constant, node distance = 0.9cm](set_top_barrier){Set top barrier};
   
     \path [line] (norm_constant) -- (set_top_barrier);
   
    \node[1Dstage, below of=set_top_barrier,  fill=myblue!20, node distance = 1cm](characterize_central){Characterize central barrier};
    
     \path [line] (set_top_barrier) -- (characterize_central);
    
        \node[action_simple, below of=characterize_central, fill=myblue!20,  node distance = 1.2cm](set_central){Set central barrier};
        
     \path [line] (characterize_central) -- (set_central);
    
  \node[1Dstage, below of=set_central, fill=mygreen!20, node distance = 1.2cm](characterise_outer_barriers){Characterize outer barriers};
  
      \path [line] (set_central) -- (characterise_outer_barriers);
  
    \node[action_simple, below of=characterise_outer_barriers, fill=mygreen!20, node distance = 1.2cm](set_outer_barriers){Set outer barriers};
    
       \path[line] (set_outer_barriers.east) -- ++(0.6, 0) node[right, align=center]{no \\ success}|-(set_top_barrier.east);
    
    \path [line] (characterise_outer_barriers) -- (set_outer_barriers);
  
    \node[1Dstage, below of=set_outer_barriers, fill=myred!20, node distance = 1.2cm](plungers){Characterize plungers};

 \path [line] (set_outer_barriers) -- (plungers);
 
   \path[line] (plungers.west) -- ++(-0.3, 0) node[left, align=center, yshift=-0.1cm]{no \\ success}|-(set_outer_barriers);
   
     \node[2Dstage, below of=plungers,  fill=myred!20, node distance = 1.6cm] (charge_diagram) {Characterize charge stability diagram};
     
 \path[line] (plungers) -- node[align=left, text width=1.5cm]{success}(charge_diagram);
    \path[line] (charge_diagram.west) -- ++(-0.3, 0) |-(set_outer_barriers);

  \node[my_decision, below of=charge_diagram, node distance = 2cm](classification){Classify \\ charge state};
  \path[line] (charge_diagram) -- (classification);

        \node[init_return, below of=classification, node distance = 2cm](success){Success};
     
      \path[line] (classification) -- node[align=left, text width=1.5cm]{good \\double dot}(success);
      
  \node (auxnode0) [text width=4cm, right of=classification, node distance = 2cm] {};
       
   \node[action_simple, right of=auxnode0, fill=myblue!20, node distance = 1.5cm](change_central_barrier){Change central barrier};
   \path[line] (classification) -- node[below, yshift=-0.3cm, align=center]{good \\single dot}(change_central_barrier);

   \path[line] (change_central_barrier) |- node[below, yshift=-4.9cm,  xshift=0.6cm, align=center]{no \\ success}(set_top_barrier.east);
      \path[line] (change_central_barrier) |- node[near end,above] {success}(plungers.east);

     
     \path[line] (classification.west) -- ++ (-1.6, 0) node[near start, below, yshift=-0.3cm, xshift=-0.4cm, align=center]{no \\ good dot} |- (set_top_barrier);
     

%
%

\end{tikzpicture}}
 \caption{The dot tuning algorithm illustrated here for targeting a double dot regime. The colour encoding serves as a guide to which gates are characterized. Characterizing steps are implemented in software modules introduced in \autoref{sec:pinchoff} and \autoref{sec:charge_diagram}.
 Individual gate characterization and charge diagram modules are combined to tune working devices. No prior knowledge of valid voltages is assumed. Time efficient 1D measurements determine suitable voltage values for barriers and narrow down valid ranges for plungers before measuring a 2D charge stability diagram. Using three binary classifiers the quality and dot regime is predicted, based on which next tuning step is chosen.
 }
    \label{fig:tuning_algorithm}
\end{figure}
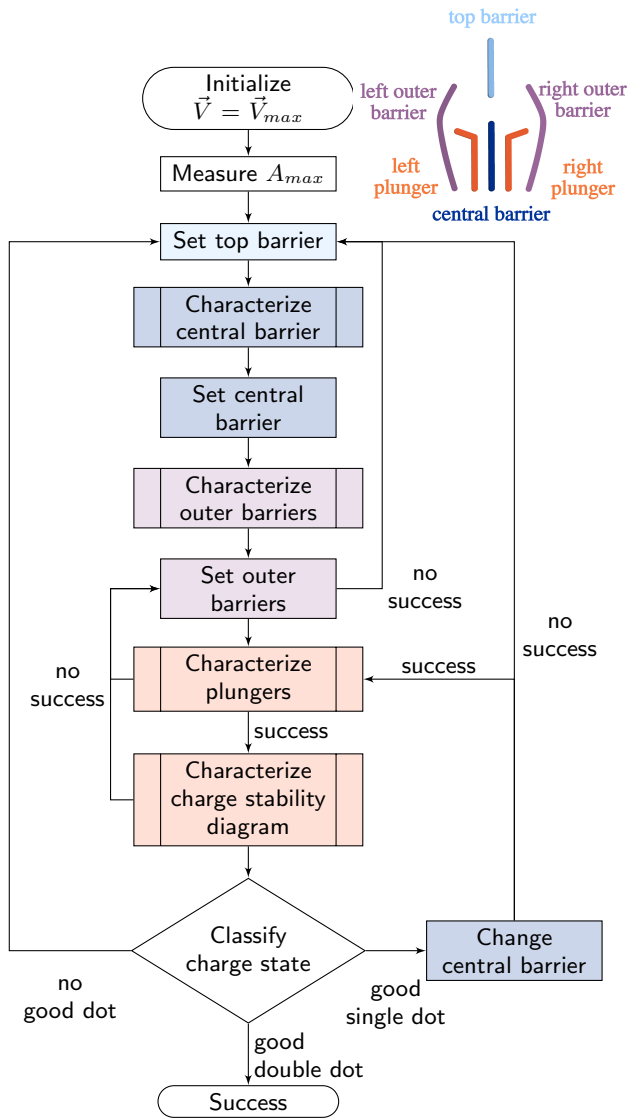
\begin{figure*}[!t]
\begin{tabular}{p{0.32\columnwidth}p{0.32\columnwidth}p{0.32\columnwidth}}
    \subfloat{%
        \noindent\stackinset{r}{16pt}{b}{22pt}{ 
		\includegraphics[width=0.055\columnwidth]{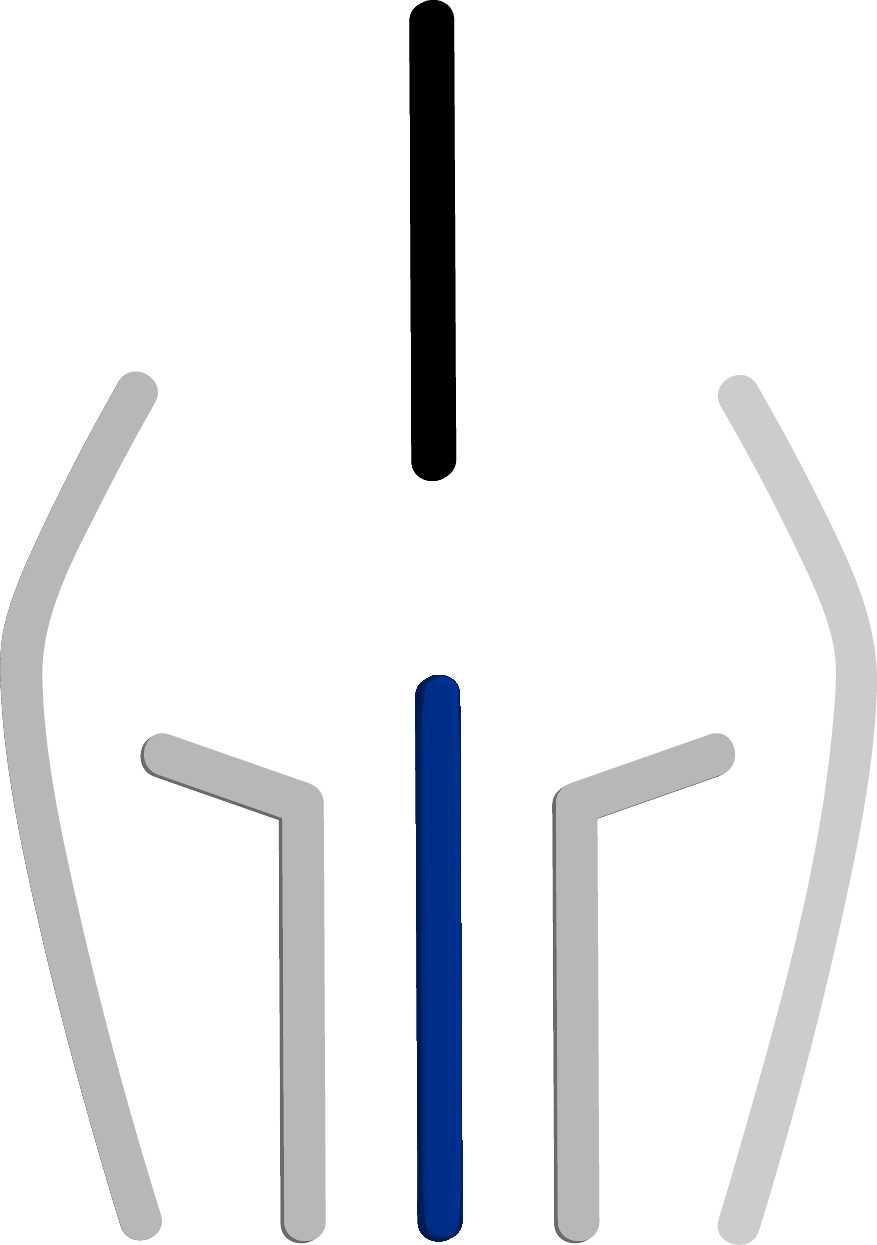}
  		}
  		{ \includegraphics[width=0.32\columnwidth]{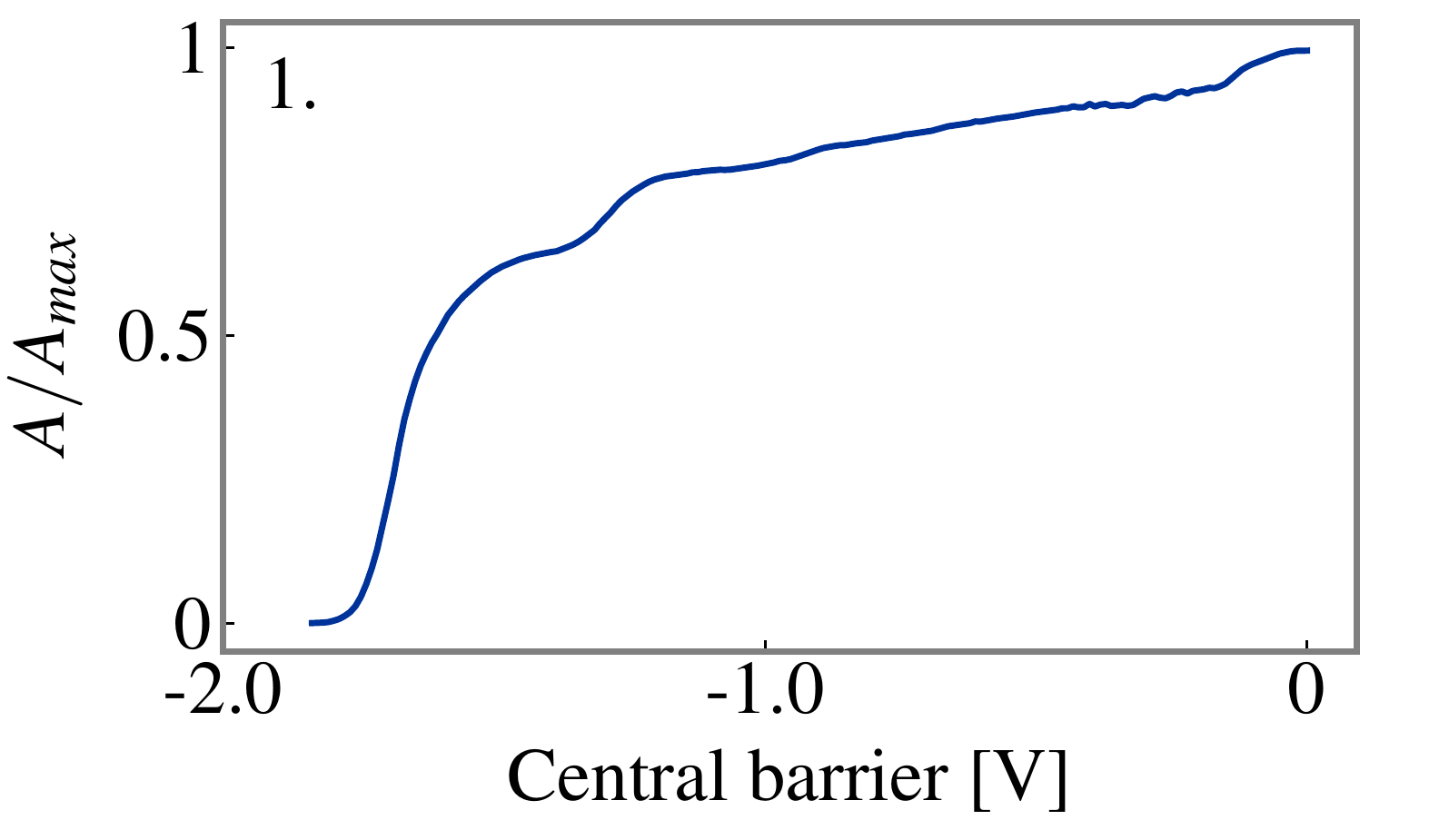} }
		 }& 
    \subfloat{%
        \noindent\stackinset{r}{16pt}{b}{22pt}{ 
		\includegraphics[width=0.055\columnwidth]{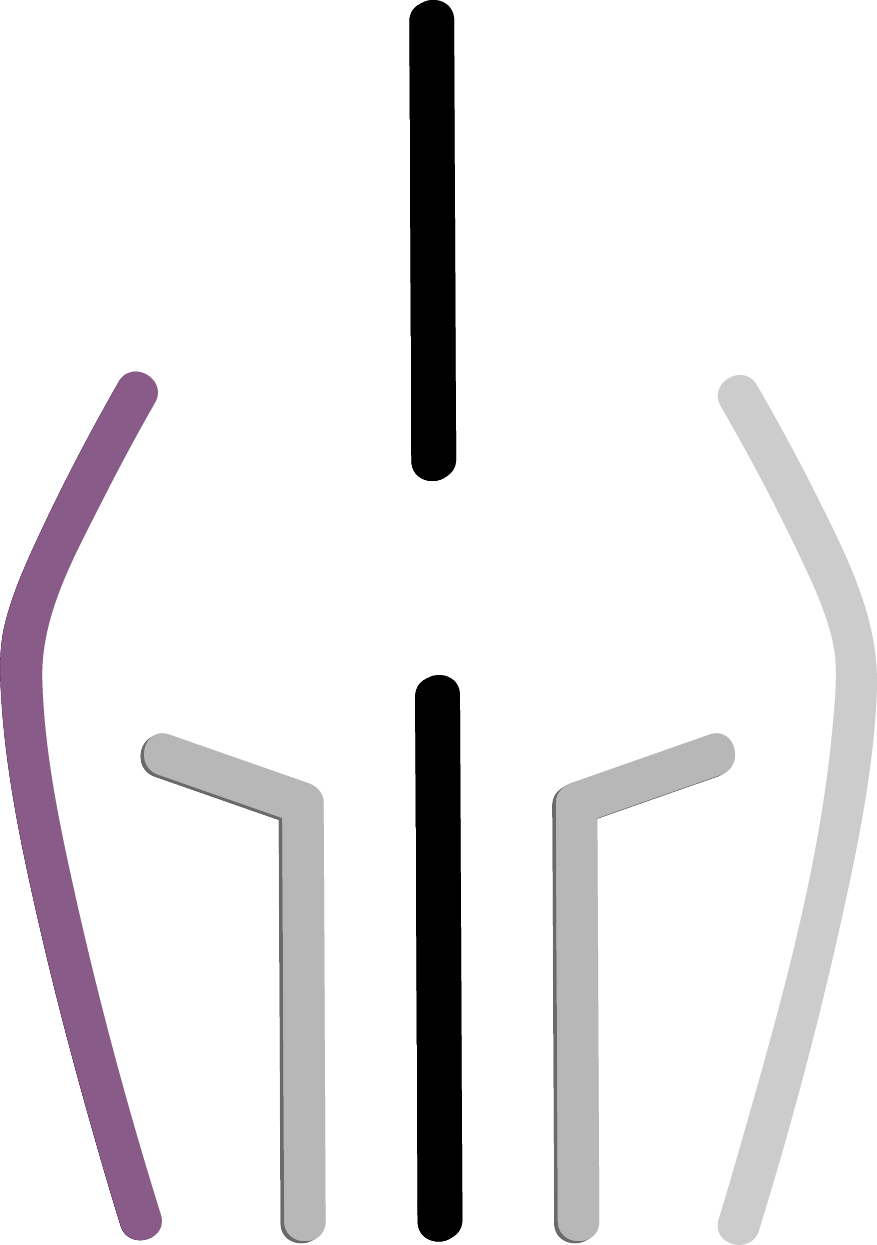}}
  		{ \includegraphics[width=0.32\columnwidth]{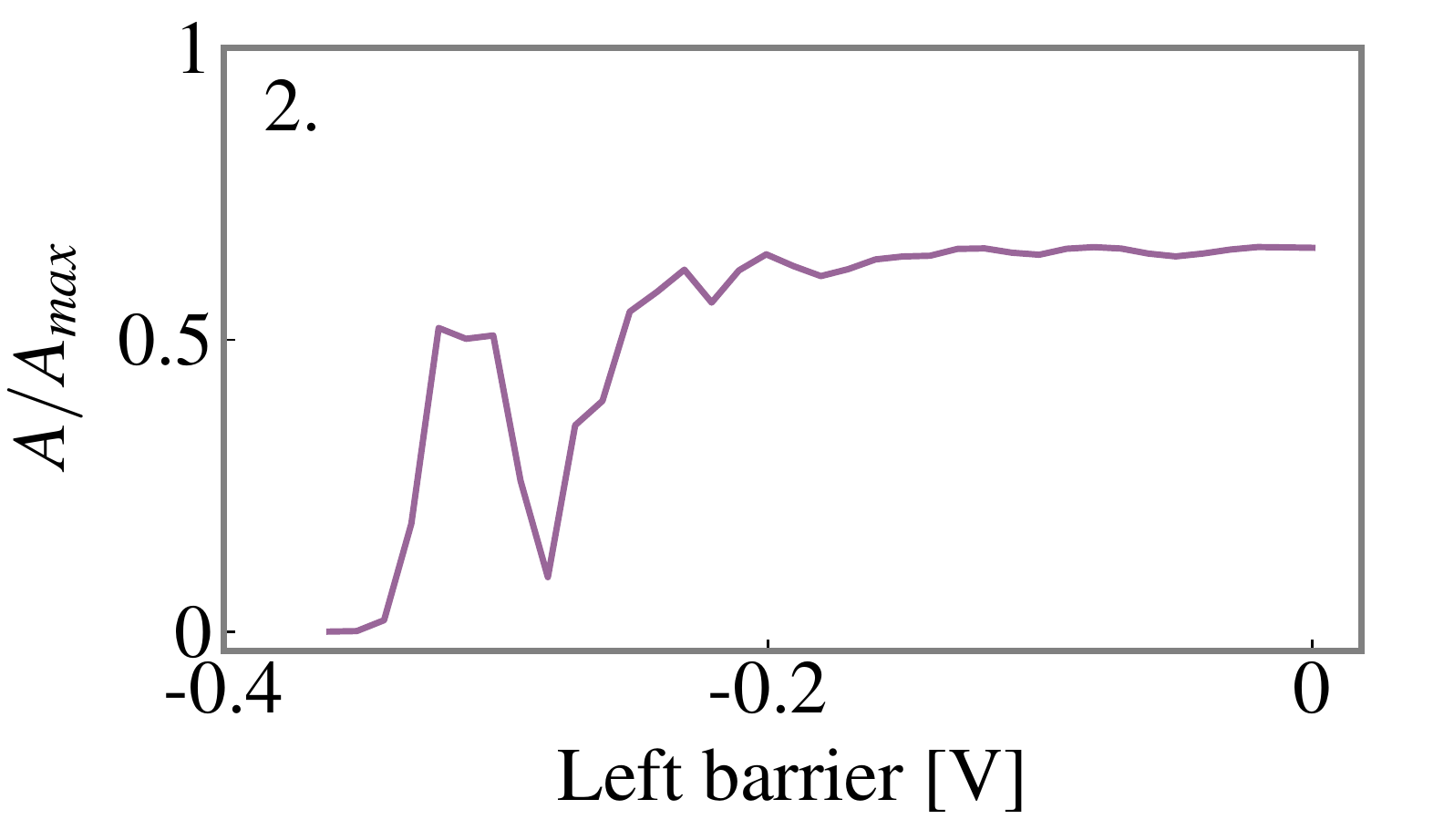} }
		 }  & 
\subfloat{%
        \noindent\stackinset{r}{16pt}{b}{22pt}{ 
		\includegraphics[width=0.055\columnwidth]{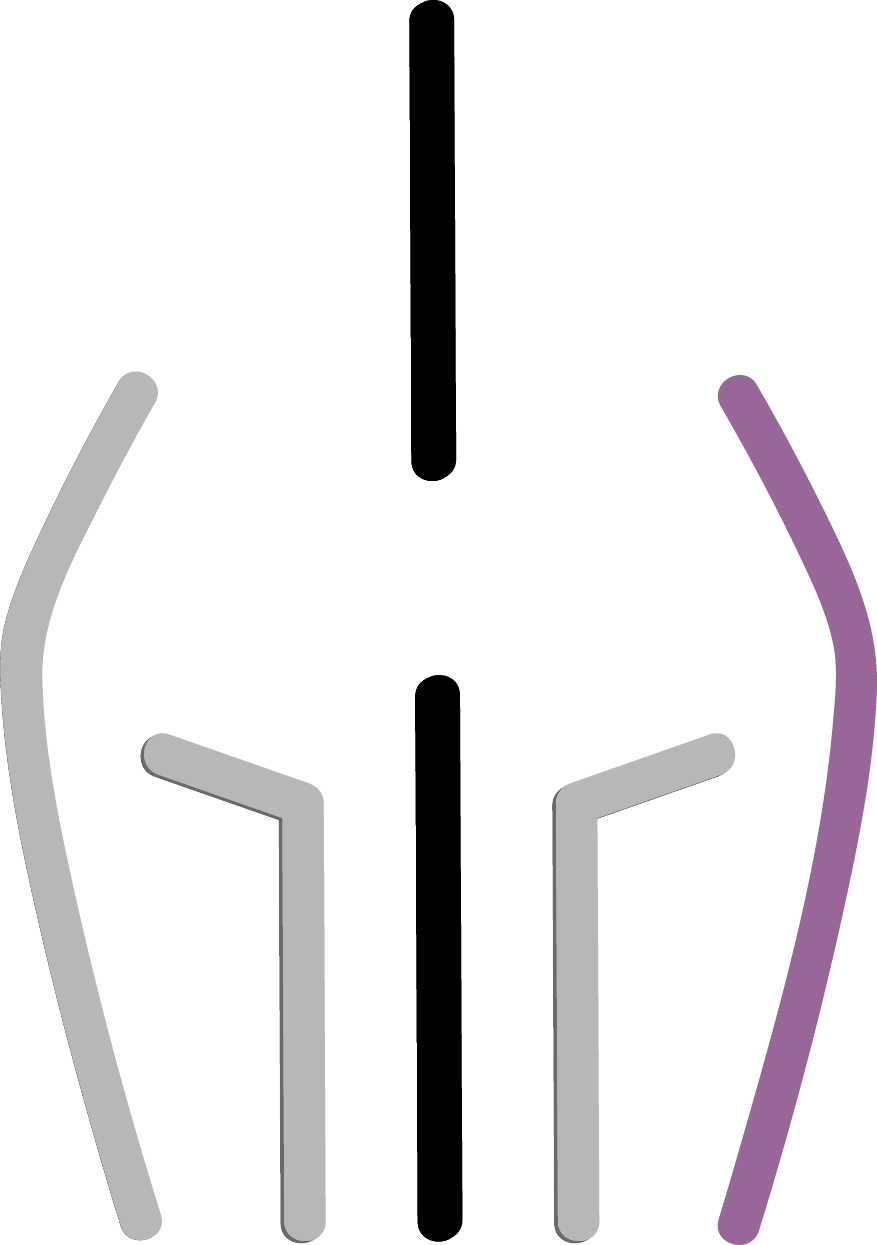}}
  		{ \includegraphics[width=0.32\columnwidth]{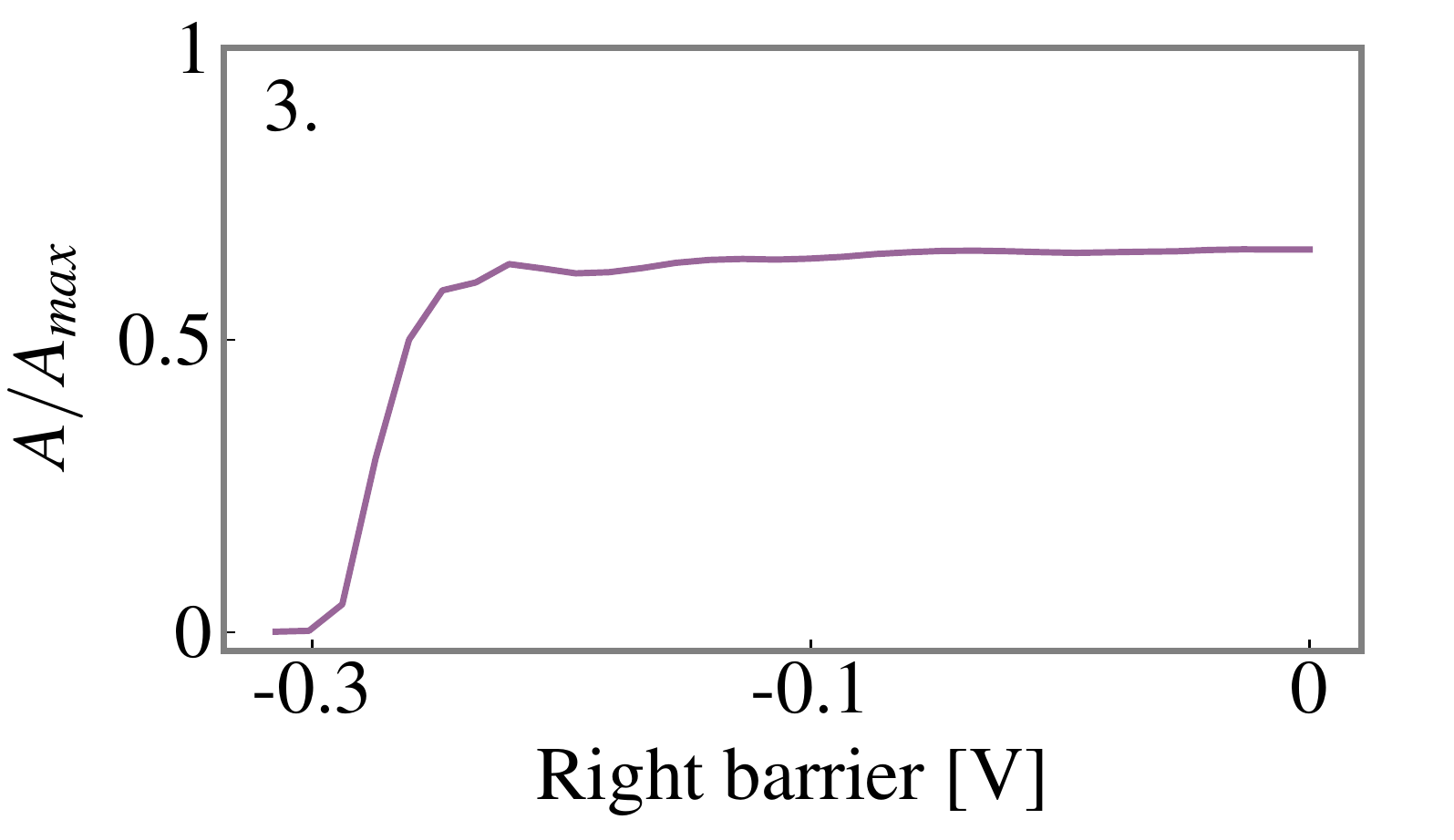} }
		 }  \\
    \subfloat{%
        \noindent\stackinset{l}{28pt}{t}{16pt}{ 
		\includegraphics[width=0.055\columnwidth]{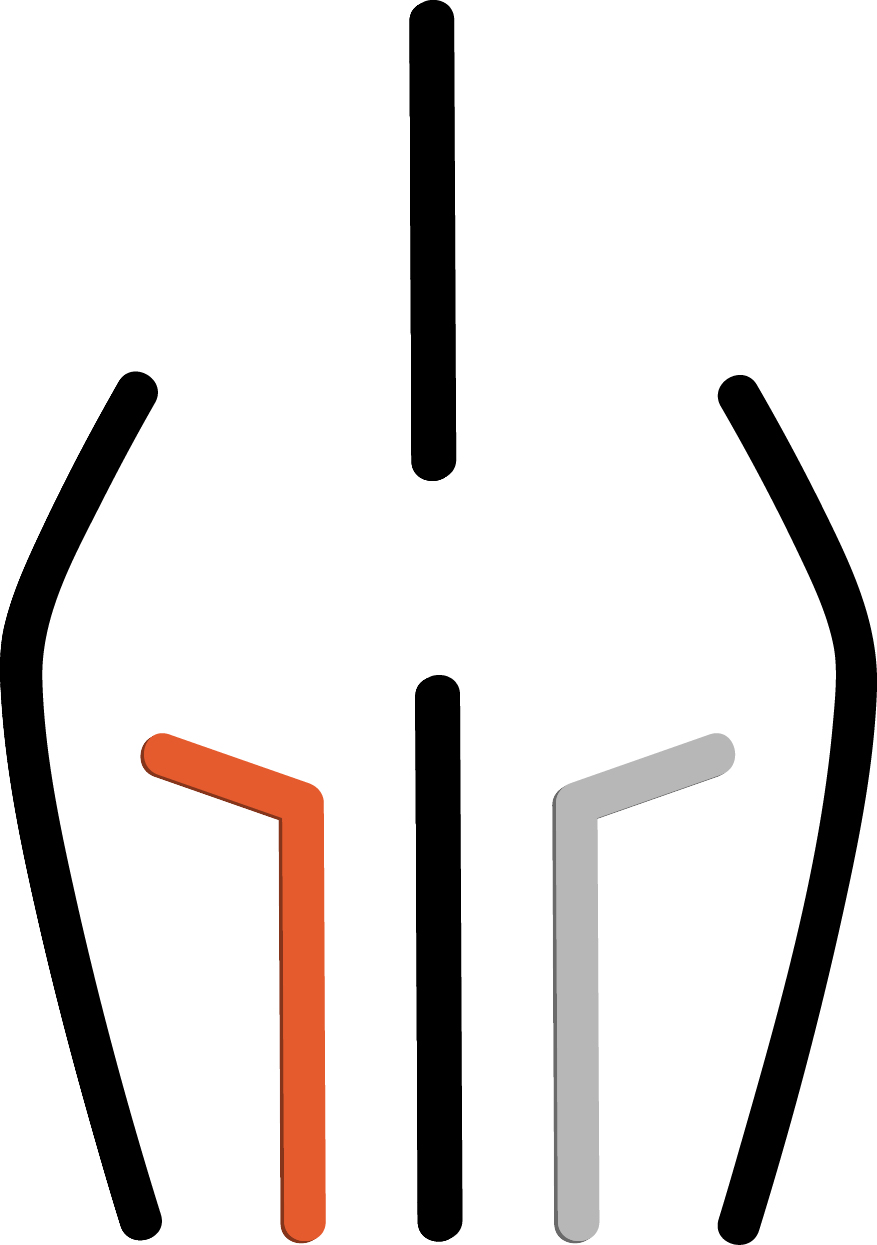}}
  		{ \includegraphics[width=0.32\columnwidth]{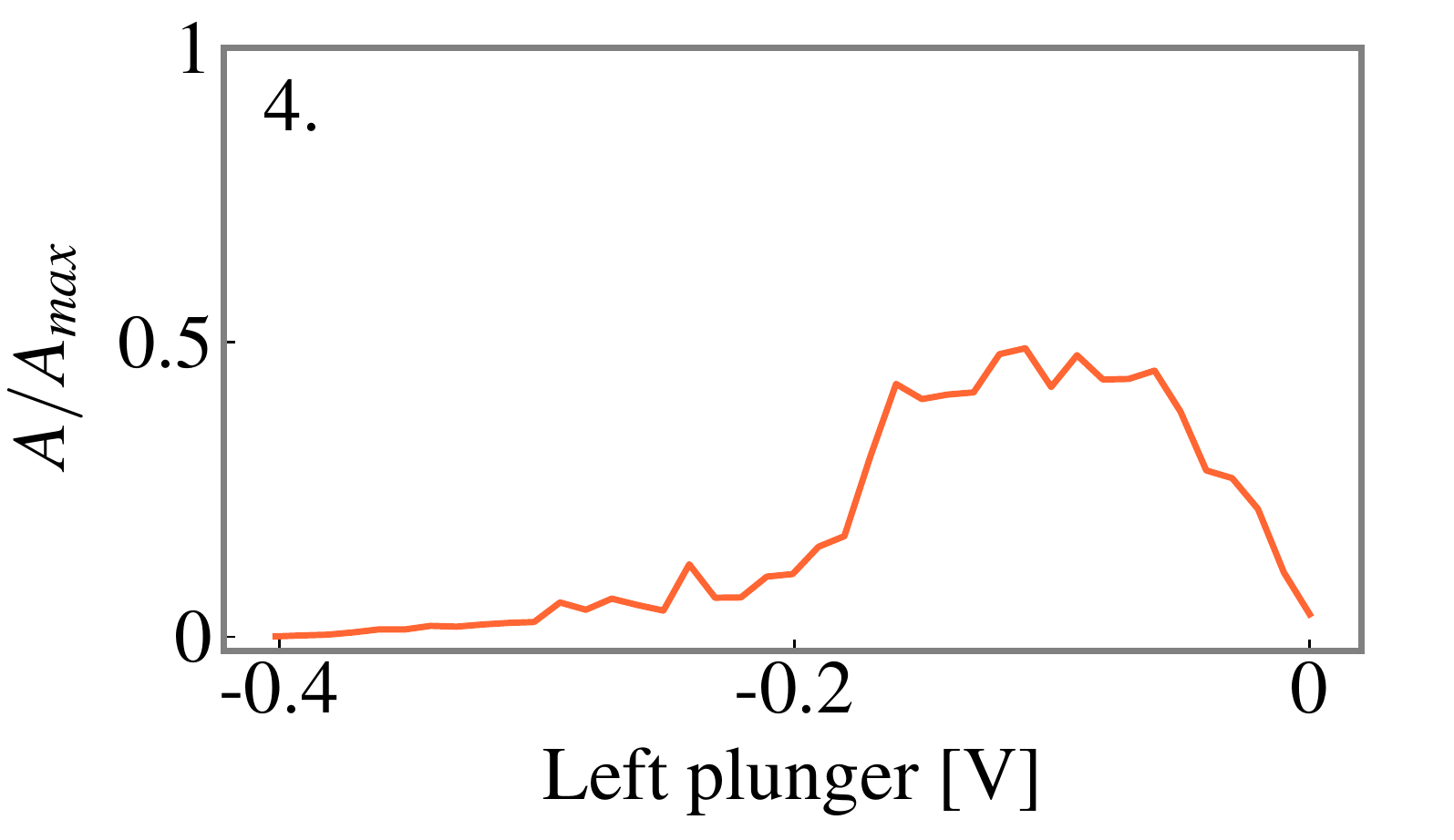} }
		 }  & 
    \subfloat{%
        \noindent\stackinset{l}{28pt}{t}{16pt}{ 
		\includegraphics[width=0.055\columnwidth]{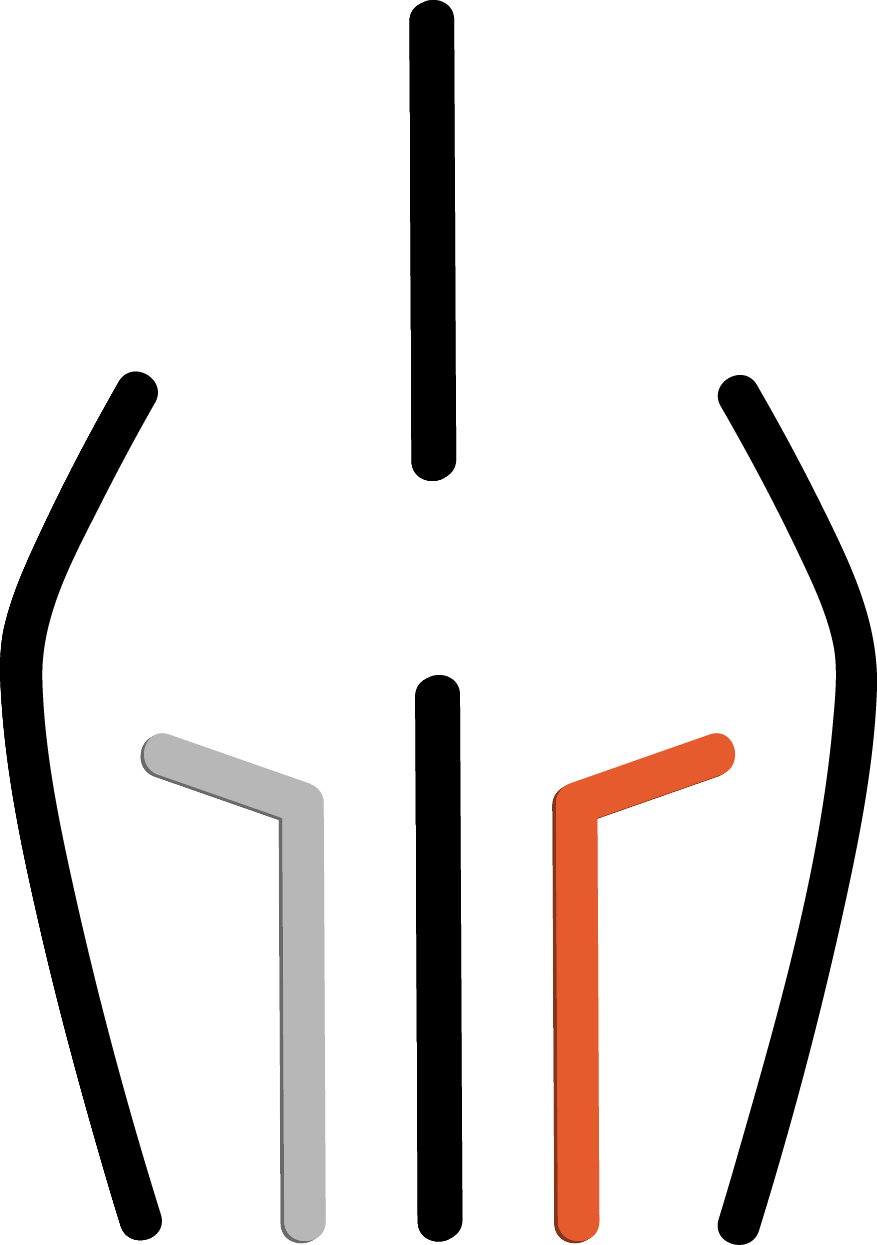}}
  		{ \includegraphics[width=0.32\columnwidth]{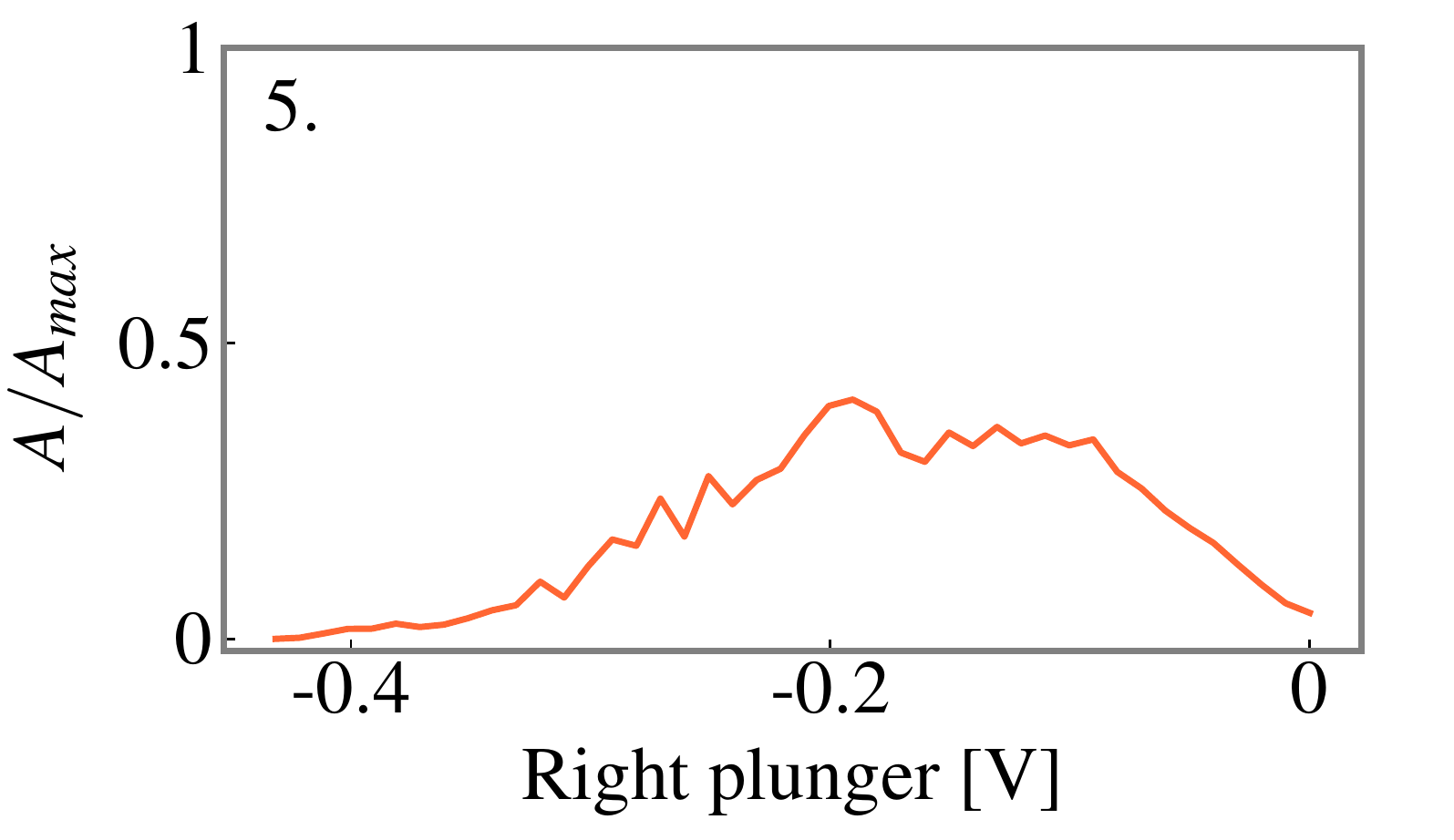} }
		 }  & 
    \subfloat{%
          \noindent\stackinset{l}{32pt}{b}{22pt}{ 
		{\setlength{\fboxsep}{1pt}\colorbox{white}{\includegraphics[width=0.055\columnwidth]{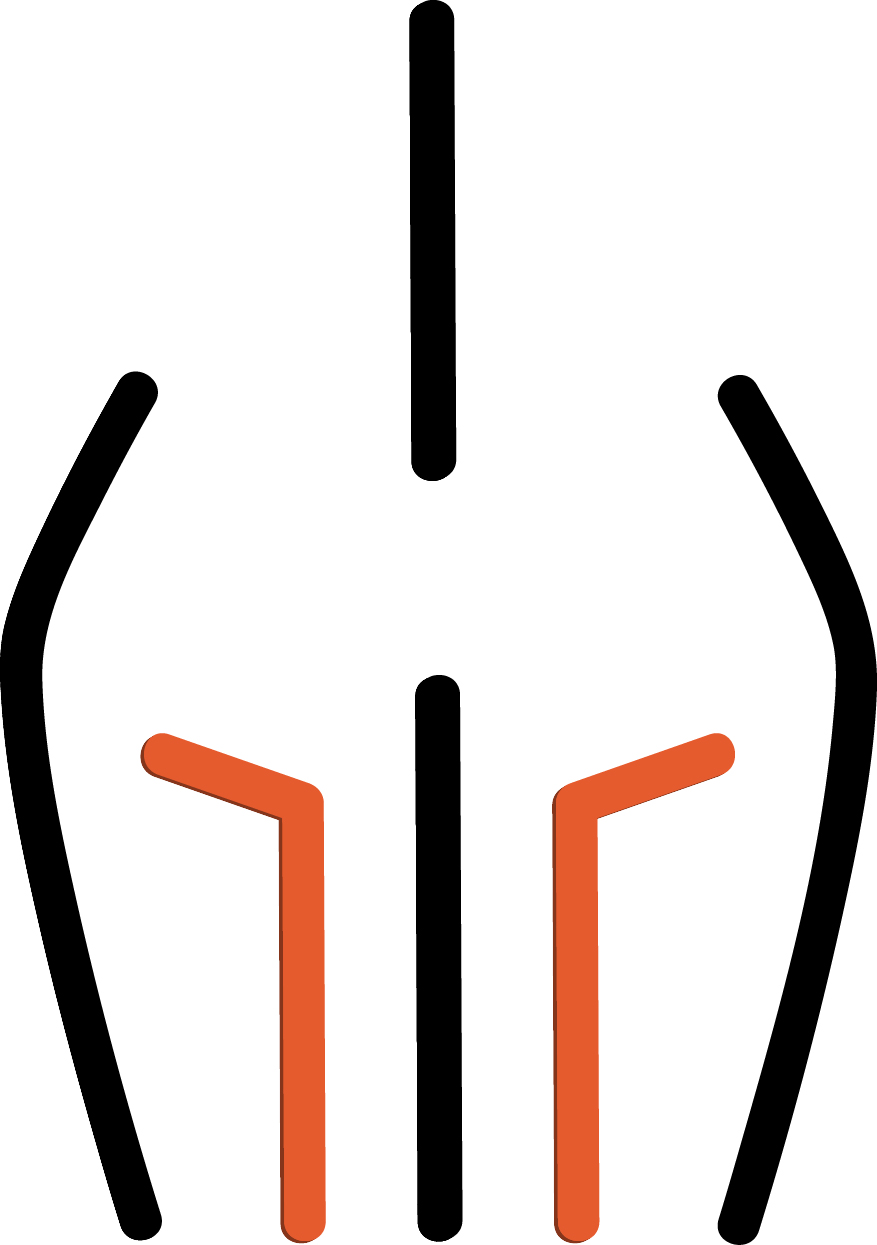}}}}
  		{ \includegraphics[width=0.32\columnwidth]{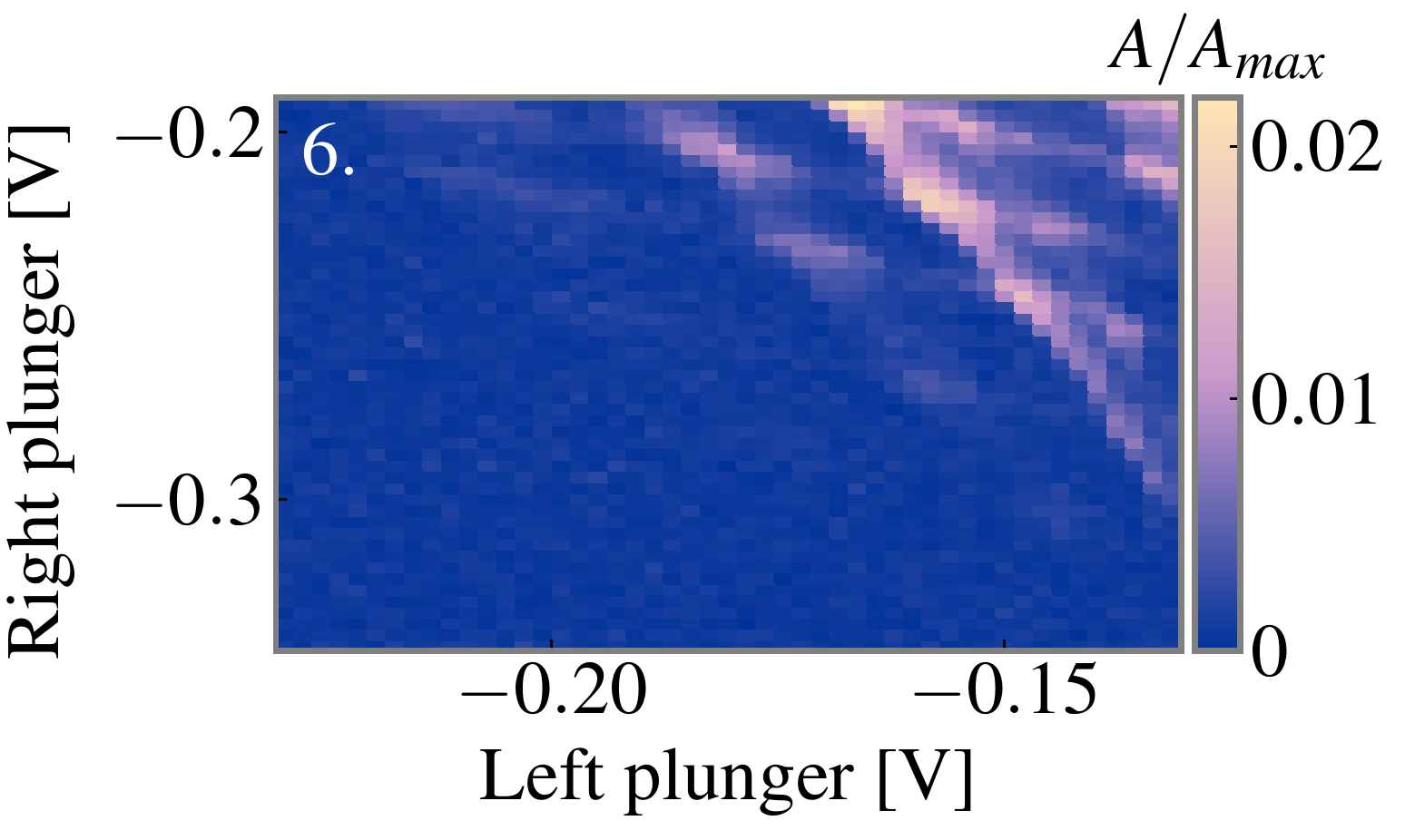} }
		 }  
  \end{tabular}
  \caption{Example of a successful tune-up. The sequence shows the tune-up of device 3.B during the second cooldown in \autoref{tab:results}, which summarizes all device characterizations and tune-ups.  The normalized current, amplitude $A/\smax$, is plotted as a function of gate voltage and the device schemes in the insets indicate which gates were swept with inactive gates set to $0 \:V$ marked in grey. Based on the top barrier's initial valid range, it is set to $\V[\text{TB}] = -1.4 \:V$ before characterizing the remaining gates. The central barrier is set to $\V[\text{CB}] = -1.3444\: V$, the left barrier to $\V[\text{LB}] = -0.2776\: V$ and the right barrier to $\V[\text{RB}] = -0.25753 \:V$. Plunger ranges are narrowed down to $\validrange[\text{LP}] = [-0.2006, -0.1304] \:V$, $\validrange[\text{RP}] = [-0.3411, -0.1906] \:V$ before measuring a charge diagram shown in the bottom right figure. Using binary classifiers it is identified as a good double dot regime, the required condition to stop the tuning. }
 \label{fig:successful_tuneup}
\end{figure*}
First, the algorithm initializes the device by setting all gates to their upper safety range $\safemax$ and measures the saturation current $\smax$. It then chooses a voltage for the top barrier. Values within $[\validmin[\text{TB}], \validmax[\text{TB}]]$ are considered promising values for a double dot regime. More positive top barrier voltages are compensated for by more negative voltages on the remaining barriers.
%
Initially, the top barrier is set to 
\be
\V[\text{TB}] = \validmax[\text{TB}] - \frac{1}{4}  (\validmax[\text{TB}] - \validmin[\text{TB}]),
\ee
at which the central barrier is characterized. Its initial value is set to the voltage corresponding to the current being at 75\% of the saturation current,
\be
\signal(\V[\text{CB}]) = 0.75 \: \smax.
\ee
The left and right barriers are characterized individually, with the respective other held at its highest allowed voltage. They are set to
\be
\V[i] = \validmin[i]+ \frac{1}{3}  (\validmax[i]- \validmin[i]),  \quad i = \text{LB, RB}
\ee
The plungers' active voltage ranges $[ \vL[\text{LP}], \vH[\text{LP}]], [ \vL[\text{LP}], \vH[\text{LP}]]$ are determined by individual gate characterizations before measuring a charge stability diagram. The charge stability module adjusts plunger ranges to keep average currents within $[0.004 \:\smax, 0.1\: \smax]$ as discussed in \autoref{sec:charge_diagram}. If this is not successful, meaning one or both plunger ranges reached the gate's safety limit, the respective neighbouring outer barrier voltage is updated. The voltage change is calculated by the following rule:
\be
\begin{split}
\label{eq:gate_change}
\widetilde{\V[i]^{\delta}} &= 
\begin{dcases*}
0.5 \: ( \validmax[i] - \V[i]) ,& \text{if current is too low} \\
0.5 \: (  \V[i] - \validmin[i]), & \text{if current is too high}
  \end{dcases*} \\[0.5ex]
\V[i]^{\delta} & = \min{\Big \{ 0.1,  \enspace \max{\left \{ \widetilde{\V[i]^{\delta}} , \enspace 0.05 \right \}} \Big  \}}.
\end{split}
\ee
The new barrier voltage $\V[i]^{\text{new}}$ is then set to:
\be
\begin{split}
\label{eq:gate_change}
\V[i]^{\text{new}} &= 
\begin{dcases*}
\V[i] + \V[i]^{\delta} ,& \text{if current is too low} \\
\V[i] - \V[i]^{\delta}, & \text{if current is too high}.
  \end{dcases*} \\[0.5ex]
\end{split}
\ee
If any of the outer barrier's new voltages are within $0.1\:V$ of their safety range, a new top barrier is chosen. Using the same rule as above, the top barrier is set more negative if a lower safety limit has been reached, more positive otherwise.

If the charge diagram module is successful in measuring a diagram with the desired current strength, classification of its segments guides further actions.
 If one or more segments have been classified as a good single dot, the algorithm adjusts the central barrier to more negative values using \autoref{eq:gate_change}. Conversely, if aiming for single dot regime a segment is classified as good double dot, the central barrier voltage is increased.
 If this new value is within $0.05\:V$ of the central barrier's safety range, the top barrier is changed instead, following the same rule as above.
Unless the top barrier has changed, the algorithm resumes with plunger characterizations followed by a new charge diagram. If a new top barrier voltage was chosen, it continues with outer barrier characterizations.
If none of the segments features a good regime the algorithm recommences by choosing a different initial top barrier. If the average current of the last charge diagram is below $0.15 \:\smax$ of the saturation current, the new value is chosen more positive using \autoref{eq:gate_change}, more negative otherwise.

The algorithm stops when at least one segment of the charge diagram is classified as the desired dot regime.
We found it useful to continue iterating after finding the right regime for the first time, as the algorithm often found even better defined double dots in subsequent iterations.

\section{Results}\label{sec:results}

To test our characterization and tuning algorithms we used 12 double quantum dot devices over two thermal cycles. An example of device characterization is shown in \autoref{fig:characterization}, and for double dot tuning in \autoref{fig:successful_tuneup}.

Of the six pairs of devices of our chip we bonded the four pairs in the corners: 1.A, 1.B, 3.A, 3.B, 4.A, 4.B, 6.A, 6.B, as shown in \autoref{fig:qdp_chip}.
 Devices 6.A and 6.B never had currents above noise level in neither of the two cool downs. These were thus identified as broken, which was either due to a contact failure within the device, or within the wiring of the experimental setup.
Each of the remaining devices was first characterized using five individual gate characterization steps and classified using a binary classifier. Device 1.B was identified as broken as no gate was able to pinch off due to an unresponsive top barrier, and hence not tuned. All remaining devices were tuned using the dot tuning algorithm and all but two tune-ups successfully reached the double dot regime within the set number of iterations. Dielectric charging impeded tunability of device 4.A and 4.B in the first cooldown and these runs were stopped before reaching a maximum number of 2D measurements used as stopping criterion. We were also not able to tune these devices manually, however both recovered after a thermal cycle.

The tuning results  as well as the number of 1D and 2D measurements required to complete the tasks are summarized in \autoref{tab:results}.
The number of 1D and 2D measurements for dot tuning vary from short tune-ups of five 1D and one 2D measurements to fifteen and six respectively. Using a lock-in excitation voltage to measure DC transport through the device, 1D measurements took 20 - 90 seconds each while 2D measurements took approximately 25 minutes each.

\begin{figure}[!t]
\includegraphics[width=\columnwidth]{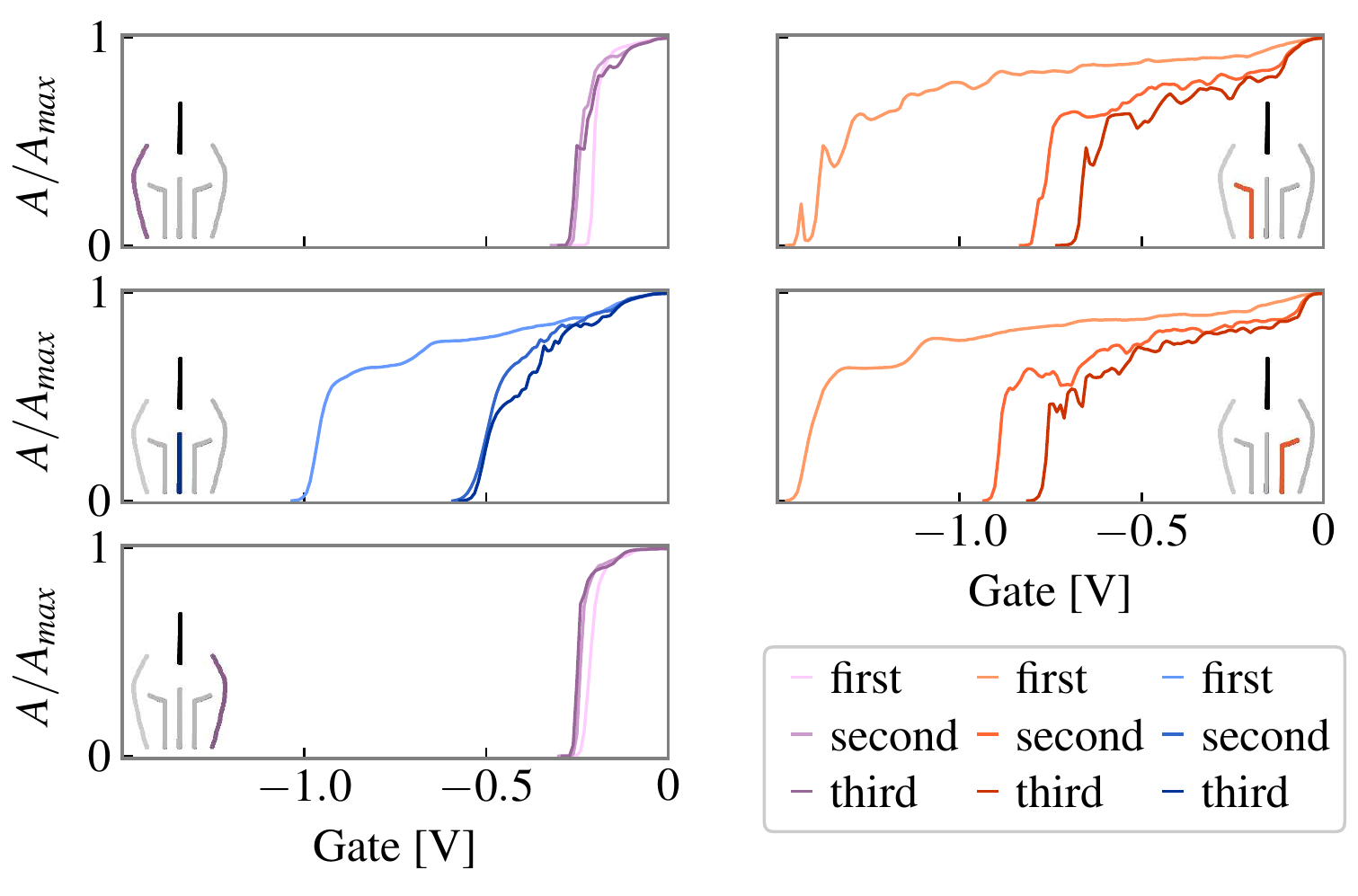}
 \caption{Evolution of individual gate characterizations over three consecutive characterization and tuning iterations. Light colours correspond to the first tuning of the cooldown, medium dark and dark to second and third respectively. The device schemes in the insets indicate which gates were swept with inactive gates set to $0 \:V$ marked in grey. The top barrier voltage was at $\safemin[\text{TB}] = -3\:V$ for all sweeps.
 Pinch-off voltages of outer barriers do not change significantly while the central barrier and plungers show variations which decrease at each consecutive run.}
    \label{fig:characterize_runs}
\end{figure}

Automation allows us to easily study the evolution of device characteristics over time, tuning iterations, thermal cycles or other experimental variables. As an example, we show in \autoref{fig:characterize_runs} the evolution of individual gate characterizations over three consecutive characterization and tuning iterations. We find that while the outer barriers do not change significantly, the central barrier and plungers show a large variation between the first and second tuning and a smaller one between the second and third. We speculate that electronic defects within the heterostructure become passivated during the first tuning iteration, and that further iterations have a smaller effect. The number of iterations needed to reach stability depends on the active area of the channel between two gates being pinched off. This explains the observed differences between outer barriers, the central barrier and plungers over successive tunings.

\begin{table*}[!t]

\setlength{\tabcolsep}{4pt}
\renewcommand{\arraystretch}{1.1}
\begin{tabular}{@{\extracolsep{4pt}}ccccccccccccc} 
 \hline
  \hline
   & \multicolumn{6}{c}{cooldown 1} & \multicolumn{6}{c}{cooldown 2} \\
    \cline{2-7} \cline{8-13}
    &  & \multicolumn{2}{c}{characterization} & \multicolumn{3}{c}{tuning} &  &\multicolumn{2}{c}{characterization} & \multicolumn{3}{c}{tuning}  \\
     \cline{3-4} \cline{5-7}  \cline{9-10} \cline{11-13}
  \multirow{1}{*}{device}  & i.q.a  & $n_{\text{1D}}$ &  quality & $n_{\text{1D}}$  & $n_{\text{2D}}$  & success &  i.q.a & $n_{\text{1D}}$ & quality & $n_{\text{1D}}$  & $n_{\text{2D}}$ & success  \\
   \hline
   \hline
    1.A & \cmark   & 5(+2) + 16 & \cmark &13 & 5 & \cmark & 
       \cmark   &  5 + 15 &  \cmark & 15 & 6 & \cmark  \\ 
    1.B & \cmark   & 5(+2) +  \hfill -  & \xmark &  - & - & - &%
         \cmark   &5 + \hfill - & \xmark  &  - & - & -  \\%
    3.A & \cmark   & 5(+2) + 19 & \cmark  &  7 & 2 & \cmark & 
        \cmark   &  5 + 13 &  \cmark & 11 & 5 & \cmark   \\ 
    3.B & \cmark   & 5(+2) + 20 & \cmark  &  7 & 2 & \cmark & 
          \cmark   &  5 + 15 &  \cmark &  5 & 1 & \cmark   \\ 
    4.A& \cmark  & 5(+2) + 11 & \cmark  &  59 & 19 & \xmark & 
       \cmark   &   5 + 14 &  \cmark &   11 & 4 & \cmark  \\ 
    4.B & \cmark   & 5(+2) + 12 & \cmark  &  21 & 9 & \xmark & 
          \cmark   &  5 + 14 &  \cmark &  15 & 6 & \cmark   \\ 
    6.A & \xmark   & - & -  &  - & - & - & %
         \xmark   &   - &  - &  - & - & -   \\
    6.B & \xmark   & - & -  &  - & - & - & 
         \xmark   &   - &  - &  - & - & -   \\
 \hline
 \hline
\end{tabular}
\caption{Summary of initial quality assessment (i.q.a), device characterization and dot tuning of all devices measured. Device characterization of the first cooldown include two additional individual gate characterizations to determine a suitable top barrier to set. These sweeps were made redundand in the final characterization algorithm by choosing the top barrier's safety limit $\safemin[\text{TB}]$.
Device 1.B was correctly identified as broken and thus not tuned. Devices 4.A and 4.B did not reach the desired dot regime in the first cooldown due to dielectric charging which impeded tuneability. These runs were stopped manually before reaching the maximum number of measurements used as stopping criterion. However, both devices recovered after a thermal cycle. Devices 6.A and 6.B did not pass the initial quality assessment due to a lack of current, hence saturation current $\smax$.
The number of measurement to establish valid top barrier ranges during the second stage of device characterization, omitted in other device designs, range from 11 to 25 gate characterizations. 
Using standard lock-in techniques, the times to measure 1D gate traces or 2D charge stability diagrams were between 20 and 90 seconds, or 25 minutes respectively.}
\label{tab:results}
\end{table*}

To determine the best classifiers we compare Logistic  Regression,  Support  Vector  Machines  (SVM), Multilayer Perceptron (MLP), Gaussian Process, Decision Tree,  Random Forest,  Quadratic Discriminant Analysis and $k$-Nearest Neighbors (KNN) performances for various input data: normalized current map and/or its Fourier transform and with or without PCA. The results are detailed in \autoref{ax:classifiers} and show that performances reach up to 95 \% test accuracy for individual gate characterization but stay below 90 \% for charge state detection. Using principle components as input data either does not effect or decreases accuracy while not significantly improving training and prediction times for any category.
Based on these results we choose the Decision Tree classifier applied to features to predict gate responses of individual gate characterizations. To assess the charge state of a charge stability diagram we use all three classifiers to determine three possible outcomes: Good single dot, good double dot or no dot. We use the redundancy to overcome low accuracy scores of individual classifiers. First, single and double dot quality classifiers assess whether a good charge state is present: if both outcomes are negative, there is no dot. If one is positive, the good regime it is indicating is present and if both predict good regimes the dot regime classifier establishes a clear outcome.

The Multilayer Perceptron classifier performs best on all three charge state classification categories. 
Single and double dot qualities are best predicted using both current maps and their Fourier transform. Dot regime classification performs equally well on FT alone, suggesting that the FT captures the charge transitions pattern differences of single and double dot regimes. This is not unexpected as as single dot give rise to one dimensional periodic structures while double dots give rise to two dimensional periodic structures.

\section{Conclusion}
The tuning algorithm presented here establishes an autonomous procedure to tune gate defined quantum dots, facilitating qubit initialization. This paves the road for quantum scale up in quantum dot systems, where the growing chip size and complexity make manual tuning procedures impractical.
By automating well established manual tuning procedures and using binary classifier to transfer scientists' knowledge of quantum dot devices, we implemented an autonomous two-stage device characterization and dot-tuning process. These algorithms enable to characterize and tune devices in parallel and on a large scale without any pre-measured input.

While we were able to successfully tune a range of devices across various scenarios, there are many opportunities for further improvements. 
Finding the double dot regime can be a starting point for a subsequent local search to optimize the quality of the double dot. Fitting to a capacitance model can also be used during the search.

In order to make semiconductor qubit initialization fully autonomous from cooldown to qubit operation, setup specific measurement initialization and calibration  needs to be addressed, as well as failure handling. Existing automation efforts, such as fine tuning of the inter-dot tunnel coupling,  and virtual gate definition \citep{vanDiepen_automation, doi:10.1063/1.5121444} could then also be integrated into our workflow.

Taking advantage of multiplexing technologies will allow us to characterize more devices without being limited by the number of control lines. 
Replacing slow lock-in measurements by high frequency charge sensing will decrease run times and noise, while frequency multiplexing will further improve throughput \citep{Hornibrook2014}. 
Turning to other measurement techniques will require appropriate feature selection and data preprocessing to ensure correct charge regime classification, similar to what has been implemented here. 

Deeper investigations into feature selection, including voltage resolution and segment size, classifier choice and hyper parameter optimization will make tuning more reliable. Understanding why some classifiers perform better than others and which features represent data best will enable improved prediction accuracy. More sophisticated models in combination with boosting \citep{Schapire1999} may lead to further improvements.

Labelling errors, the partially due to ambiguities between experts, and the vast variety of noise and charge states demands extensive data to capture most measurement outcomes. As this noise is difficult to simulate, classifiers need to be trained with real, not simulated data. 
This enables us to distinguish between noisy and intermediate charge states, a distinction not captured by classifiers trained with ideal single and double dot charge diagrams.
This is especially relevant if we wish to design new classifiers assessing specific tuning failures or  fabrication defects.

Our procedure can be applied to other quantum dot designs, including ones based on nanowires by neglecting the top barrier. The structure of both algorithms is general and can be applied to different materials by making minor changes to the parameters of the algorithms or inverting the polarity of the tuning procedure for holes \citep{Volk2019FastCS, Kawakami:vg}.

The characterization procedure can also be extended to enable material and fabrication optimization, for example by comparing pinch-off voltages or adding gate characterizations sweeping voltages in opposite direction to assess hysteresis. The tuning procedure can be modified to account for more complex devices, such as multi-dot arrays and topological quantum dot structures \citep{ Mills:2019ud, doi:10.1063/1.5025928, PhysRevB.95.235305} or generalized to account for new physical phenomena such as zero bias peaks or finding the topological phase \citep{Albrecht:vq, Mourik1003}.

As this field grows and more measurement data becomes available, there is an opportunity to take advantage of more complex machine learning techniques such as reinforcement learning and Bayesian inference \citep{sutton2018reinforcement, liu2001combined, granade2012robust}.

Going beyond device tuning, machine learning may have applications in areas as diverse as device design, fabrication optimization, measurement and readout improvements or to optimize qubit control and feedback.

\section{Acknowledgments}
We thank Alice C. Mahoney, John M. Hornibrook, Rachpon Kalra and Xanthe G. Croot for technical assistance and  Christopher E. Granade, John K. Gamble, Charles M. Marcus and David J. Reilly for helpful discussions and critical feedback.  This research was supported by the Microsoft Corporation and the Australian Research Council Centre of Excellence for Engineered Quantum Systems (EQUS, CE170100009). The authors acknowledge the facilities as well as the scientific and technical assistance of the Research \& Prototype Foundry Core Research Facility at the University of Sydney, part of the Australian National Fabrication Facility.
\newpage

\bibliographystyle{apsrev4-1}
\bibliography{literature}

\begin{thebibliography}{52}%
\makeatletter
\providecommand \@ifxundefined [1]{%
 \@ifx{#1\undefined}
}%
\providecommand \@ifnum [1]{%
 \ifnum #1\expandafter \@firstoftwo
 \else \expandafter \@secondoftwo
 \fi
}%
\providecommand \@ifx [1]{%
 \ifx #1\expandafter \@firstoftwo
 \else \expandafter \@secondoftwo
 \fi
}%
\providecommand \natexlab [1]{#1}%
\providecommand \enquote  [1]{``#1''}%
\providecommand \bibnamefont  [1]{#1}%
\providecommand \bibfnamefont [1]{#1}%
\providecommand \citenamefont [1]{#1}%
\providecommand \href@noop [0]{\@secondoftwo}%
\providecommand \href [0]{\begingroup \@sanitize@url \@href}%
\providecommand \@href[1]{\@@startlink{#1}\@@href}%
\providecommand \@@href[1]{\endgroup#1\@@endlink}%
\providecommand \@sanitize@url [0]{\catcode `\\12\catcode `\$12\catcode
  `\&12\catcode `\#12\catcode `\^12\catcode `\_12\catcode `\%12\relax}%
\providecommand \@@startlink[1]{}%
\providecommand \@@endlink[0]{}%
\providecommand \url  [0]{\begingroup\@sanitize@url \@url }%
\providecommand \@url [1]{\endgroup\@href {#1}{\urlprefix }}%
\providecommand \urlprefix  [0]{URL }%
\providecommand \Eprint [0]{\href }%
\providecommand \doibase [0]{http://dx.doi.org/}%
\providecommand \selectlanguage [0]{\@gobble}%
\providecommand \bibinfo  [0]{\@secondoftwo}%
\providecommand \bibfield  [0]{\@secondoftwo}%
\providecommand \translation [1]{[#1]}%
\providecommand \BibitemOpen [0]{}%
\providecommand \bibitemStop [0]{}%
\providecommand \bibitemNoStop [0]{.\EOS\space}%
\providecommand \EOS [0]{\spacefactor3000\relax}%
\providecommand \BibitemShut  [1]{\csname bibitem#1\endcsname}%
\let\auto@bib@innerbib\@empty
\bibitem [{\citenamefont {Shor}(1997)}]{Shor:1997:PAP:264393.264406}%
  \BibitemOpen
  \bibfield  {author} {\bibinfo {author} {\bibfnamefont {P.~W.}\ \bibnamefont
  {Shor}},\ }\href {\doibase 10.1137/S0097539795293172} {\bibfield  {journal}
  {\bibinfo  {journal} {SIAM J. Comput.}\ }\textbf {\bibinfo {volume} {26}},\
  \bibinfo {pages} {1484} (\bibinfo {year} {1997})}\BibitemShut {NoStop}%
\bibitem [{\citenamefont {Reiher}\ \emph {et~al.}(2017)\citenamefont {Reiher},
  \citenamefont {Wiebe}, \citenamefont {Svore}, \citenamefont {Wecker},\ and\
  \citenamefont {Troyer}}]{Reiher201619152}%
  \BibitemOpen
  \bibfield  {author} {\bibinfo {author} {\bibfnamefont {M.}~\bibnamefont
  {Reiher}}, \bibinfo {author} {\bibfnamefont {N.}~\bibnamefont {Wiebe}},
  \bibinfo {author} {\bibfnamefont {K.~M.}\ \bibnamefont {Svore}}, \bibinfo
  {author} {\bibfnamefont {D.}~\bibnamefont {Wecker}}, \ and\ \bibinfo {author}
  {\bibfnamefont {M.}~\bibnamefont {Troyer}},\ }\href {\doibase
  10.1073/pnas.1619152114} {\bibfield  {journal} {\bibinfo  {journal} {Proc.
  Natl. Acad. Sci. U.S.A.}\ } (\bibinfo {year} {2017}),\
  10.1073/pnas.1619152114}\BibitemShut {NoStop}%
\bibitem [{\citenamefont {Wecker}\ \emph {et~al.}(2015)\citenamefont {Wecker},
  \citenamefont {Hastings}, \citenamefont {Wiebe}, \citenamefont {Clark},
  \citenamefont {Nayak},\ and\ \citenamefont {Troyer}}]{Wecker2015}%
  \BibitemOpen
  \bibfield  {author} {\bibinfo {author} {\bibfnamefont {D.}~\bibnamefont
  {Wecker}}, \bibinfo {author} {\bibfnamefont {M.~B.}\ \bibnamefont
  {Hastings}}, \bibinfo {author} {\bibfnamefont {N.}~\bibnamefont {Wiebe}},
  \bibinfo {author} {\bibfnamefont {B.~K.}\ \bibnamefont {Clark}}, \bibinfo
  {author} {\bibfnamefont {C.}~\bibnamefont {Nayak}}, \ and\ \bibinfo {author}
  {\bibfnamefont {M.}~\bibnamefont {Troyer}},\ }\href {\doibase
  10.1103/PhysRevA.92.062318} {\bibfield  {journal} {\bibinfo  {journal} {Phys.
  Rev. A}\ }\textbf {\bibinfo {volume} {92}},\ \bibinfo {pages} {062318}
  (\bibinfo {year} {2015})}\BibitemShut {NoStop}%
\bibitem [{\citenamefont {Rungger}\ \emph {et~al.}(2019)\citenamefont
  {Rungger}, \citenamefont {Fitzpatrick}, \citenamefont {Chen}, \citenamefont
  {Alderete}, \citenamefont {Apel}, \citenamefont {Cowtan}, \citenamefont
  {Patterson}, \citenamefont {Ramo}, \citenamefont {Zhu}, \citenamefont
  {Nguyen}, \citenamefont {Grant}, \citenamefont {Chretien}, \citenamefont
  {Wossnig}, \citenamefont {Linke},\ and\ \citenamefont
  {Duncan}}]{Rungger2019}%
  \BibitemOpen
  \bibfield  {author} {\bibinfo {author} {\bibfnamefont {I.}~\bibnamefont
  {Rungger}}, \bibinfo {author} {\bibfnamefont {N.}~\bibnamefont
  {Fitzpatrick}}, \bibinfo {author} {\bibfnamefont {H.}~\bibnamefont {Chen}},
  \bibinfo {author} {\bibfnamefont {C.~H.}\ \bibnamefont {Alderete}}, \bibinfo
  {author} {\bibfnamefont {H.}~\bibnamefont {Apel}}, \bibinfo {author}
  {\bibfnamefont {A.}~\bibnamefont {Cowtan}}, \bibinfo {author} {\bibfnamefont
  {A.}~\bibnamefont {Patterson}}, \bibinfo {author} {\bibfnamefont {D.~M.}\
  \bibnamefont {Ramo}}, \bibinfo {author} {\bibfnamefont {Y.}~\bibnamefont
  {Zhu}}, \bibinfo {author} {\bibfnamefont {N.~H.}\ \bibnamefont {Nguyen}},
  \bibinfo {author} {\bibfnamefont {E.}~\bibnamefont {Grant}}, \bibinfo
  {author} {\bibfnamefont {S.}~\bibnamefont {Chretien}}, \bibinfo {author}
  {\bibfnamefont {L.}~\bibnamefont {Wossnig}}, \bibinfo {author} {\bibfnamefont
  {N.~M.}\ \bibnamefont {Linke}}, \ and\ \bibinfo {author} {\bibfnamefont
  {R.}~\bibnamefont {Duncan}},\ }\href {http://arxiv.org/abs/1910.04735} {\
  (\bibinfo {year} {2019})},\ \Eprint {http://arxiv.org/abs/1910.04735}
  {arXiv:1910.04735} \BibitemShut {NoStop}%
\bibitem [{\citenamefont {Zhang}\ \emph {et~al.}(2017)\citenamefont {Zhang},
  \citenamefont {Pagano}, \citenamefont {Hess}, \citenamefont {Kyprianidis},
  \citenamefont {Becker}, \citenamefont {Kaplan}, \citenamefont {Gorshkov},
  \citenamefont {Gong},\ and\ \citenamefont {Monroe}}]{Zhang2017}%
  \BibitemOpen
  \bibfield  {author} {\bibinfo {author} {\bibfnamefont {J.}~\bibnamefont
  {Zhang}}, \bibinfo {author} {\bibfnamefont {G.}~\bibnamefont {Pagano}},
  \bibinfo {author} {\bibfnamefont {P.~W.}\ \bibnamefont {Hess}}, \bibinfo
  {author} {\bibfnamefont {A.}~\bibnamefont {Kyprianidis}}, \bibinfo {author}
  {\bibfnamefont {P.}~\bibnamefont {Becker}}, \bibinfo {author} {\bibfnamefont
  {H.}~\bibnamefont {Kaplan}}, \bibinfo {author} {\bibfnamefont {A.~V.}\
  \bibnamefont {Gorshkov}}, \bibinfo {author} {\bibfnamefont {Z.-X.}\
  \bibnamefont {Gong}}, \ and\ \bibinfo {author} {\bibfnamefont
  {C.}~\bibnamefont {Monroe}},\ }\href {\doibase 10.1038/nature24654}
  {\bibfield  {journal} {\bibinfo  {journal} {Nature}\ }\textbf {\bibinfo
  {volume} {551}},\ \bibinfo {pages} {601} (\bibinfo {year}
  {2017})}\BibitemShut {NoStop}%
\bibitem [{\citenamefont {Chiesa}\ \emph {et~al.}(2019)\citenamefont {Chiesa},
  \citenamefont {Tacchino}, \citenamefont {Grossi}, \citenamefont {Santini},
  \citenamefont {Tavernelli}, \citenamefont {Gerace},\ and\ \citenamefont
  {Carretta}}]{Chiesa:2019uc}%
  \BibitemOpen
  \bibfield  {author} {\bibinfo {author} {\bibfnamefont {A.}~\bibnamefont
  {Chiesa}}, \bibinfo {author} {\bibfnamefont {F.}~\bibnamefont {Tacchino}},
  \bibinfo {author} {\bibfnamefont {M.}~\bibnamefont {Grossi}}, \bibinfo
  {author} {\bibfnamefont {P.}~\bibnamefont {Santini}}, \bibinfo {author}
  {\bibfnamefont {I.}~\bibnamefont {Tavernelli}}, \bibinfo {author}
  {\bibfnamefont {D.}~\bibnamefont {Gerace}}, \ and\ \bibinfo {author}
  {\bibfnamefont {S.}~\bibnamefont {Carretta}},\ }\href {\doibase
  10.1038/s41567-019-0437-4} {\bibfield  {journal} {\bibinfo  {journal} {Nat.
  Phys}\ }\textbf {\bibinfo {volume} {15}},\ \bibinfo {pages} {455} (\bibinfo
  {year} {2019})}\BibitemShut {NoStop}%
\bibitem [{\citenamefont {Hamilton}\ \emph {et~al.}(2017)\citenamefont
  {Hamilton}, \citenamefont {Kruse}, \citenamefont {Sansoni}, \citenamefont
  {Barkhofen}, \citenamefont {Silberhorn},\ and\ \citenamefont
  {Jex}}]{PhysRevLett.119.170501}%
  \BibitemOpen
  \bibfield  {author} {\bibinfo {author} {\bibfnamefont {C.~S.}\ \bibnamefont
  {Hamilton}}, \bibinfo {author} {\bibfnamefont {R.}~\bibnamefont {Kruse}},
  \bibinfo {author} {\bibfnamefont {L.}~\bibnamefont {Sansoni}}, \bibinfo
  {author} {\bibfnamefont {S.}~\bibnamefont {Barkhofen}}, \bibinfo {author}
  {\bibfnamefont {C.}~\bibnamefont {Silberhorn}}, \ and\ \bibinfo {author}
  {\bibfnamefont {I.}~\bibnamefont {Jex}},\ }\href {\doibase
  10.1103/PhysRevLett.119.170501} {\bibfield  {journal} {\bibinfo  {journal}
  {Phys. Rev. Lett.}\ }\textbf {\bibinfo {volume} {119}},\ \bibinfo {pages}
  {170501} (\bibinfo {year} {2017})}\BibitemShut {NoStop}%
\bibitem [{\citenamefont {Preskill}(2018)}]{Preskill2018quantumcomputingin}%
  \BibitemOpen
  \bibfield  {author} {\bibinfo {author} {\bibfnamefont {J.}~\bibnamefont
  {Preskill}},\ }\href {\doibase 10.22331/q-2018-08-06-79} {\bibfield
  {journal} {\bibinfo  {journal} {{Quantum}}\ }\textbf {\bibinfo {volume}
  {2}},\ \bibinfo {pages} {79} (\bibinfo {year} {2018})}\BibitemShut {NoStop}%
\bibitem [{\citenamefont {Arute}\ \emph {et~al.}(2019)\citenamefont {Arute},
  \citenamefont {Arya}, \citenamefont {Babbush}, \citenamefont {Bacon},
  \citenamefont {Bardin}, \citenamefont {Barends}, \citenamefont {Biswas},
  \citenamefont {Boixo}, \citenamefont {Brandao}, \citenamefont {Buell},
  \citenamefont {Burkett}, \citenamefont {Chen}, \citenamefont {Chen},
  \citenamefont {Chiaro}, \citenamefont {Collins}, \citenamefont {Courtney},
  \citenamefont {Dunsworth}, \citenamefont {Farhi}, \citenamefont {Foxen},
  \citenamefont {Fowler}, \citenamefont {Gidney}, \citenamefont {Giustina},
  \citenamefont {Graff}, \citenamefont {Guerin}, \citenamefont {Habegger},
  \citenamefont {Harrigan}, \citenamefont {Hartmann}, \citenamefont {Ho},
  \citenamefont {Hoffmann}, \citenamefont {Huang}, \citenamefont {Humble},
  \citenamefont {Isakov}, \citenamefont {Jeffrey}, \citenamefont {Jiang},
  \citenamefont {Kafri}, \citenamefont {Kechedzhi}, \citenamefont {Kelly},
  \citenamefont {Klimov}, \citenamefont {Knysh}, \citenamefont {Korotkov},
  \citenamefont {Kostritsa}, \citenamefont {Landhuis}, \citenamefont
  {Lindmark}, \citenamefont {Lucero}, \citenamefont {Lyakh}, \citenamefont
  {Mandr{\`{a}}}, \citenamefont {McClean}, \citenamefont {McEwen},
  \citenamefont {Megrant}, \citenamefont {Mi}, \citenamefont {Michielsen},
  \citenamefont {Mohseni}, \citenamefont {Mutus}, \citenamefont {Naaman},
  \citenamefont {Neeley}, \citenamefont {Neill}, \citenamefont {Niu},
  \citenamefont {Ostby}, \citenamefont {Petukhov}, \citenamefont {Platt},
  \citenamefont {Quintana}, \citenamefont {Rieffel}, \citenamefont {Roushan},
  \citenamefont {Rubin}, \citenamefont {Sank}, \citenamefont {Satzinger},
  \citenamefont {Smelyanskiy}, \citenamefont {Sung}, \citenamefont
  {Trevithick}, \citenamefont {Vainsencher}, \citenamefont {Villalonga},
  \citenamefont {White}, \citenamefont {Yao}, \citenamefont {Yeh},
  \citenamefont {Zalcman}, \citenamefont {Neven},\ and\ \citenamefont
  {Martinis}}]{Arute:2019fg}%
  \BibitemOpen
  \bibfield  {author} {\bibinfo {author} {\bibfnamefont {F.}~\bibnamefont
  {Arute}}, \bibinfo {author} {\bibfnamefont {K.}~\bibnamefont {Arya}},
  \bibinfo {author} {\bibfnamefont {R.}~\bibnamefont {Babbush}}, \bibinfo
  {author} {\bibfnamefont {D.}~\bibnamefont {Bacon}}, \bibinfo {author}
  {\bibfnamefont {J.~C.}\ \bibnamefont {Bardin}}, \bibinfo {author}
  {\bibfnamefont {R.}~\bibnamefont {Barends}}, \bibinfo {author} {\bibfnamefont
  {R.}~\bibnamefont {Biswas}}, \bibinfo {author} {\bibfnamefont
  {S.}~\bibnamefont {Boixo}}, \bibinfo {author} {\bibfnamefont {F.~G. S.~L.}\
  \bibnamefont {Brandao}}, \bibinfo {author} {\bibfnamefont {D.~A.}\
  \bibnamefont {Buell}}, \bibinfo {author} {\bibfnamefont {B.}~\bibnamefont
  {Burkett}}, \bibinfo {author} {\bibfnamefont {Y.}~\bibnamefont {Chen}},
  \bibinfo {author} {\bibfnamefont {Z.}~\bibnamefont {Chen}}, \bibinfo {author}
  {\bibfnamefont {B.}~\bibnamefont {Chiaro}}, \bibinfo {author} {\bibfnamefont
  {R.}~\bibnamefont {Collins}}, \bibinfo {author} {\bibfnamefont
  {W.}~\bibnamefont {Courtney}}, \bibinfo {author} {\bibfnamefont
  {A.}~\bibnamefont {Dunsworth}}, \bibinfo {author} {\bibfnamefont
  {E.}~\bibnamefont {Farhi}}, \bibinfo {author} {\bibfnamefont
  {B.}~\bibnamefont {Foxen}}, \bibinfo {author} {\bibfnamefont
  {A.}~\bibnamefont {Fowler}}, \bibinfo {author} {\bibfnamefont
  {C.}~\bibnamefont {Gidney}}, \bibinfo {author} {\bibfnamefont
  {M.}~\bibnamefont {Giustina}}, \bibinfo {author} {\bibfnamefont
  {R.}~\bibnamefont {Graff}}, \bibinfo {author} {\bibfnamefont
  {K.}~\bibnamefont {Guerin}}, \bibinfo {author} {\bibfnamefont
  {S.}~\bibnamefont {Habegger}}, \bibinfo {author} {\bibfnamefont {M.~P.}\
  \bibnamefont {Harrigan}}, \bibinfo {author} {\bibfnamefont {M.~J.}\
  \bibnamefont {Hartmann}}, \bibinfo {author} {\bibfnamefont {A.}~\bibnamefont
  {Ho}}, \bibinfo {author} {\bibfnamefont {M.}~\bibnamefont {Hoffmann}},
  \bibinfo {author} {\bibfnamefont {T.}~\bibnamefont {Huang}}, \bibinfo
  {author} {\bibfnamefont {T.~S.}\ \bibnamefont {Humble}}, \bibinfo {author}
  {\bibfnamefont {S.~V.}\ \bibnamefont {Isakov}}, \bibinfo {author}
  {\bibfnamefont {E.}~\bibnamefont {Jeffrey}}, \bibinfo {author} {\bibfnamefont
  {Z.}~\bibnamefont {Jiang}}, \bibinfo {author} {\bibfnamefont
  {D.}~\bibnamefont {Kafri}}, \bibinfo {author} {\bibfnamefont
  {K.}~\bibnamefont {Kechedzhi}}, \bibinfo {author} {\bibfnamefont
  {J.}~\bibnamefont {Kelly}}, \bibinfo {author} {\bibfnamefont {P.~V.}\
  \bibnamefont {Klimov}}, \bibinfo {author} {\bibfnamefont {S.}~\bibnamefont
  {Knysh}}, \bibinfo {author} {\bibfnamefont {A.}~\bibnamefont {Korotkov}},
  \bibinfo {author} {\bibfnamefont {F.}~\bibnamefont {Kostritsa}}, \bibinfo
  {author} {\bibfnamefont {D.}~\bibnamefont {Landhuis}}, \bibinfo {author}
  {\bibfnamefont {M.}~\bibnamefont {Lindmark}}, \bibinfo {author}
  {\bibfnamefont {E.}~\bibnamefont {Lucero}}, \bibinfo {author} {\bibfnamefont
  {D.}~\bibnamefont {Lyakh}}, \bibinfo {author} {\bibfnamefont
  {S.}~\bibnamefont {Mandr{\`{a}}}}, \bibinfo {author} {\bibfnamefont {J.~R.}\
  \bibnamefont {McClean}}, \bibinfo {author} {\bibfnamefont {M.}~\bibnamefont
  {McEwen}}, \bibinfo {author} {\bibfnamefont {A.}~\bibnamefont {Megrant}},
  \bibinfo {author} {\bibfnamefont {X.}~\bibnamefont {Mi}}, \bibinfo {author}
  {\bibfnamefont {K.}~\bibnamefont {Michielsen}}, \bibinfo {author}
  {\bibfnamefont {M.}~\bibnamefont {Mohseni}}, \bibinfo {author} {\bibfnamefont
  {J.}~\bibnamefont {Mutus}}, \bibinfo {author} {\bibfnamefont
  {O.}~\bibnamefont {Naaman}}, \bibinfo {author} {\bibfnamefont
  {M.}~\bibnamefont {Neeley}}, \bibinfo {author} {\bibfnamefont
  {C.}~\bibnamefont {Neill}}, \bibinfo {author} {\bibfnamefont {M.~Y.}\
  \bibnamefont {Niu}}, \bibinfo {author} {\bibfnamefont {E.}~\bibnamefont
  {Ostby}}, \bibinfo {author} {\bibfnamefont {A.}~\bibnamefont {Petukhov}},
  \bibinfo {author} {\bibfnamefont {J.~C.}\ \bibnamefont {Platt}}, \bibinfo
  {author} {\bibfnamefont {C.}~\bibnamefont {Quintana}}, \bibinfo {author}
  {\bibfnamefont {E.~G.}\ \bibnamefont {Rieffel}}, \bibinfo {author}
  {\bibfnamefont {P.}~\bibnamefont {Roushan}}, \bibinfo {author} {\bibfnamefont
  {N.~C.}\ \bibnamefont {Rubin}}, \bibinfo {author} {\bibfnamefont
  {D.}~\bibnamefont {Sank}}, \bibinfo {author} {\bibfnamefont {K.~J.}\
  \bibnamefont {Satzinger}}, \bibinfo {author} {\bibfnamefont {V.}~\bibnamefont
  {Smelyanskiy}}, \bibinfo {author} {\bibfnamefont {K.~J.}\ \bibnamefont
  {Sung}}, \bibinfo {author} {\bibfnamefont {M.~D.}\ \bibnamefont
  {Trevithick}}, \bibinfo {author} {\bibfnamefont {A.}~\bibnamefont
  {Vainsencher}}, \bibinfo {author} {\bibfnamefont {B.}~\bibnamefont
  {Villalonga}}, \bibinfo {author} {\bibfnamefont {T.}~\bibnamefont {White}},
  \bibinfo {author} {\bibfnamefont {Z.~J.}\ \bibnamefont {Yao}}, \bibinfo
  {author} {\bibfnamefont {P.}~\bibnamefont {Yeh}}, \bibinfo {author}
  {\bibfnamefont {A.}~\bibnamefont {Zalcman}}, \bibinfo {author} {\bibfnamefont
  {H.}~\bibnamefont {Neven}}, \ and\ \bibinfo {author} {\bibfnamefont {J.~M.}\
  \bibnamefont {Martinis}},\ }\href {\doibase 10.1038/s41586-019-1666-5}
  {\bibfield  {journal} {\bibinfo  {journal} {Nature}\ }\textbf {\bibinfo
  {volume} {574}},\ \bibinfo {pages} {505} (\bibinfo {year}
  {2019})}\BibitemShut {NoStop}%
\bibitem [{\citenamefont {Petersson}\ \emph {et~al.}(2010)\citenamefont
  {Petersson}, \citenamefont {Petta}, \citenamefont {Lu},\ and\ \citenamefont
  {Gossard}}]{PhysRevLett.105.246804}%
  \BibitemOpen
  \bibfield  {author} {\bibinfo {author} {\bibfnamefont {K.~D.}\ \bibnamefont
  {Petersson}}, \bibinfo {author} {\bibfnamefont {J.~R.}\ \bibnamefont
  {Petta}}, \bibinfo {author} {\bibfnamefont {H.}~\bibnamefont {Lu}}, \ and\
  \bibinfo {author} {\bibfnamefont {A.~C.}\ \bibnamefont {Gossard}},\ }\href
  {\doibase 10.1103/PhysRevLett.105.246804} {\bibfield  {journal} {\bibinfo
  {journal} {Phys. Rev. Lett.}\ }\textbf {\bibinfo {volume} {105}},\ \bibinfo
  {pages} {246804} (\bibinfo {year} {2010})}\BibitemShut {NoStop}%
\bibitem [{\citenamefont {Gorman}\ \emph {et~al.}(2005)\citenamefont {Gorman},
  \citenamefont {Hasko},\ and\ \citenamefont
  {Williams}}]{PhysRevLett.95.090502}%
  \BibitemOpen
  \bibfield  {author} {\bibinfo {author} {\bibfnamefont {J.}~\bibnamefont
  {Gorman}}, \bibinfo {author} {\bibfnamefont {D.~G.}\ \bibnamefont {Hasko}}, \
  and\ \bibinfo {author} {\bibfnamefont {D.~A.}\ \bibnamefont {Williams}},\
  }\href {\doibase 10.1103/PhysRevLett.95.090502} {\bibfield  {journal}
  {\bibinfo  {journal} {Phys. Rev. Lett.}\ }\textbf {\bibinfo {volume} {95}},\
  \bibinfo {pages} {090502} (\bibinfo {year} {2005})}\BibitemShut {NoStop}%
\bibitem [{\citenamefont {Yang}\ \emph {et~al.}(2019)\citenamefont {Yang},
  \citenamefont {Coppersmith},\ and\ \citenamefont {Friesen}}]{Yang:2019wu}%
  \BibitemOpen
  \bibfield  {author} {\bibinfo {author} {\bibfnamefont {Y.-C.}\ \bibnamefont
  {Yang}}, \bibinfo {author} {\bibfnamefont {S.~N.}\ \bibnamefont
  {Coppersmith}}, \ and\ \bibinfo {author} {\bibfnamefont {M.}~\bibnamefont
  {Friesen}},\ }\href {\doibase 10.1038/s41534-019-0127-1} {\bibfield
  {journal} {\bibinfo  {journal} {npj Quantum Information}\ }\textbf {\bibinfo
  {volume} {5}},\ \bibinfo {pages} {12} (\bibinfo {year} {2019})}\BibitemShut
  {NoStop}%
\bibitem [{\citenamefont {Loss}\ and\ \citenamefont
  {DiVincenzo}(1998)}]{PhysRevA.57.120}%
  \BibitemOpen
  \bibfield  {author} {\bibinfo {author} {\bibfnamefont {D.}~\bibnamefont
  {Loss}}\ and\ \bibinfo {author} {\bibfnamefont {D.~P.}\ \bibnamefont
  {DiVincenzo}},\ }\href {\doibase 10.1103/PhysRevA.57.120} {\bibfield
  {journal} {\bibinfo  {journal} {Phys. Rev. A}\ }\textbf {\bibinfo {volume}
  {57}},\ \bibinfo {pages} {120} (\bibinfo {year} {1998})}\BibitemShut
  {NoStop}%
\bibitem [{\citenamefont {Hanson}\ \emph {et~al.}(2007)\citenamefont {Hanson},
  \citenamefont {Kouwenhoven}, \citenamefont {Petta}, \citenamefont {Tarucha},\
  and\ \citenamefont {Vandersypen}}]{RevModPhys.79.1217}%
  \BibitemOpen
  \bibfield  {author} {\bibinfo {author} {\bibfnamefont {R.}~\bibnamefont
  {Hanson}}, \bibinfo {author} {\bibfnamefont {L.~P.}\ \bibnamefont
  {Kouwenhoven}}, \bibinfo {author} {\bibfnamefont {J.~R.}\ \bibnamefont
  {Petta}}, \bibinfo {author} {\bibfnamefont {S.}~\bibnamefont {Tarucha}}, \
  and\ \bibinfo {author} {\bibfnamefont {L.~M.~K.}\ \bibnamefont
  {Vandersypen}},\ }\href {\doibase 10.1103/RevModPhys.79.1217} {\bibfield
  {journal} {\bibinfo  {journal} {Rev. Mod. Phys.}\ }\textbf {\bibinfo {volume}
  {79}},\ \bibinfo {pages} {1217} (\bibinfo {year} {2007})}\BibitemShut
  {NoStop}%
\bibitem [{\citenamefont {Petta}\ \emph {et~al.}(2005)\citenamefont {Petta},
  \citenamefont {Johnson}, \citenamefont {Taylor}, \citenamefont {Laird},
  \citenamefont {Yacoby}, \citenamefont {Lukin}, \citenamefont {Marcus},
  \citenamefont {Hanson},\ and\ \citenamefont {Gossard}}]{Petta2180}%
  \BibitemOpen
  \bibfield  {author} {\bibinfo {author} {\bibfnamefont {J.~R.}\ \bibnamefont
  {Petta}}, \bibinfo {author} {\bibfnamefont {A.~C.}\ \bibnamefont {Johnson}},
  \bibinfo {author} {\bibfnamefont {J.~M.}\ \bibnamefont {Taylor}}, \bibinfo
  {author} {\bibfnamefont {E.~A.}\ \bibnamefont {Laird}}, \bibinfo {author}
  {\bibfnamefont {A.}~\bibnamefont {Yacoby}}, \bibinfo {author} {\bibfnamefont
  {M.~D.}\ \bibnamefont {Lukin}}, \bibinfo {author} {\bibfnamefont {C.~M.}\
  \bibnamefont {Marcus}}, \bibinfo {author} {\bibfnamefont {M.~P.}\
  \bibnamefont {Hanson}}, \ and\ \bibinfo {author} {\bibfnamefont {A.~C.}\
  \bibnamefont {Gossard}},\ }\href {\doibase 10.1126/science.1116955}
  {\bibfield  {journal} {\bibinfo  {journal} {Science}\ }\textbf {\bibinfo
  {volume} {309}},\ \bibinfo {pages} {2180} (\bibinfo {year}
  {2005})}\BibitemShut {NoStop}%
\bibitem [{\citenamefont {Veldhorst}\ \emph {et~al.}(2015)\citenamefont
  {Veldhorst}, \citenamefont {Yang}, \citenamefont {Hwang}, \citenamefont
  {Huang}, \citenamefont {Dehollain}, \citenamefont {Muhonen}, \citenamefont
  {Simmons}, \citenamefont {Laucht}, \citenamefont {Hudson}, \citenamefont
  {Itoh}, \citenamefont {Morello},\ and\ \citenamefont
  {Dzurak}}]{Veldhorst:tr}%
  \BibitemOpen
  \bibfield  {author} {\bibinfo {author} {\bibfnamefont {M.}~\bibnamefont
  {Veldhorst}}, \bibinfo {author} {\bibfnamefont {C.~H.}\ \bibnamefont {Yang}},
  \bibinfo {author} {\bibfnamefont {J.~C.~C.}\ \bibnamefont {Hwang}}, \bibinfo
  {author} {\bibfnamefont {W.}~\bibnamefont {Huang}}, \bibinfo {author}
  {\bibfnamefont {J.~P.}\ \bibnamefont {Dehollain}}, \bibinfo {author}
  {\bibfnamefont {J.~T.}\ \bibnamefont {Muhonen}}, \bibinfo {author}
  {\bibfnamefont {S.}~\bibnamefont {Simmons}}, \bibinfo {author} {\bibfnamefont
  {A.}~\bibnamefont {Laucht}}, \bibinfo {author} {\bibfnamefont {F.~E.}\
  \bibnamefont {Hudson}}, \bibinfo {author} {\bibfnamefont {K.~M.}\
  \bibnamefont {Itoh}}, \bibinfo {author} {\bibfnamefont {A.}~\bibnamefont
  {Morello}}, \ and\ \bibinfo {author} {\bibfnamefont {A.~S.}\ \bibnamefont
  {Dzurak}},\ }\href {\doibase 10.1038/nature15263} {\bibfield  {journal}
  {\bibinfo  {journal} {Nature}\ }\textbf {\bibinfo {volume} {526}},\ \bibinfo
  {pages} {410} (\bibinfo {year} {2015})}\BibitemShut {NoStop}%
\bibitem [{\citenamefont {Kitaev}(2001)}]{Kitaev:2001gb}%
  \BibitemOpen
  \bibfield  {author} {\bibinfo {author} {\bibfnamefont {A.~Y.}\ \bibnamefont
  {Kitaev}},\ }\href {\doibase 10.1070/1063-7869/44/10S/S29} {\bibfield
  {journal} {\bibinfo  {journal} {Physics-Uspekhi}\ }\textbf {\bibinfo {volume}
  {44}},\ \bibinfo {pages} {131} (\bibinfo {year} {2001})}\BibitemShut
  {NoStop}%
\bibitem [{\citenamefont {Karzig}\ \emph
  {et~al.}(2017{\natexlab{a}})\citenamefont {Karzig}, \citenamefont {Knapp},
  \citenamefont {Lutchyn}, \citenamefont {Bonderson}, \citenamefont {Hastings},
  \citenamefont {Nayak}, \citenamefont {Alicea}, \citenamefont {Flensberg},
  \citenamefont {Plugge}, \citenamefont {Oreg}, \citenamefont {Marcus},\ and\
  \citenamefont {Freedman}}]{Karzig:2017if}%
  \BibitemOpen
  \bibfield  {author} {\bibinfo {author} {\bibfnamefont {T.}~\bibnamefont
  {Karzig}}, \bibinfo {author} {\bibfnamefont {C.}~\bibnamefont {Knapp}},
  \bibinfo {author} {\bibfnamefont {R.~M.}\ \bibnamefont {Lutchyn}}, \bibinfo
  {author} {\bibfnamefont {P.}~\bibnamefont {Bonderson}}, \bibinfo {author}
  {\bibfnamefont {M.~B.}\ \bibnamefont {Hastings}}, \bibinfo {author}
  {\bibfnamefont {C.}~\bibnamefont {Nayak}}, \bibinfo {author} {\bibfnamefont
  {J.}~\bibnamefont {Alicea}}, \bibinfo {author} {\bibfnamefont
  {K.}~\bibnamefont {Flensberg}}, \bibinfo {author} {\bibfnamefont
  {S.}~\bibnamefont {Plugge}}, \bibinfo {author} {\bibfnamefont
  {Y.}~\bibnamefont {Oreg}}, \bibinfo {author} {\bibfnamefont {C.~M.}\
  \bibnamefont {Marcus}}, \ and\ \bibinfo {author} {\bibfnamefont {M.~H.}\
  \bibnamefont {Freedman}},\ }\href {\doibase 10.1103/PhysRevB.95.235305}
  {\bibfield  {journal} {\bibinfo  {journal} {Phys. Rev. B}\ }\textbf {\bibinfo
  {volume} {95}},\ \bibinfo {pages} {235305} (\bibinfo {year}
  {2017}{\natexlab{a}})}\BibitemShut {NoStop}%
\bibitem [{\citenamefont {Alicea}\ \emph {et~al.}(2011)\citenamefont {Alicea},
  \citenamefont {Oreg}, \citenamefont {Refael}, \citenamefont {von Oppen},\
  and\ \citenamefont {Fisher}}]{Alicea:2011fe}%
  \BibitemOpen
  \bibfield  {author} {\bibinfo {author} {\bibfnamefont {J.}~\bibnamefont
  {Alicea}}, \bibinfo {author} {\bibfnamefont {Y.}~\bibnamefont {Oreg}},
  \bibinfo {author} {\bibfnamefont {G.}~\bibnamefont {Refael}}, \bibinfo
  {author} {\bibfnamefont {F.}~\bibnamefont {von Oppen}}, \ and\ \bibinfo
  {author} {\bibfnamefont {M.~P.~A.}\ \bibnamefont {Fisher}},\ }\href {\doibase
  10.1038/nphys1915} {\bibfield  {journal} {\bibinfo  {journal} {Nat. Phys.}\
  }\textbf {\bibinfo {volume} {7}},\ \bibinfo {pages} {412} (\bibinfo {year}
  {2011})}\BibitemShut {NoStop}%
\bibitem [{\citenamefont {Kalantre}\ \emph {et~al.}(2019)\citenamefont
  {Kalantre}, \citenamefont {Zwolak}, \citenamefont {Ragole}, \citenamefont
  {Wu}, \citenamefont {Zimmerman}, \citenamefont {Stewart},\ and\ \citenamefont
  {Taylor}}]{Kalantre_dnn}%
  \BibitemOpen
  \bibfield  {author} {\bibinfo {author} {\bibfnamefont {S.~S.}\ \bibnamefont
  {Kalantre}}, \bibinfo {author} {\bibfnamefont {J.~P.}\ \bibnamefont
  {Zwolak}}, \bibinfo {author} {\bibfnamefont {S.}~\bibnamefont {Ragole}},
  \bibinfo {author} {\bibfnamefont {X.}~\bibnamefont {Wu}}, \bibinfo {author}
  {\bibfnamefont {N.~M.}\ \bibnamefont {Zimmerman}}, \bibinfo {author}
  {\bibfnamefont {M.~D.}\ \bibnamefont {Stewart}}, \ and\ \bibinfo {author}
  {\bibfnamefont {J.~M.}\ \bibnamefont {Taylor}},\ }\href {\doibase
  10.1038/s41534-018-0118-7} {\bibfield  {journal} {\bibinfo  {journal} {npj
  Quantum Inf.}\ } (\bibinfo {year} {2019}),\ 10.1038/s41534-018-0118-7},\
  \Eprint {http://arxiv.org/abs/1712.04914} {arXiv:1712.04914} \BibitemShut
  {NoStop}%
\bibitem [{\citenamefont {Zwolak}\ \emph {et~al.}(2018)\citenamefont {Zwolak},
  \citenamefont {Kalantre}, \citenamefont {Wu}, \citenamefont {Ragole},\ and\
  \citenamefont {Taylor}}]{zwolak_qlite}%
  \BibitemOpen
  \bibfield  {author} {\bibinfo {author} {\bibfnamefont {J.~P.}\ \bibnamefont
  {Zwolak}}, \bibinfo {author} {\bibfnamefont {S.~S.}\ \bibnamefont
  {Kalantre}}, \bibinfo {author} {\bibfnamefont {X.}~\bibnamefont {Wu}},
  \bibinfo {author} {\bibfnamefont {S.}~\bibnamefont {Ragole}}, \ and\ \bibinfo
  {author} {\bibfnamefont {J.~M.}\ \bibnamefont {Taylor}},\ }\href {\doibase
  10.1371/journal.pone.0205844} {\bibfield  {journal} {\bibinfo  {journal}
  {PLOS ONE}\ }\textbf {\bibinfo {volume} {13}},\ \bibinfo {pages} {1}
  (\bibinfo {year} {2018})}\BibitemShut {NoStop}%
\bibitem [{\citenamefont {Zwolak}\ \emph {et~al.}(2019)\citenamefont {Zwolak},
  \citenamefont {McJunkin}, \citenamefont {Kalantre}, \citenamefont {Dodson},
  \citenamefont {MacQuarrie}, \citenamefont {Savage}, \citenamefont {Lagally},
  \citenamefont {Coppersmith}, \citenamefont {Eriksson},\ and\ \citenamefont
  {Taylor}}]{Zwolak:2019tx}%
  \BibitemOpen
  \bibfield  {author} {\bibinfo {author} {\bibfnamefont {J.~P.}\ \bibnamefont
  {Zwolak}}, \bibinfo {author} {\bibfnamefont {T.}~\bibnamefont {McJunkin}},
  \bibinfo {author} {\bibfnamefont {S.~S.}\ \bibnamefont {Kalantre}}, \bibinfo
  {author} {\bibfnamefont {J.~P.}\ \bibnamefont {Dodson}}, \bibinfo {author}
  {\bibfnamefont {E.~R.}\ \bibnamefont {MacQuarrie}}, \bibinfo {author}
  {\bibfnamefont {D.~E.}\ \bibnamefont {Savage}}, \bibinfo {author}
  {\bibfnamefont {M.~G.}\ \bibnamefont {Lagally}}, \bibinfo {author}
  {\bibfnamefont {S.~N.}\ \bibnamefont {Coppersmith}}, \bibinfo {author}
  {\bibfnamefont {M.~A.}\ \bibnamefont {Eriksson}}, \ and\ \bibinfo {author}
  {\bibfnamefont {J.~M.}\ \bibnamefont {Taylor}},\ }\href
  {http://arxiv.org/abs/1909.08030} {\  (\bibinfo {year} {2019})},\ \Eprint
  {http://arxiv.org/abs/1909.08030} {arXiv:1909.08030} \BibitemShut {NoStop}%
\bibitem [{\citenamefont {Botzem}\ \emph {et~al.}(2018)\citenamefont {Botzem},
  \citenamefont {Shulman}, \citenamefont {Foletti}, \citenamefont {Harvey},
  \citenamefont {Dial}, \citenamefont {Bethke}, \citenamefont {Cerfontaine},
  \citenamefont {McNeil}, \citenamefont {Mahalu}, \citenamefont {Umansky},
  \citenamefont {Ludwig}, \citenamefont {Wieck}, \citenamefont {Schuh},
  \citenamefont {Bougeard}, \citenamefont {Yacoby},\ and\ \citenamefont
  {Bluhm}}]{Botzem}%
  \BibitemOpen
  \bibfield  {author} {\bibinfo {author} {\bibfnamefont {T.}~\bibnamefont
  {Botzem}}, \bibinfo {author} {\bibfnamefont {M.~D.}\ \bibnamefont {Shulman}},
  \bibinfo {author} {\bibfnamefont {S.}~\bibnamefont {Foletti}}, \bibinfo
  {author} {\bibfnamefont {S.~P.}\ \bibnamefont {Harvey}}, \bibinfo {author}
  {\bibfnamefont {O.~E.}\ \bibnamefont {Dial}}, \bibinfo {author}
  {\bibfnamefont {P.}~\bibnamefont {Bethke}}, \bibinfo {author} {\bibfnamefont
  {P.}~\bibnamefont {Cerfontaine}}, \bibinfo {author} {\bibfnamefont
  {R.~P.~G.}\ \bibnamefont {McNeil}}, \bibinfo {author} {\bibfnamefont
  {D.}~\bibnamefont {Mahalu}}, \bibinfo {author} {\bibfnamefont
  {V.}~\bibnamefont {Umansky}}, \bibinfo {author} {\bibfnamefont
  {A.}~\bibnamefont {Ludwig}}, \bibinfo {author} {\bibfnamefont
  {A.}~\bibnamefont {Wieck}}, \bibinfo {author} {\bibfnamefont
  {D.}~\bibnamefont {Schuh}}, \bibinfo {author} {\bibfnamefont
  {D.}~\bibnamefont {Bougeard}}, \bibinfo {author} {\bibfnamefont
  {A.}~\bibnamefont {Yacoby}}, \ and\ \bibinfo {author} {\bibfnamefont
  {H.}~\bibnamefont {Bluhm}},\ }\href {\doibase
  10.1103/PhysRevApplied.10.054026} {\bibfield  {journal} {\bibinfo  {journal}
  {Phys. Rev. Appl.}\ }\textbf {\bibinfo {volume} {10}},\ \bibinfo {pages}
  {054026} (\bibinfo {year} {2018})}\BibitemShut {NoStop}%
\bibitem [{\citenamefont {van Diepen}\ \emph {et~al.}(2018)\citenamefont {van
  Diepen}, \citenamefont {Eendebak}, \citenamefont {Buijtendorp}, \citenamefont
  {Mukhopadhyay}, \citenamefont {Fujita}, \citenamefont {Reichl}, \citenamefont
  {Wegscheider},\ and\ \citenamefont {Vandersypen}}]{vanDiepen_automation}%
  \BibitemOpen
  \bibfield  {author} {\bibinfo {author} {\bibfnamefont {C.~J.}\ \bibnamefont
  {van Diepen}}, \bibinfo {author} {\bibfnamefont {P.~T.}\ \bibnamefont
  {Eendebak}}, \bibinfo {author} {\bibfnamefont {B.~T.}\ \bibnamefont
  {Buijtendorp}}, \bibinfo {author} {\bibfnamefont {U.}~\bibnamefont
  {Mukhopadhyay}}, \bibinfo {author} {\bibfnamefont {T.}~\bibnamefont
  {Fujita}}, \bibinfo {author} {\bibfnamefont {C.}~\bibnamefont {Reichl}},
  \bibinfo {author} {\bibfnamefont {W.}~\bibnamefont {Wegscheider}}, \ and\
  \bibinfo {author} {\bibfnamefont {L.~M.~K.}\ \bibnamefont {Vandersypen}},\
  }\href {\doibase 10.1063/1.5031034} {\bibfield  {journal} {\bibinfo
  {journal} {Appl. Phys. Lett.}\ }\textbf {\bibinfo {volume} {113}},\ \bibinfo
  {pages} {033101} (\bibinfo {year} {2018})}\BibitemShut {NoStop}%
\bibitem [{\citenamefont {Teske}\ \emph {et~al.}(2019)\citenamefont {Teske},
  \citenamefont {Humpohl}, \citenamefont {Otten}, \citenamefont {Bethke},
  \citenamefont {Cerfontaine}, \citenamefont {Dedden}, \citenamefont {Ludwig},
  \citenamefont {Wieck},\ and\ \citenamefont {Bluhm}}]{Teske:760139}%
  \BibitemOpen
  \bibfield  {author} {\bibinfo {author} {\bibfnamefont {J.~D.}\ \bibnamefont
  {Teske}}, \bibinfo {author} {\bibfnamefont {S.~S.}\ \bibnamefont {Humpohl}},
  \bibinfo {author} {\bibfnamefont {R.}~\bibnamefont {Otten}}, \bibinfo
  {author} {\bibfnamefont {P.}~\bibnamefont {Bethke}}, \bibinfo {author}
  {\bibfnamefont {P.}~\bibnamefont {Cerfontaine}}, \bibinfo {author}
  {\bibfnamefont {J.}~\bibnamefont {Dedden}}, \bibinfo {author} {\bibfnamefont
  {A.}~\bibnamefont {Ludwig}}, \bibinfo {author} {\bibfnamefont {A.~D.}\
  \bibnamefont {Wieck}}, \ and\ \bibinfo {author} {\bibfnamefont {J.~H.}\
  \bibnamefont {Bluhm}},\ }\href {\doibase 10.1063/1.5088412} {\bibfield
  {journal} {\bibinfo  {journal} {Appl. Phys. Lett.}\ }\textbf {\bibinfo
  {volume} {114}},\ \bibinfo {pages} {133102} (\bibinfo {year}
  {2019})}\BibitemShut {NoStop}%
\bibitem [{\citenamefont {Mills}\ \emph
  {et~al.}(2019{\natexlab{a}})\citenamefont {Mills}, \citenamefont {Feldman},
  \citenamefont {Monical}, \citenamefont {Lewis}, \citenamefont {Larson},
  \citenamefont {Mounce},\ and\ \citenamefont {Petta}}]{Mills:2019fy}%
  \BibitemOpen
  \bibfield  {author} {\bibinfo {author} {\bibfnamefont {A.~R.}\ \bibnamefont
  {Mills}}, \bibinfo {author} {\bibfnamefont {M.~M.}\ \bibnamefont {Feldman}},
  \bibinfo {author} {\bibfnamefont {C.}~\bibnamefont {Monical}}, \bibinfo
  {author} {\bibfnamefont {P.~J.}\ \bibnamefont {Lewis}}, \bibinfo {author}
  {\bibfnamefont {K.~W.}\ \bibnamefont {Larson}}, \bibinfo {author}
  {\bibfnamefont {A.~M.}\ \bibnamefont {Mounce}}, \ and\ \bibinfo {author}
  {\bibfnamefont {J.~R.}\ \bibnamefont {Petta}},\ }\href {\doibase
  10.1063/1.5121444} {\bibfield  {journal} {\bibinfo  {journal} {Appl. Phys.
  Lett.}\ }\textbf {\bibinfo {volume} {115}},\ \bibinfo {pages} {113501}
  (\bibinfo {year} {2019}{\natexlab{a}})},\ \Eprint
  {http://arxiv.org/abs/1907.10775} {arXiv:1907.10775} \BibitemShut {NoStop}%
\bibitem [{\citenamefont {Lennon}\ \emph {et~al.}(2019)\citenamefont {Lennon},
  \citenamefont {Moon}, \citenamefont {Camenzind}, \citenamefont {Yu},
  \citenamefont {Zumb{\"u}hl}, \citenamefont {Briggs}, \citenamefont {Osborne},
  \citenamefont {Laird},\ and\ \citenamefont {Ares}}]{Lennon:2019uq}%
  \BibitemOpen
  \bibfield  {author} {\bibinfo {author} {\bibfnamefont {D.~T.}\ \bibnamefont
  {Lennon}}, \bibinfo {author} {\bibfnamefont {H.}~\bibnamefont {Moon}},
  \bibinfo {author} {\bibfnamefont {L.~C.}\ \bibnamefont {Camenzind}}, \bibinfo
  {author} {\bibfnamefont {L.}~\bibnamefont {Yu}}, \bibinfo {author}
  {\bibfnamefont {D.~M.}\ \bibnamefont {Zumb{\"u}hl}}, \bibinfo {author}
  {\bibfnamefont {G.~A.~D.}\ \bibnamefont {Briggs}}, \bibinfo {author}
  {\bibfnamefont {M.~A.}\ \bibnamefont {Osborne}}, \bibinfo {author}
  {\bibfnamefont {E.~A.}\ \bibnamefont {Laird}}, \ and\ \bibinfo {author}
  {\bibfnamefont {N.}~\bibnamefont {Ares}},\ }\href {\doibase
  10.1038/s41534-019-0193-4} {\bibfield  {journal} {\bibinfo  {journal} {npj
  Quantum Inf.}\ }\textbf {\bibinfo {volume} {5}} (\bibinfo {year} {2019}),\
  10.1038/s41534-019-0193-4},\ \Eprint {http://arxiv.org/abs/1810.10042}
  {arXiv:1810.10042} \BibitemShut {NoStop}%
\bibitem [{\citenamefont {Baart}\ \emph {et~al.}(2016)\citenamefont {Baart},
  \citenamefont {Eendebak}, \citenamefont {Reichl}, \citenamefont
  {Wegscheider},\ and\ \citenamefont {Vandersypen}}]{Baart_automation}%
  \BibitemOpen
  \bibfield  {author} {\bibinfo {author} {\bibfnamefont {T.~A.}\ \bibnamefont
  {Baart}}, \bibinfo {author} {\bibfnamefont {P.~T.}\ \bibnamefont {Eendebak}},
  \bibinfo {author} {\bibfnamefont {C.}~\bibnamefont {Reichl}}, \bibinfo
  {author} {\bibfnamefont {W.}~\bibnamefont {Wegscheider}}, \ and\ \bibinfo
  {author} {\bibfnamefont {L.~M.}\ \bibnamefont {Vandersypen}},\ }\href
  {\doibase 10.1063/1.4952624} {\bibfield  {journal} {\bibinfo  {journal}
  {Appl. Phys. Lett.}\ } (\bibinfo {year} {2016}),\ 10.1063/1.4952624},\
  \Eprint {http://arxiv.org/abs/1603.02274} {arXiv:1603.02274} \BibitemShut
  {NoStop}%
\bibitem [{qco(16  )}]{qcodes}%
  \BibitemOpen
  \href {https://github.com/QCoDeS/Qcodes} {\enquote {\bibinfo {title}
  {{QCoDeS}: Python-based data acquisition framework},}\ } (\bibinfo {year}
  {2016--}),\ \bibinfo {note} {[Online; accessed October 2019]}\BibitemShut
  {NoStop}%
\bibitem [{\citenamefont {van~der Wiel}\ \emph {et~al.}(2002)\citenamefont
  {van~der Wiel}, \citenamefont {De~Franceschi}, \citenamefont {Elzerman},
  \citenamefont {Fujisawa}, \citenamefont {Tarucha},\ and\ \citenamefont
  {Kouwenhoven}}]{vanderWiel:2002gr}%
  \BibitemOpen
  \bibfield  {author} {\bibinfo {author} {\bibfnamefont {W.~G.}\ \bibnamefont
  {van~der Wiel}}, \bibinfo {author} {\bibfnamefont {S.}~\bibnamefont
  {De~Franceschi}}, \bibinfo {author} {\bibfnamefont {J.~M.}\ \bibnamefont
  {Elzerman}}, \bibinfo {author} {\bibfnamefont {T.}~\bibnamefont {Fujisawa}},
  \bibinfo {author} {\bibfnamefont {S.}~\bibnamefont {Tarucha}}, \ and\
  \bibinfo {author} {\bibfnamefont {L.~P.}\ \bibnamefont {Kouwenhoven}},\
  }\href {\doibase 10.1103/RevModPhys.75.1} {\bibfield  {journal} {\bibinfo
  {journal} {Rev. Mod. Phys.}\ }\textbf {\bibinfo {volume} {75}},\ \bibinfo
  {pages} {1} (\bibinfo {year} {2002})}\BibitemShut {NoStop}%
\bibitem [{\citenamefont {Carleo}\ and\ \citenamefont
  {Troyer}(2017)}]{Carleo602}%
  \BibitemOpen
  \bibfield  {author} {\bibinfo {author} {\bibfnamefont {G.}~\bibnamefont
  {Carleo}}\ and\ \bibinfo {author} {\bibfnamefont {M.}~\bibnamefont
  {Troyer}},\ }\href {\doibase 10.1126/science.aag2302} {\bibfield  {journal}
  {\bibinfo  {journal} {Science}\ }\textbf {\bibinfo {volume} {355}},\ \bibinfo
  {pages} {602} (\bibinfo {year} {2017})}\BibitemShut {NoStop}%
\bibitem [{\citenamefont {Carleo}\ \emph {et~al.}(2018)\citenamefont {Carleo},
  \citenamefont {Nomura},\ and\ \citenamefont {Imada}}]{Carleo:2018tj}%
  \BibitemOpen
  \bibfield  {author} {\bibinfo {author} {\bibfnamefont {G.}~\bibnamefont
  {Carleo}}, \bibinfo {author} {\bibfnamefont {Y.}~\bibnamefont {Nomura}}, \
  and\ \bibinfo {author} {\bibfnamefont {M.}~\bibnamefont {Imada}},\ }\href
  {\doibase 10.1038/s41467-018-07520-3} {\bibfield  {journal} {\bibinfo
  {journal} {Nat. Commun.}\ }\textbf {\bibinfo {volume} {9}},\ \bibinfo {pages}
  {5322} (\bibinfo {year} {2018})}\BibitemShut {NoStop}%
\bibitem [{\citenamefont {Torlai}\ \emph {et~al.}(2018)\citenamefont {Torlai},
  \citenamefont {Mazzola}, \citenamefont {Carrasquilla}, \citenamefont
  {Troyer}, \citenamefont {Melko},\ and\ \citenamefont
  {Carleo}}]{Torlai:2018wn}%
  \BibitemOpen
  \bibfield  {author} {\bibinfo {author} {\bibfnamefont {G.}~\bibnamefont
  {Torlai}}, \bibinfo {author} {\bibfnamefont {G.}~\bibnamefont {Mazzola}},
  \bibinfo {author} {\bibfnamefont {J.}~\bibnamefont {Carrasquilla}}, \bibinfo
  {author} {\bibfnamefont {M.}~\bibnamefont {Troyer}}, \bibinfo {author}
  {\bibfnamefont {R.}~\bibnamefont {Melko}}, \ and\ \bibinfo {author}
  {\bibfnamefont {G.}~\bibnamefont {Carleo}},\ }\href {\doibase
  10.1038/s41567-018-0048-5} {\bibfield  {journal} {\bibinfo  {journal} {Nat.
  Phys.}\ }\textbf {\bibinfo {volume} {14}},\ \bibinfo {pages} {447} (\bibinfo
  {year} {2018})}\BibitemShut {NoStop}%
\bibitem [{\citenamefont {Nautrup}\ \emph {et~al.}(2018)\citenamefont
  {Nautrup}, \citenamefont {Delfosse}, \citenamefont {Dunjko}, \citenamefont
  {Briegel},\ and\ \citenamefont {Friis}}]{nautrup2018optimizing}%
  \BibitemOpen
  \bibfield  {author} {\bibinfo {author} {\bibfnamefont {H.~P.}\ \bibnamefont
  {Nautrup}}, \bibinfo {author} {\bibfnamefont {N.}~\bibnamefont {Delfosse}},
  \bibinfo {author} {\bibfnamefont {V.}~\bibnamefont {Dunjko}}, \bibinfo
  {author} {\bibfnamefont {H.~J.}\ \bibnamefont {Briegel}}, \ and\ \bibinfo
  {author} {\bibfnamefont {N.}~\bibnamefont {Friis}},\ }\href
  {http://arxiv.org/abs/1812.08451} {\  (\bibinfo {year} {2018})},\ \Eprint
  {http://arxiv.org/abs/1812.08451} {arXiv:1812.08451} \BibitemShut {NoStop}%
\bibitem [{\citenamefont {Sweke}\ \emph {et~al.}(2018)\citenamefont {Sweke},
  \citenamefont {Kesselring}, \citenamefont {van Nieuwenburg},\ and\
  \citenamefont {Eisert}}]{sweke2018reinforcement}%
  \BibitemOpen
  \bibfield  {author} {\bibinfo {author} {\bibfnamefont {R.}~\bibnamefont
  {Sweke}}, \bibinfo {author} {\bibfnamefont {M.~S.}\ \bibnamefont
  {Kesselring}}, \bibinfo {author} {\bibfnamefont {E.~P.~L.}\ \bibnamefont {van
  Nieuwenburg}}, \ and\ \bibinfo {author} {\bibfnamefont {J.}~\bibnamefont
  {Eisert}},\ }\href {http://arxiv.org/abs/1810.07207} {\  (\bibinfo {year}
  {2018})},\ \Eprint {http://arxiv.org/abs/1810.07207} {arXiv:1810.07207}
  \BibitemShut {NoStop}%
\bibitem [{\citenamefont {Croot}\ \emph {et~al.}(2018)\citenamefont {Croot},
  \citenamefont {Pauka}, \citenamefont {Watson}, \citenamefont {Gardner},
  \citenamefont {Fallahi}, \citenamefont {Manfra},\ and\ \citenamefont
  {Reilly}}]{Croot:2018iq}%
  \BibitemOpen
  \bibfield  {author} {\bibinfo {author} {\bibfnamefont {X.}~\bibnamefont
  {Croot}}, \bibinfo {author} {\bibfnamefont {S.}~\bibnamefont {Pauka}},
  \bibinfo {author} {\bibfnamefont {J.}~\bibnamefont {Watson}}, \bibinfo
  {author} {\bibfnamefont {G.}~\bibnamefont {Gardner}}, \bibinfo {author}
  {\bibfnamefont {S.}~\bibnamefont {Fallahi}}, \bibinfo {author} {\bibfnamefont
  {M.}~\bibnamefont {Manfra}}, \ and\ \bibinfo {author} {\bibfnamefont
  {D.}~\bibnamefont {Reilly}},\ }\href {\doibase
  10.1103/PhysRevApplied.10.044058} {\bibfield  {journal} {\bibinfo  {journal}
  {Phys. Rev. Appl.}\ }\textbf {\bibinfo {volume} {10}},\ \bibinfo {pages}
  {044058} (\bibinfo {year} {2018})}\BibitemShut {NoStop}%
\bibitem [{\citenamefont {Ortiz-Conde}\ \emph {et~al.}(2002)\citenamefont
  {Ortiz-Conde}, \citenamefont {{Garc\'{i}a S{\'{a}}nchez}}, \citenamefont
  {Liou}, \citenamefont {Cerdeira}, \citenamefont {Estrada},\ and\
  \citenamefont {Yue}}]{OrtizConde:2002jg}%
  \BibitemOpen
  \bibfield  {author} {\bibinfo {author} {\bibfnamefont {A.}~\bibnamefont
  {Ortiz-Conde}}, \bibinfo {author} {\bibfnamefont {F.}~\bibnamefont
  {{Garc\'{i}a S{\'{a}}nchez}}}, \bibinfo {author} {\bibfnamefont
  {J.}~\bibnamefont {Liou}}, \bibinfo {author} {\bibfnamefont {A.}~\bibnamefont
  {Cerdeira}}, \bibinfo {author} {\bibfnamefont {M.}~\bibnamefont {Estrada}}, \
  and\ \bibinfo {author} {\bibfnamefont {Y.}~\bibnamefont {Yue}},\ }\href
  {\doibase 10.1016/S0026-2714(02)00027-6} {\bibfield  {journal} {\bibinfo
  {journal} {Microelectronics Reliability}\ }\textbf {\bibinfo {volume} {42}},\
  \bibinfo {pages} {583} (\bibinfo {year} {2002})}\BibitemShut {NoStop}%
\bibitem [{\citenamefont {van~der Walt}\ \emph {et~al.}(2014)\citenamefont
  {van~der Walt}, \citenamefont {{S}ch\"onberger}, \citenamefont
  {{Nunez-Iglesias}}, \citenamefont {{B}oulogne}, \citenamefont {{W}arner},
  \citenamefont {{Y}ager}, \citenamefont {{G}ouillart}, \citenamefont {{Y}u},\
  and\ \citenamefont {the scikit-image contributors}}]{scikit-image}%
  \BibitemOpen
  \bibfield  {author} {\bibinfo {author} {\bibfnamefont {S.}~\bibnamefont
  {van~der Walt}}, \bibinfo {author} {\bibfnamefont {J.~L.}\ \bibnamefont
  {{S}ch\"onberger}}, \bibinfo {author} {\bibfnamefont {J.}~\bibnamefont
  {{Nunez-Iglesias}}}, \bibinfo {author} {\bibfnamefont {F.}~\bibnamefont
  {{B}oulogne}}, \bibinfo {author} {\bibfnamefont {J.~D.}\ \bibnamefont
  {{W}arner}}, \bibinfo {author} {\bibfnamefont {N.}~\bibnamefont {{Y}ager}},
  \bibinfo {author} {\bibfnamefont {E.}~\bibnamefont {{G}ouillart}}, \bibinfo
  {author} {\bibfnamefont {T.}~\bibnamefont {{Y}u}}, \ and\ \bibinfo {author}
  {\bibnamefont {the scikit-image contributors}},\ }\href {\doibase
  10.7717/peerj.453} {\bibfield  {journal} {\bibinfo  {journal} {PeerJ}\
  }\textbf {\bibinfo {volume} {2}},\ \bibinfo {pages} {e453} (\bibinfo {year}
  {2014})}\BibitemShut {NoStop}%
\bibitem [{\citenamefont {Pedregosa}\ \emph {et~al.}(2011)\citenamefont
  {Pedregosa}, \citenamefont {Varoquaux}, \citenamefont {Gramfort},
  \citenamefont {Michel}, \citenamefont {Thirion}, \citenamefont {Grisel},
  \citenamefont {Blondel}, \citenamefont {Prettenhofer}, \citenamefont {Weiss},
  \citenamefont {Dubourg}, \citenamefont {Vanderplas}, \citenamefont {Passos},
  \citenamefont {Cournapeau}, \citenamefont {Brucher}, \citenamefont {Perrot},\
  and\ \citenamefont {Duchesnay}}]{scikit-learn}%
  \BibitemOpen
  \bibfield  {author} {\bibinfo {author} {\bibfnamefont {F.}~\bibnamefont
  {Pedregosa}}, \bibinfo {author} {\bibfnamefont {G.}~\bibnamefont
  {Varoquaux}}, \bibinfo {author} {\bibfnamefont {A.}~\bibnamefont {Gramfort}},
  \bibinfo {author} {\bibfnamefont {V.}~\bibnamefont {Michel}}, \bibinfo
  {author} {\bibfnamefont {B.}~\bibnamefont {Thirion}}, \bibinfo {author}
  {\bibfnamefont {O.}~\bibnamefont {Grisel}}, \bibinfo {author} {\bibfnamefont
  {M.}~\bibnamefont {Blondel}}, \bibinfo {author} {\bibfnamefont
  {P.}~\bibnamefont {Prettenhofer}}, \bibinfo {author} {\bibfnamefont
  {R.}~\bibnamefont {Weiss}}, \bibinfo {author} {\bibfnamefont
  {V.}~\bibnamefont {Dubourg}}, \bibinfo {author} {\bibfnamefont
  {J.}~\bibnamefont {Vanderplas}}, \bibinfo {author} {\bibfnamefont
  {A.}~\bibnamefont {Passos}}, \bibinfo {author} {\bibfnamefont
  {D.}~\bibnamefont {Cournapeau}}, \bibinfo {author} {\bibfnamefont
  {M.}~\bibnamefont {Brucher}}, \bibinfo {author} {\bibfnamefont
  {M.}~\bibnamefont {Perrot}}, \ and\ \bibinfo {author} {\bibfnamefont
  {E.}~\bibnamefont {Duchesnay}},\ }\href@noop {} {\bibfield  {journal}
  {\bibinfo  {journal} {J. Mach. Learn. Res.}\ }\textbf {\bibinfo {volume}
  {12}},\ \bibinfo {pages} {2825} (\bibinfo {year} {2011})}\BibitemShut
  {NoStop}%
\bibitem [{\citenamefont {Mills}\ \emph
  {et~al.}(2019{\natexlab{b}})\citenamefont {Mills}, \citenamefont {Feldman},
  \citenamefont {Monical}, \citenamefont {Lewis}, \citenamefont {Larson},
  \citenamefont {Mounce},\ and\ \citenamefont {Petta}}]{doi:10.1063/1.5121444}%
  \BibitemOpen
  \bibfield  {author} {\bibinfo {author} {\bibfnamefont {A.~R.}\ \bibnamefont
  {Mills}}, \bibinfo {author} {\bibfnamefont {M.~M.}\ \bibnamefont {Feldman}},
  \bibinfo {author} {\bibfnamefont {C.}~\bibnamefont {Monical}}, \bibinfo
  {author} {\bibfnamefont {P.~J.}\ \bibnamefont {Lewis}}, \bibinfo {author}
  {\bibfnamefont {K.~W.}\ \bibnamefont {Larson}}, \bibinfo {author}
  {\bibfnamefont {A.~M.}\ \bibnamefont {Mounce}}, \ and\ \bibinfo {author}
  {\bibfnamefont {J.~R.}\ \bibnamefont {Petta}},\ }\href {\doibase
  10.1063/1.5121444} {\bibfield  {journal} {\bibinfo  {journal} {Appl. Phys.
  Lett.}\ }\textbf {\bibinfo {volume} {115}},\ \bibinfo {pages} {113501}
  (\bibinfo {year} {2019}{\natexlab{b}})}\BibitemShut {NoStop}%
\bibitem [{\citenamefont {Hornibrook}\ \emph {et~al.}(2014)\citenamefont
  {Hornibrook}, \citenamefont {Colless}, \citenamefont {Mahoney}, \citenamefont
  {Croot}, \citenamefont {Blanvillain}, \citenamefont {Lu}, \citenamefont
  {Gossard},\ and\ \citenamefont {Reilly}}]{Hornibrook2014}%
  \BibitemOpen
  \bibfield  {author} {\bibinfo {author} {\bibfnamefont {J.~M.}\ \bibnamefont
  {Hornibrook}}, \bibinfo {author} {\bibfnamefont {J.~I.}\ \bibnamefont
  {Colless}}, \bibinfo {author} {\bibfnamefont {A.~C.}\ \bibnamefont
  {Mahoney}}, \bibinfo {author} {\bibfnamefont {X.~G.}\ \bibnamefont {Croot}},
  \bibinfo {author} {\bibfnamefont {S.}~\bibnamefont {Blanvillain}}, \bibinfo
  {author} {\bibfnamefont {H.}~\bibnamefont {Lu}}, \bibinfo {author}
  {\bibfnamefont {A.~C.}\ \bibnamefont {Gossard}}, \ and\ \bibinfo {author}
  {\bibfnamefont {D.~J.}\ \bibnamefont {Reilly}},\ }\href {\doibase
  10.1063/1.4868107} {\bibfield  {journal} {\bibinfo  {journal} {Appl. Phys.
  Lett.}\ }\textbf {\bibinfo {volume} {104}},\ \bibinfo {pages} {103108}
  (\bibinfo {year} {2014})}\BibitemShut {NoStop}%
\bibitem [{\citenamefont {Schapire}(1999)}]{Schapire1999}%
  \BibitemOpen
  \bibfield  {author} {\bibinfo {author} {\bibfnamefont {R.~E.}\ \bibnamefont
  {Schapire}},\ }in\ \href {http://dl.acm.org/citation.cfm?id=1624312.1624417}
  {\emph {\bibinfo {booktitle} {Proceedings of the 16th International Joint
  Conference on Artificial Intelligence - Volume 2}}},\ \bibinfo {series and
  number} {IJCAI'99}\ (\bibinfo  {publisher} {Morgan Kaufmann Publishers
  Inc.},\ \bibinfo {address} {San Francisco, CA, USA},\ \bibinfo {year}
  {1999})\ pp.\ \bibinfo {pages} {1401--1406}\BibitemShut {NoStop}%
\bibitem [{\citenamefont {Volk}\ \emph {et~al.}(2019)\citenamefont {Volk},
  \citenamefont {Chatterjee}, \citenamefont {Ansaloni}, \citenamefont
  {Marcus},\ and\ \citenamefont {Kuemmeth}}]{Volk2019FastCS}%
  \BibitemOpen
  \bibfield  {author} {\bibinfo {author} {\bibfnamefont {C.}~\bibnamefont
  {Volk}}, \bibinfo {author} {\bibfnamefont {A.}~\bibnamefont {Chatterjee}},
  \bibinfo {author} {\bibfnamefont {F.}~\bibnamefont {Ansaloni}}, \bibinfo
  {author} {\bibfnamefont {C.~M.}\ \bibnamefont {Marcus}}, \ and\ \bibinfo
  {author} {\bibfnamefont {F.}~\bibnamefont {Kuemmeth}},\ }\href {\doibase
  10.1021/acs.nanolett.9b02149} {\bibfield  {journal} {\bibinfo  {journal}
  {Nano Lett.}\ }\textbf {\bibinfo {volume} {19}},\ \bibinfo {pages} {5628}
  (\bibinfo {year} {2019})}\BibitemShut {NoStop}%
\bibitem [{\citenamefont {Kawakami}\ \emph {et~al.}(2014)\citenamefont
  {Kawakami}, \citenamefont {Scarlino}, \citenamefont {Ward}, \citenamefont
  {Braakman}, \citenamefont {Savage}, \citenamefont {Lagally}, \citenamefont
  {Friesen}, \citenamefont {Coppersmith}, \citenamefont {Eriksson},\ and\
  \citenamefont {Vandersypen}}]{Kawakami:vg}%
  \BibitemOpen
  \bibfield  {author} {\bibinfo {author} {\bibfnamefont {E.}~\bibnamefont
  {Kawakami}}, \bibinfo {author} {\bibfnamefont {P.}~\bibnamefont {Scarlino}},
  \bibinfo {author} {\bibfnamefont {D.~R.}\ \bibnamefont {Ward}}, \bibinfo
  {author} {\bibfnamefont {F.~R.}\ \bibnamefont {Braakman}}, \bibinfo {author}
  {\bibfnamefont {D.~E.}\ \bibnamefont {Savage}}, \bibinfo {author}
  {\bibfnamefont {M.~G.}\ \bibnamefont {Lagally}}, \bibinfo {author}
  {\bibfnamefont {M.}~\bibnamefont {Friesen}}, \bibinfo {author} {\bibfnamefont
  {S.~N.}\ \bibnamefont {Coppersmith}}, \bibinfo {author} {\bibfnamefont
  {M.~A.}\ \bibnamefont {Eriksson}}, \ and\ \bibinfo {author} {\bibfnamefont
  {L.~M.~K.}\ \bibnamefont {Vandersypen}},\ }\href {\doibase
  10.1038/nnano.2014.153} {\bibfield  {journal} {\bibinfo  {journal} {Nat.
  Nanotechnol.}\ }\textbf {\bibinfo {volume} {9}},\ \bibinfo {pages} {666}
  (\bibinfo {year} {2014})}\BibitemShut {NoStop}%
\bibitem [{\citenamefont {Mills}\ \emph
  {et~al.}(2019{\natexlab{c}})\citenamefont {Mills}, \citenamefont {Zajac},
  \citenamefont {Gullans}, \citenamefont {Schupp}, \citenamefont {Hazard},\
  and\ \citenamefont {Petta}}]{Mills:2019ud}%
  \BibitemOpen
  \bibfield  {author} {\bibinfo {author} {\bibfnamefont {A.~R.}\ \bibnamefont
  {Mills}}, \bibinfo {author} {\bibfnamefont {D.~M.}\ \bibnamefont {Zajac}},
  \bibinfo {author} {\bibfnamefont {M.~J.}\ \bibnamefont {Gullans}}, \bibinfo
  {author} {\bibfnamefont {F.~J.}\ \bibnamefont {Schupp}}, \bibinfo {author}
  {\bibfnamefont {T.~M.}\ \bibnamefont {Hazard}}, \ and\ \bibinfo {author}
  {\bibfnamefont {J.~R.}\ \bibnamefont {Petta}},\ }\href {\doibase
  10.1038/s41467-019-08970-z} {\bibfield  {journal} {\bibinfo  {journal}
  {Nature Commun.}\ }\textbf {\bibinfo {volume} {10}},\ \bibinfo {pages} {1063}
  (\bibinfo {year} {2019}{\natexlab{c}})}\BibitemShut {NoStop}%
\bibitem [{\citenamefont {Mukhopadhyay}\ \emph {et~al.}(2018)\citenamefont
  {Mukhopadhyay}, \citenamefont {Dehollain}, \citenamefont {Reichl},
  \citenamefont {Wegscheider},\ and\ \citenamefont
  {Vandersypen}}]{doi:10.1063/1.5025928}%
  \BibitemOpen
  \bibfield  {author} {\bibinfo {author} {\bibfnamefont {U.}~\bibnamefont
  {Mukhopadhyay}}, \bibinfo {author} {\bibfnamefont {J.~P.}\ \bibnamefont
  {Dehollain}}, \bibinfo {author} {\bibfnamefont {C.}~\bibnamefont {Reichl}},
  \bibinfo {author} {\bibfnamefont {W.}~\bibnamefont {Wegscheider}}, \ and\
  \bibinfo {author} {\bibfnamefont {L.~M.~K.}\ \bibnamefont {Vandersypen}},\
  }\href {\doibase 10.1063/1.5025928} {\bibfield  {journal} {\bibinfo
  {journal} {Appl. Phys. Lett.}\ }\textbf {\bibinfo {volume} {112}},\ \bibinfo
  {pages} {183505} (\bibinfo {year} {2018})}\BibitemShut {NoStop}%
\bibitem [{\citenamefont {Karzig}\ \emph
  {et~al.}(2017{\natexlab{b}})\citenamefont {Karzig}, \citenamefont {Knapp},
  \citenamefont {Lutchyn}, \citenamefont {Bonderson}, \citenamefont {Hastings},
  \citenamefont {Nayak}, \citenamefont {Alicea}, \citenamefont {Flensberg},
  \citenamefont {Plugge}, \citenamefont {Oreg}, \citenamefont {Marcus},\ and\
  \citenamefont {Freedman}}]{PhysRevB.95.235305}%
  \BibitemOpen
  \bibfield  {author} {\bibinfo {author} {\bibfnamefont {T.}~\bibnamefont
  {Karzig}}, \bibinfo {author} {\bibfnamefont {C.}~\bibnamefont {Knapp}},
  \bibinfo {author} {\bibfnamefont {R.~M.}\ \bibnamefont {Lutchyn}}, \bibinfo
  {author} {\bibfnamefont {P.}~\bibnamefont {Bonderson}}, \bibinfo {author}
  {\bibfnamefont {M.~B.}\ \bibnamefont {Hastings}}, \bibinfo {author}
  {\bibfnamefont {C.}~\bibnamefont {Nayak}}, \bibinfo {author} {\bibfnamefont
  {J.}~\bibnamefont {Alicea}}, \bibinfo {author} {\bibfnamefont
  {K.}~\bibnamefont {Flensberg}}, \bibinfo {author} {\bibfnamefont
  {S.}~\bibnamefont {Plugge}}, \bibinfo {author} {\bibfnamefont
  {Y.}~\bibnamefont {Oreg}}, \bibinfo {author} {\bibfnamefont {C.~M.}\
  \bibnamefont {Marcus}}, \ and\ \bibinfo {author} {\bibfnamefont {M.~H.}\
  \bibnamefont {Freedman}},\ }\href {\doibase 10.1103/PhysRevB.95.235305}
  {\bibfield  {journal} {\bibinfo  {journal} {Phys. Rev. B}\ }\textbf {\bibinfo
  {volume} {95}},\ \bibinfo {pages} {235305} (\bibinfo {year}
  {2017}{\natexlab{b}})}\BibitemShut {NoStop}%
\bibitem [{\citenamefont {Albrecht}\ \emph {et~al.}(2016)\citenamefont
  {Albrecht}, \citenamefont {Higginbotham}, \citenamefont {Madsen},
  \citenamefont {Kuemmeth}, \citenamefont {Jespersen}, \citenamefont
  {Nyg{\aa}rd}, \citenamefont {Krogstrup},\ and\ \citenamefont
  {Marcus}}]{Albrecht:vq}%
  \BibitemOpen
  \bibfield  {author} {\bibinfo {author} {\bibfnamefont {S.~M.}\ \bibnamefont
  {Albrecht}}, \bibinfo {author} {\bibfnamefont {A.~P.}\ \bibnamefont
  {Higginbotham}}, \bibinfo {author} {\bibfnamefont {M.}~\bibnamefont
  {Madsen}}, \bibinfo {author} {\bibfnamefont {F.}~\bibnamefont {Kuemmeth}},
  \bibinfo {author} {\bibfnamefont {T.~S.}\ \bibnamefont {Jespersen}}, \bibinfo
  {author} {\bibfnamefont {J.}~\bibnamefont {Nyg{\aa}rd}}, \bibinfo {author}
  {\bibfnamefont {P.}~\bibnamefont {Krogstrup}}, \ and\ \bibinfo {author}
  {\bibfnamefont {C.~M.}\ \bibnamefont {Marcus}},\ }\href {\doibase
  10.1038/nature17162} {\bibfield  {journal} {\bibinfo  {journal} {Nature}\
  }\textbf {\bibinfo {volume} {531}},\ \bibinfo {pages} {206} (\bibinfo {year}
  {2016})}\BibitemShut {NoStop}%
\bibitem [{\citenamefont {Mourik}\ \emph {et~al.}(2012)\citenamefont {Mourik},
  \citenamefont {Zuo}, \citenamefont {Frolov}, \citenamefont {Plissard},
  \citenamefont {Bakkers},\ and\ \citenamefont {Kouwenhoven}}]{Mourik1003}%
  \BibitemOpen
  \bibfield  {author} {\bibinfo {author} {\bibfnamefont {V.}~\bibnamefont
  {Mourik}}, \bibinfo {author} {\bibfnamefont {K.}~\bibnamefont {Zuo}},
  \bibinfo {author} {\bibfnamefont {S.~M.}\ \bibnamefont {Frolov}}, \bibinfo
  {author} {\bibfnamefont {S.~R.}\ \bibnamefont {Plissard}}, \bibinfo {author}
  {\bibfnamefont {E.~P. A.~M.}\ \bibnamefont {Bakkers}}, \ and\ \bibinfo
  {author} {\bibfnamefont {L.~P.}\ \bibnamefont {Kouwenhoven}},\ }\href
  {\doibase 10.1126/science.1222360} {\bibfield  {journal} {\bibinfo  {journal}
  {Science}\ }\textbf {\bibinfo {volume} {336}},\ \bibinfo {pages} {1003}
  (\bibinfo {year} {2012})}\BibitemShut {NoStop}%
\bibitem [{\citenamefont {Sutton}\ and\ \citenamefont
  {Barto}(2018)}]{sutton2018reinforcement}%
  \BibitemOpen
  \bibfield  {author} {\bibinfo {author} {\bibfnamefont {R.~S.}\ \bibnamefont
  {Sutton}}\ and\ \bibinfo {author} {\bibfnamefont {A.~G.}\ \bibnamefont
  {Barto}},\ }\href@noop {} {\emph {\bibinfo {title} {Reinforcement learning:
  An introduction}}}\ (\bibinfo  {publisher} {MIT press},\ \bibinfo {year}
  {2018})\BibitemShut {NoStop}%
\bibitem [{\citenamefont {Liu}\ and\ \citenamefont
  {West}(2001)}]{liu2001combined}%
  \BibitemOpen
  \bibfield  {author} {\bibinfo {author} {\bibfnamefont {J.}~\bibnamefont
  {Liu}}\ and\ \bibinfo {author} {\bibfnamefont {M.}~\bibnamefont {West}},\
  }in\ \href@noop {} {\emph {\bibinfo {booktitle} {Sequential Monte Carlo
  methods in practice}}}\ (\bibinfo  {publisher} {Springer},\ \bibinfo {year}
  {2001})\ pp.\ \bibinfo {pages} {197--223}\BibitemShut {NoStop}%
\bibitem [{\citenamefont {Granade}\ \emph {et~al.}(2012)\citenamefont
  {Granade}, \citenamefont {Ferrie}, \citenamefont {Wiebe},\ and\ \citenamefont
  {Cory}}]{granade2012robust}%
  \BibitemOpen
  \bibfield  {author} {\bibinfo {author} {\bibfnamefont {C.~E.}\ \bibnamefont
  {Granade}}, \bibinfo {author} {\bibfnamefont {C.}~\bibnamefont {Ferrie}},
  \bibinfo {author} {\bibfnamefont {N.}~\bibnamefont {Wiebe}}, \ and\ \bibinfo
  {author} {\bibfnamefont {D.~G.}\ \bibnamefont {Cory}},\ }\href {\doibase
  10.1088/1367-2630/14/10/103013} {\bibfield  {journal} {\bibinfo  {journal}
  {New J. Phys.}\ }\textbf {\bibinfo {volume} {14}},\ \bibinfo {pages} {103013}
  (\bibinfo {year} {2012})}\BibitemShut {NoStop}%
\end{thebibliography}%
\clearpage

\setcounter{figure}{0} \renewcommand{\thefigure}{A.\arabic{figure}} 
\setcounter{figure}{0} \renewcommand{\thetable}{A.\arabic{figure}} 
\appendix 


\section{Metric fluctuations}\label{ax:data_fluctuation}
\begin{figure}[!t]
\includegraphics[width=\columnwidth]{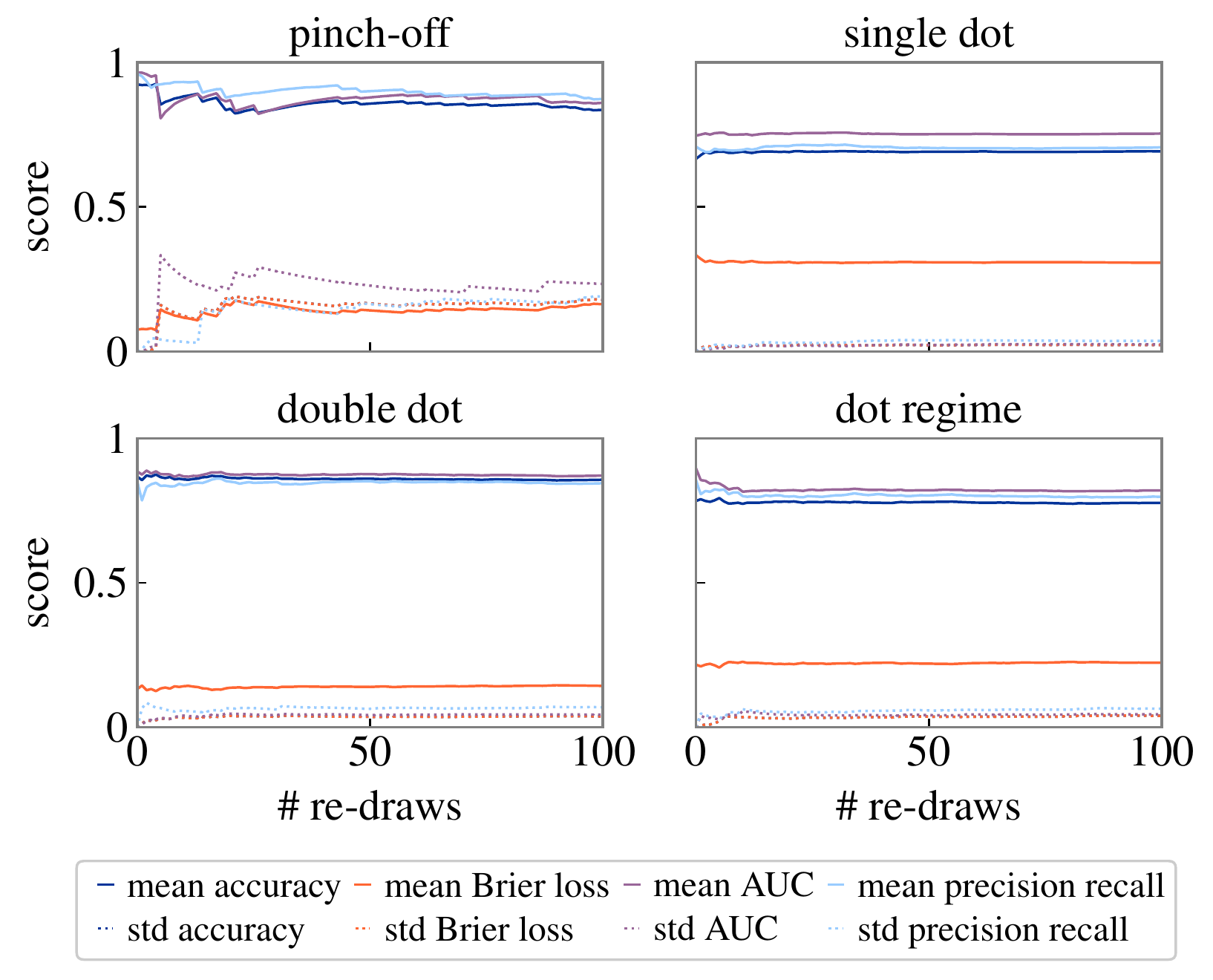}
 \caption{Classifier performance dependance on the number of redraws during training and testing.  Performance of Support Vector Machines on pinch-off, single dot and double dot current maps are evaluated over an increasing number of train and test data selection among the available data. Performances vary for a small number of iterations for individual gate characterization data while they do not change much for charge stability diagram data.  We choose to average individual gate characterizations quality classification over $n = 20$ times and charge stability diagram quality and state prediction over $n = 10$.}
    \label{fig:metric_fluctuations}
\end{figure}

In \autoref{fig:metric_fluctuations} we show the dependance of the classifier performance on the number of redraws during training and testing. Based on this data we have choose to average over $n = 20$ redraws for individual gate characterization and $n = 20$ for charge state  classification.

\section{Classifier choice}\label{ax:classifiers}

We studied classifier performance in three stages. First we compare performances of 12 classifiers, namely Decision Tree, Gaussian Process, Quadratic Discriminant Analysis, Random Forest Multi-layer Perceptron, Logistic Regression, Support Vector Machines and $k$-Nearest Neighbours. We compute performances by comparing accuracy which is defined as the total number of samples (true positive (TP) and true negative (TN)) over the total number of samples (all positive (P) and all negtive(N)):
\be
\text{ACC} = \frac{\text{TP} + \text{TN}}{\text{P} + \text{N}}.
\ee
The accuracy is determined over $n$ train and test splits of all available training data, selected randomly and with equal population sizes, i.e equal numbers of good and poor results, where e choose $n = 20$ for 1D data and $n = 10$ for 2D.
We compare how these classifiers perform with default hyper parameters and on different feature vectors, i.e. normalized current, Fourier transform, both or on extracted features if they are available. Based on this initial analysis we discard two poorly performing classifiers, Quadratic Discriminant Analysis and Gaussian Process. All remaining classifiers are then optimized using scikit-learn's selection.GridSearchCV \citep{scikit-learn} method to determine their hyper parameters $h$.  The grid search method evaluates all possible combinations of a supplied range of possible hyper parameters to select the optimal one. Given the long run times of this method we optimize each classifier for a single split of available data into train and test subsets. This slightly overestimates each classifier's performance as it optimizes hyper parameters to one train and test data split. We therefore compute performances of classifiers with their respective hyper parameters again and over $n = 20$ and  $n = 10$ train and test data splits for pinch-off and charge stability diagram data respectively. Based on these results we choose the Decision Tree classifier to predict pinch-off curves and the Multilayer Perceptron to determine charge stability diagram quality and charge state.  Lastly, we look at the confusion matrix of each of these classifiers which is defined by the number of false positives (FP), false negatives (FN), true positives (TP), and true negatives (TN):
\be
CM =
\begin{pmatrix}
 TP & FP\\
 FN & TN\\
\end{pmatrix}.
\ee

For pinch-off curves classified using a Decision Tree classifier we find
\be
\begin{split}
CM_{\text{DT,  pinch-off}}  & =
\begin{pmatrix}
70.98 & 8.24 \\
4.71 & 74.07
\end{pmatrix} \\
\text{ACC}_{\text{DT,  pinch-off}} & = 0.9181 \pm 0.0442,
\end{split}
\ee
for single dot charge stability diagram quality classified using a Multi-layer Perceptron
\be
\begin{split}
CM_{\text{MLP, single dot}} &=
\begin{pmatrix}
129.5 & 29.25 \\
28.05 & 128.2
\end{pmatrix} \\
\text{ACC}_{\text{MLP, single dot}}   &= 0.8181 \pm 0.0178,
\end{split}
\ee
for double dot stability diagram quality classified using a Multi-layer Perceptron
\be
\begin{split}
CM_{\text{MLP, single dot}} &=
\begin{pmatrix}
37.72 & 3.84 \\
5.6 & 35.84
\end{pmatrix} \\
 \text{ACC}_{\text{MLP, single dot}}  &= 0.8863 \pm 0.0427,
\end{split}
\ee
and for dot regime of charge stability diagrams classified using a Multi-layer Perceptron:
\be
\begin{split}
CM_{\text{MLP, single dot}} & =
\begin{pmatrix}
33.95 & 7.45 \\
5.65 & 35.95
\end{pmatrix} \\
 \text{ACC}_{\text{MLP, single dot}}   &= 0.8422 \pm 0.0363.
\end{split}
\ee

\begin{figure}[!t]
\includegraphics[width=\columnwidth]{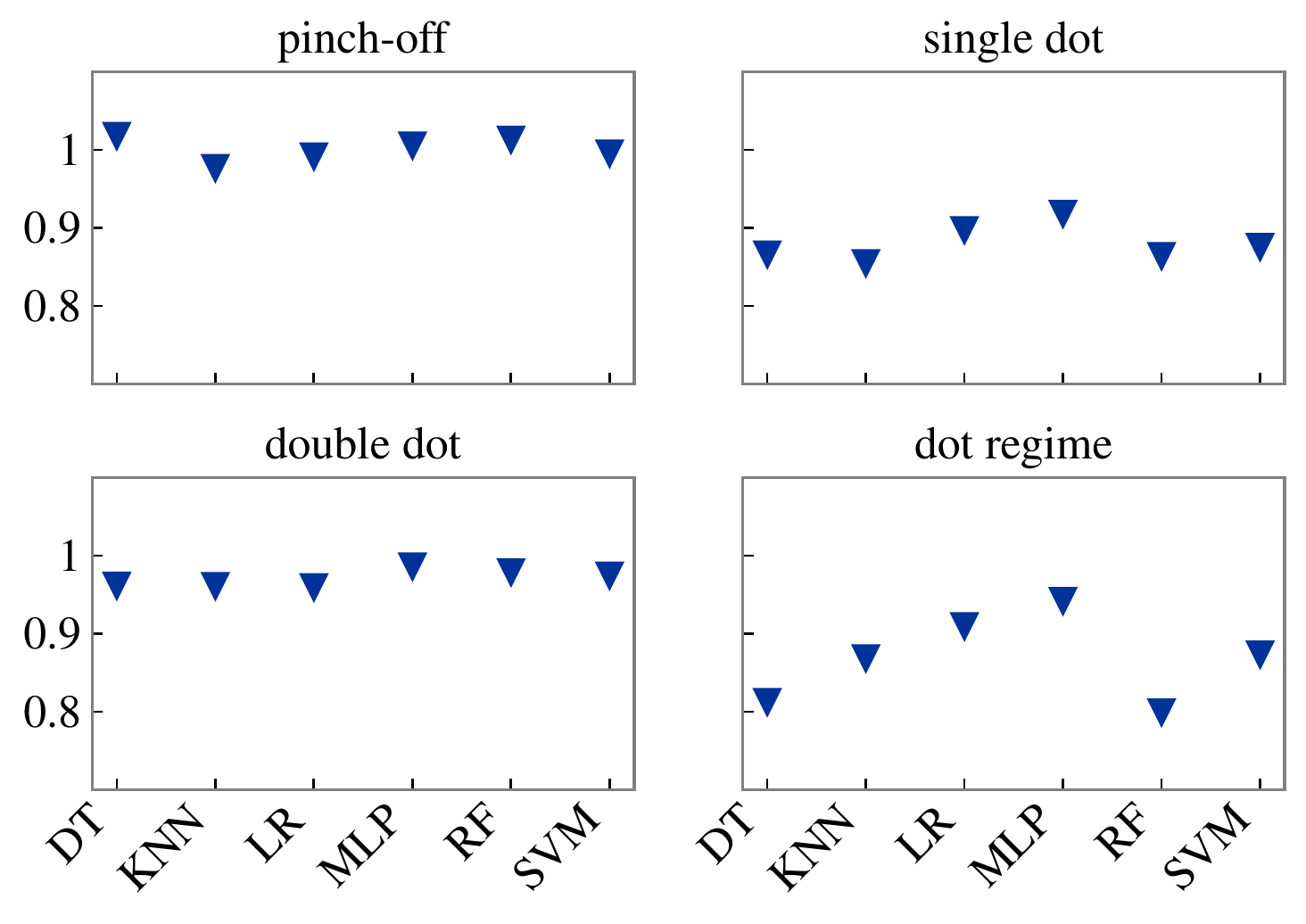}
 \caption{Accuracies of optimized classifiers for individual gate characterizations (pinch-off), single dot, double dot and dot regime. Performances are high for pinch-off curves and double dot quality and relatively low for single dot quality and dot regime. }
    \label{fig:best_performances}
\end{figure}

The tables below summarize all intermediate performance evaluations as well as hyper parameter optimization results.
Final performances are illustrated in \autoref{fig:best_performances}  and optimized hyper parameters with corresponding accuracy over one train and test data split in \autoref{tab:hyperparameters_DecisionTreeClassifier} to \ref{tab:hyperparameters_SVC},  for  the Decision Tree,  $k$-earest Neighbor classifier,   Logistic regression,   Multi-layer Perceptron,  Random Forest and Support Vector Machine classifier respectively. Performances of classifiers with default hyper parameters are listed in \autoref{tab:po_clf_metrics} for individual gate characterizations, \autoref{tab:sd_clf_metrics} for single dot quality, \autoref{tab:dd_clf_metrics} for double dot quality and \autoref{tab:dr_clf_metrics} for dot regime.

\begin{table}[H]
\resizebox{1\linewidth}{!}{\begin{tabular}{lcccc}
\hline \hline 
   hyper parameter & pinch-off & single dot & double dot & dot regime \\
\hline
         criterion &      gini &    entropy &    entropy &    entropy \\
      max\ features &      auto &       None &       sqrt &       auto \\
  min\ samples\ leaf &         2 &          3 &          3 &          1 \\
 min\ samples\ split &         6 &          2 &          6 &          2 \\
          splitter &    random &       best &       best &     random \\
          accuracy &  0.928571 &   0.755768 &   0.861027 &   0.767372 \\
\hline \hline
\end{tabular}
}
\caption{Optimized hyper parameters for the Decision Tree classifier determined by performing a grid search over one train and test split.}
\label{tab:hyperparameters_DecisionTreeClassifier}
\end{table}

\begin{table}[H]
\resizebox{1\linewidth}{!}{\begin{tabular}{lcccc}
\hline \hline 
hyper parameter & pinch-off & single dot & double dot & dot regime \\
\hline
      algorithm &      auto &       auto &       auto &       auto \\
      leaf\ size &        10 &         10 &         10 &         10 \\
         n\ jobs &        -1 &         -1 &         -1 &         -1 \\
    n\ neighbors &         2 &          2 &          2 &          2 \\
              p &         3 &          2 &          1 &          2 \\
        weights &  distance &   distance &   distance &    uniform \\
       accuracy &  0.912698 &   0.790772 &   0.903323 &    0.81571 \\
\hline \hline
\end{tabular}
}
\caption{Optimized hyper parameters for the $k$-Nearest Neighbor classifier determined by performing a grid search over one train and test split.}
\label{tab:hyperparameters_KNeighborsClassifier}
\end{table}

\begin{table}[H]
\resizebox{1\linewidth}{!}{\begin{tabular}{lcccc}
\hline \hline 
hyper parameter &  pinch-off & single dot & double dot & dot regime \\
\hline
              C &        100 &        0.1 &       1000 &        0.1 \\
   class\ weight &   balanced &       None &   balanced &   balanced \\
  fit\ intercept &       True &       True &       True &       True \\
       max\ iter &       1000 &       1000 &       1000 &       1000 \\
         n\ jobs &         -1 &         -1 &         -1 &         -1 \\
        penalty &         l1 &         l1 &         l2 &         l2 \\
         solver &  liblinear &  liblinear &  newton-cg &        sag \\
       accuracy &    0.92381 &   0.820207 &   0.897281 &   0.836858 \\
\hline \hline
\end{tabular}
}
\caption{Optimized hyper parameters for the Logistic Regression classifier determined by performing a grid search over one train and test split.}
\label{tab:hyperparameters_LogisticRegression}
\end{table}

\begin{table}[H]
\resizebox{1\linewidth}{!}{\begin{tabular}{lcccc}
\hline \hline 
    hyper parameter & pinch-off & single dot &  double dot & dot regime \\
\hline
         activation &      relu &       relu &    logistic &       relu \\
              alpha &    0.0001 &        0.1 &       0.001 &      0.001 \\
         batch\ size &       100 &        300 &         200 &        200 \\
 hidden\ layer\ sizes &     [100] &      [100] &       [200] &      [300] \\
      learning\ rate &  constant &   adaptive &  invscaling &   constant \\
           max\ iter &      3000 &       3000 &        3000 &       3000 \\
            power\ t &       0.1 &        0.6 &         0.6 &        0.4 \\
             solver &     lbfgs &        sgd &       lbfgs &        sgd \\
           accuracy &   0.94127 &   0.845664 &    0.936556 &   0.876133 \\
\hline \hline
\end{tabular}
}
\caption{Optimized hyper parameters for the Multi-layer Perceptron classifier determined by performing a grid search over one train and test split.}
\label{tab:hyperparameters_MLPClassifier}
\end{table}

\begin{table}[!ht]
\resizebox{1\linewidth}{!}{\begin{tabular}{lcccc}
\hline \hline 
   hyper parameter & pinch-off & single dot & double dot & dot regime \\
\hline
         criterion &      gini &    entropy &    entropy &       gini \\
      max\ features &      auto &       sqrt &       auto &       log2 \\
  min\ samples\ leaf &         1 &          2 &          1 &          2 \\
 min\ samples\ split &         6 &          4 &          2 &          2 \\
      n\ estimators &        10 &        500 &        100 &        100 \\
            n\ jobs &        -1 &         -1 &         -1 &         -1 \\
          accuracy &  0.950794 &   0.833731 &   0.918429 &   0.818731 \\
\hline \hline
\end{tabular}
}
\caption{Optimized hyper parameters for the Random Forest classifier determined by performing a grid search over one train and test split.}
\label{tab:hyperparameters_RandomForestClassifier}
\end{table}

\begin{table}[H]
\resizebox{1\linewidth}{!}{\begin{tabular}{lcccc}
\hline \hline 
hyper parameter & pinch-off & single dot & double dot & dot regime \\
\hline
              C &      1000 &        0.1 &        0.1 &         10 \\
          gamma &         1 &        0.1 &        0.1 &        0.1 \\
         kernel &       rbf &       poly &       poly &     linear \\
       accuracy &   0.94127 &   0.786794 &   0.879154 &   0.776435 \\
\hline \hline
\end{tabular}
}
\caption{Optimized hyper parameters for the Support Vector Machine determined by performing a grid search over one train and test split.}
\label{tab:hyperparameters_SVC}
\end{table}

\begin{table*}[!h]
\setlength{\tabcolsep}{8pt}
\renewcommand{\arraystretch}{1.2}
\small
\begin{tabular}{@{\extracolsep{6pt}}lcccc}
 \\ \hline  \hline & \multicolumn{2}{c}{PCA} & \multicolumn{2}{c}{no PCA} \\ \cline{2-3} \cline{4-5} 
classifier & accuracy & evaluation time [s] & accuracy & evaluation time [s] \\ 
\hline \multicolumn{5}{c}{ Normalized current}  \\
 \hline 
                   Decision Tree &  0.889 $\pm$ 0.139 &  0.0001 $\pm$ 0.0000 &  0.849 $\pm$ 0.155 &  0.0001 $\pm$ 0.0000 \\
                Gaussiam Process &  0.854 $\pm$ 0.180 &  0.0266 $\pm$ 0.0012 &  0.847 $\pm$ 0.188 &  0.0052 $\pm$ 0.0010 \\
           $k$-Nearest Neighbors &  0.925 $\pm$ 0.013 &  0.0222 $\pm$ 0.0022 &  0.875 $\pm$ 0.085 &  0.0075 $\pm$ 0.0003 \\
             Logistic Regression &  0.879 $\pm$ 0.136 &  0.0001 $\pm$ 0.0000 &  0.877 $\pm$ 0.136 &  0.0001 $\pm$ 0.0000 \\
           Multilayer Perceptron &  0.887 $\pm$ 0.140 &  0.0005 $\pm$ 0.0001 &  0.833 $\pm$ 0.187 &  0.0003 $\pm$ 0.0001 \\
 Quadratic Discriminant Analysis &  0.857 $\pm$ 0.025 &  0.0009 $\pm$ 0.0001 &  0.803 $\pm$ 0.184 &  0.0002 $\pm$ 0.0000 \\
                   Random Forest &  0.860 $\pm$ 0.155 &  0.0010 $\pm$ 0.0001 &  0.882 $\pm$ 0.108 &  0.0010 $\pm$ 0.0001 \\
          Support Vector Machine &  0.846 $\pm$ 0.169 &  0.0085 $\pm$ 0.0074 &  0.884 $\pm$ 0.133 &  0.0007 $\pm$ 0.0003 \\
\\\hline  \hline\multicolumn{5}{c}{ Fourier transform}  \\
 \hline 
                   Decision Tree &  0.909 $\pm$ 0.018 &  0.0001 $\pm$ 0.0000 &  0.891 $\pm$ 0.029 &  0.0001 $\pm$ 0.0000 \\
                Gaussiam Process &  0.846 $\pm$ 0.162 &  0.0263 $\pm$ 0.0007 &  0.854 $\pm$ 0.153 &  0.0109 $\pm$ 0.0029 \\
           $k$-Nearest Neighbors &  0.867 $\pm$ 0.035 &  0.0277 $\pm$ 0.0012 &  0.871 $\pm$ 0.046 &  0.0104 $\pm$ 0.0012 \\
             Logistic Regression &  0.866 $\pm$ 0.129 &  0.0001 $\pm$ 0.0001 &  0.884 $\pm$ 0.101 &  0.0001 $\pm$ 0.0000 \\
           Multilayer Perceptron &  0.811 $\pm$ 0.184 &  0.0005 $\pm$ 0.0001 &  0.856 $\pm$ 0.124 &  0.0004 $\pm$ 0.0001 \\
 Quadratic Discriminant Analysis &  0.810 $\pm$ 0.029 &  0.0009 $\pm$ 0.0001 &  0.722 $\pm$ 0.093 &  0.0004 $\pm$ 0.0001 \\
                   Random Forest &  0.890 $\pm$ 0.019 &  0.0012 $\pm$ 0.0001 &  0.852 $\pm$ 0.038 &  0.0010 $\pm$ 0.0001 \\
          Support Vector Machine &  0.853 $\pm$ 0.151 &  0.0077 $\pm$ 0.0062 &  0.863 $\pm$ 0.144 &  0.0018 $\pm$ 0.0003 \\
\\\hline  \hline\multicolumn{5}{c}{ Normalized current \& Fourier transform}  \\
 \hline 
                   Decision Tree &  0.929 $\pm$ 0.021 &  0.0001 $\pm$ 0.0000 &  0.904 $\pm$ 0.031 &  0.0001 $\pm$ 0.0000 \\
                Gaussiam Process &  0.849 $\pm$ 0.177 &  0.0523 $\pm$ 0.0010 &  0.868 $\pm$ 0.156 &  0.0094 $\pm$ 0.0027 \\
           $k$-Nearest Neighbors &  0.899 $\pm$ 0.033 &  0.0483 $\pm$ 0.0033 &  0.902 $\pm$ 0.052 &  0.0103 $\pm$ 0.0014 \\
             Logistic Regression &  0.865 $\pm$ 0.142 &  0.0001 $\pm$ 0.0000 &  0.851 $\pm$ 0.160 &  0.0001 $\pm$ 0.0000 \\
           Multilayer Perceptron &  0.899 $\pm$ 0.111 &  0.0007 $\pm$ 0.0001 &  0.856 $\pm$ 0.153 &  0.0004 $\pm$ 0.0001 \\
 Quadratic Discriminant Analysis &  0.810 $\pm$ 0.022 &  0.0018 $\pm$ 0.0001 &  0.843 $\pm$ 0.115 &  0.0004 $\pm$ 0.0001 \\
                   Random Forest &  0.911 $\pm$ 0.030 &  0.0011 $\pm$ 0.0002 &  0.862 $\pm$ 0.039 &  0.0010 $\pm$ 0.0001 \\
          Support Vector Machine &  0.798 $\pm$ 0.192 &  0.0214 $\pm$ 0.0186 &  0.849 $\pm$ 0.171 &  0.0018 $\pm$ 0.0004 \\
\\\hline  \hline\multicolumn{5}{c}{ Selected features}  \\
 \hline 
                   Decision Tree &  0.928 $\pm$ 0.020 &  0.0001 $\pm$ 0.0000 &  0.921 $\pm$ 0.028 &  0.0001 $\pm$ 0.0000 \\
                Gaussiam Process &  0.922 $\pm$ 0.024 &  0.0023 $\pm$ 0.0002 &  0.919 $\pm$ 0.027 &  0.0025 $\pm$ 0.0002 \\
           $k$-Nearest Neighbors &  0.896 $\pm$ 0.040 &  0.0064 $\pm$ 0.0005 &  0.898 $\pm$ 0.033 &  0.0052 $\pm$ 0.0002 \\
             Logistic Regression &  0.916 $\pm$ 0.021 &  0.0001 $\pm$ 0.0000 &  0.904 $\pm$ 0.037 &  0.0001 $\pm$ 0.0000 \\
           Multilayer Perceptron &  0.927 $\pm$ 0.023 &  0.0002 $\pm$ 0.0000 &  0.917 $\pm$ 0.029 &  0.0002 $\pm$ 0.0000 \\
 Quadratic Discriminant Analysis &  0.872 $\pm$ 0.037 &  0.0002 $\pm$ 0.0000 &  0.870 $\pm$ 0.034 &  0.0001 $\pm$ 0.0000 \\
                   Random Forest &  0.939 $\pm$ 0.024 &  0.0010 $\pm$ 0.0001 &  0.933 $\pm$ 0.018 &  0.0009 $\pm$ 0.0001 \\
          Support Vector Machine &  0.895 $\pm$ 0.084 &  0.0012 $\pm$ 0.0028 &  0.914 $\pm$ 0.028 &  0.0005 $\pm$ 0.0002 \\
\hline \hline
\end{tabular}

\caption{Classifier performances on individual gate characterisations performed (from bottom to top)  on normalized current, Fourier transform, both and selected features and with and without PCA. Accuracies were computed over $n = 20$  train and test splits of the available data. }
\label{tab:po_clf_metrics}
\end{table*}

\begin{table*}[!h]
\setlength{\tabcolsep}{8pt}
\renewcommand{\arraystretch}{1.2}
\small
\begin{tabular}{@{\extracolsep{6pt}}lcccc}
 \\ \hline  \hline & \multicolumn{2}{c}{PCA} & \multicolumn{2}{c}{no PCA} \\ \cline{2-3} \cline{4-5} 
classifier & accuracy & evaluation time [s] & accuracy & evaluation time [s] \\ 
\hline \multicolumn{5}{c}{ Normalized current}  \\
 \hline 
                   Decision Tree &  0.743 $\pm$ 0.030 &  0.0015 $\pm$ 0.0001 &  0.737 $\pm$ 0.025 &  0.0001 $\pm$ 0.0000 \\
                Gaussiam Process &  0.732 $\pm$ 0.032 &  1.1919 $\pm$ 0.0082 &  0.741 $\pm$ 0.020 &  0.0261 $\pm$ 0.0013 \\
           $k$-Nearest Neighbors &  0.676 $\pm$ 0.026 &  1.0636 $\pm$ 0.0255 &  0.727 $\pm$ 0.030 &  0.0182 $\pm$ 0.0004 \\
             Logistic Regression &  0.717 $\pm$ 0.021 &  0.0006 $\pm$ 0.0000 &  0.753 $\pm$ 0.021 &  0.0001 $\pm$ 0.0000 \\
           Multilayer Perceptron &  0.752 $\pm$ 0.027 &  0.0038 $\pm$ 0.0002 &  0.755 $\pm$ 0.022 &  0.0004 $\pm$ 0.0000 \\
 Quadratic Discriminant Analysis &  0.559 $\pm$ 0.020 &  0.0418 $\pm$ 0.0017 &  0.766 $\pm$ 0.019 &  0.0004 $\pm$ 0.0001 \\
                   Random Forest &  0.763 $\pm$ 0.025 &  0.0026 $\pm$ 0.0001 &  0.697 $\pm$ 0.027 &  0.0011 $\pm$ 0.0000 \\
          Support Vector Machine &  0.693 $\pm$ 0.023 &  0.6249 $\pm$ 0.0188 &  0.759 $\pm$ 0.022 &  0.0108 $\pm$ 0.0005 \\
\\\hline  \hline\multicolumn{5}{c}{ Fourier transform}  \\
 \hline 
                   Decision Tree &  0.689 $\pm$ 0.020 &  0.0016 $\pm$ 0.0001 &  0.661 $\pm$ 0.024 &  0.0002 $\pm$ 0.0000 \\
                Gaussiam Process &  0.637 $\pm$ 0.022 &  1.2200 $\pm$ 0.0110 &  0.661 $\pm$ 0.030 &  0.2101 $\pm$ 0.0085 \\
           $k$-Nearest Neighbors &  0.688 $\pm$ 0.023 &  1.5608 $\pm$ 0.0284 &  0.682 $\pm$ 0.034 &  0.2085 $\pm$ 0.0083 \\
             Logistic Regression &  0.699 $\pm$ 0.030 &  0.0006 $\pm$ 0.0001 &  0.707 $\pm$ 0.024 &  0.0001 $\pm$ 0.0000 \\
           Multilayer Perceptron &  0.741 $\pm$ 0.023 &  0.0037 $\pm$ 0.0003 &  0.733 $\pm$ 0.021 &  0.0009 $\pm$ 0.0000 \\
 Quadratic Discriminant Analysis &  0.541 $\pm$ 0.026 &  0.0403 $\pm$ 0.0013 &  0.696 $\pm$ 0.020 &  0.0060 $\pm$ 0.0004 \\
                   Random Forest &  0.686 $\pm$ 0.024 &  0.0031 $\pm$ 0.0000 &  0.585 $\pm$ 0.026 &  0.0013 $\pm$ 0.0001 \\
          Support Vector Machine &  0.665 $\pm$ 0.025 &  0.5690 $\pm$ 0.0185 &  0.693 $\pm$ 0.026 &  0.0869 $\pm$ 0.0042 \\
\\\hline  \hline\multicolumn{5}{c}{ Normalized current \& Fourier transform}  \\
 \hline 
                   Decision Tree &  0.752 $\pm$ 0.027 &  0.0025 $\pm$ 0.0003 &  0.738 $\pm$ 0.035 &  0.0003 $\pm$ 0.0000 \\
                Gaussiam Process &  0.679 $\pm$ 0.025 &  2.4434 $\pm$ 0.0123 &  0.700 $\pm$ 0.021 &  0.1919 $\pm$ 0.0071 \\
           $k$-Nearest Neighbors &  0.748 $\pm$ 0.019 &  3.1041 $\pm$ 0.0306 &  0.757 $\pm$ 0.021 &  0.1542 $\pm$ 0.0084 \\
             Logistic Regression &  0.797 $\pm$ 0.016 &  0.0011 $\pm$ 0.0001 &  0.797 $\pm$ 0.022 &  0.0001 $\pm$ 0.0000 \\
           Multilayer Perceptron &  0.820 $\pm$ 0.017 &  0.0070 $\pm$ 0.0003 &  0.795 $\pm$ 0.021 &  0.0009 $\pm$ 0.0001 \\
 Quadratic Discriminant Analysis &  0.553 $\pm$ 0.028 &  0.0819 $\pm$ 0.0015 &  0.757 $\pm$ 0.025 &  0.0059 $\pm$ 0.0005 \\
                   Random Forest &  0.765 $\pm$ 0.024 &  0.0042 $\pm$ 0.0002 &  0.567 $\pm$ 0.041 &  0.0013 $\pm$ 0.0001 \\
          Support Vector Machine &  0.773 $\pm$ 0.023 &  0.8536 $\pm$ 0.0259 &  0.773 $\pm$ 0.018 &  0.0577 $\pm$ 0.0029 \\
\hline \hline
\end{tabular}

\caption{Classifier performances on single dot charge stability diagrams performed (from bottom to top)  on normalized current, Fourier transform and both and with and without PCA. Accuracies were computed over $n = 10$  train and test splits of the available data. }
\label{tab:sd_clf_metrics}
\end{table*}

\begin{table*}[!h]
\setlength{\tabcolsep}{8pt}
\renewcommand{\arraystretch}{1.2}
\small
\begin{tabular}{@{\extracolsep{6pt}}lcccc}
 \\ \hline  \hline & \multicolumn{2}{c}{PCA} & \multicolumn{2}{c}{no PCA} \\ \cline{2-3} \cline{4-5} 
classifier & accuracy & evaluation time [s] & accuracy & evaluation time [s] \\ 
\hline \multicolumn{5}{c}{ Normalized current}  \\
 \hline 
                   Decision Tree &  0.848 $\pm$ 0.052 &  0.0005 $\pm$ 0.0001 &  0.873 $\pm$ 0.043 &  0.0001 $\pm$ 0.0000 \\
                Gaussiam Process &  0.829 $\pm$ 0.044 &  0.0924 $\pm$ 0.0122 &  0.854 $\pm$ 0.039 &  0.0071 $\pm$ 0.0019 \\
           $k$-Nearest Neighbors &  0.802 $\pm$ 0.065 &  0.0962 $\pm$ 0.0079 &  0.830 $\pm$ 0.040 &  0.0062 $\pm$ 0.0011 \\
             Logistic Regression &  0.707 $\pm$ 0.063 &  0.0002 $\pm$ 0.0000 &  0.704 $\pm$ 0.072 &  0.0001 $\pm$ 0.0000 \\
           Multilayer Perceptron &  0.786 $\pm$ 0.041 &  0.0021 $\pm$ 0.0008 &  0.848 $\pm$ 0.027 &  0.0006 $\pm$ 0.0013 \\
 Quadratic Discriminant Analysis &  0.575 $\pm$ 0.080 &  0.0720 $\pm$ 0.0064 &  0.743 $\pm$ 0.062 &  0.0002 $\pm$ 0.0000 \\
                   Random Forest &  0.886 $\pm$ 0.030 &  0.0019 $\pm$ 0.0001 &  0.810 $\pm$ 0.032 &  0.0011 $\pm$ 0.0001 \\
          Support Vector Machine &  0.686 $\pm$ 0.059 &  0.0666 $\pm$ 0.0041 &  0.674 $\pm$ 0.057 &  0.0011 $\pm$ 0.0004 \\
\\\hline  \hline\multicolumn{5}{c}{ Fourier transform}  \\
 \hline 
                   Decision Tree &  0.767 $\pm$ 0.046 &  0.0005 $\pm$ 0.0001 &  0.753 $\pm$ 0.045 &  0.0001 $\pm$ 0.0000 \\
                Gaussiam Process &  0.736 $\pm$ 0.037 &  0.0863 $\pm$ 0.0008 &  0.774 $\pm$ 0.040 &  0.0128 $\pm$ 0.0015 \\
           $k$-Nearest Neighbors &  0.780 $\pm$ 0.043 &  0.1230 $\pm$ 0.0059 &  0.785 $\pm$ 0.050 &  0.0099 $\pm$ 0.0012 \\
             Logistic Regression &  0.799 $\pm$ 0.040 &  0.0002 $\pm$ 0.0001 &  0.798 $\pm$ 0.051 &  0.0011 $\pm$ 0.0009 \\
           Multilayer Perceptron &  0.822 $\pm$ 0.048 &  0.0019 $\pm$ 0.0002 &  0.827 $\pm$ 0.041 &  0.0041 $\pm$ 0.0032 \\
 Quadratic Discriminant Analysis &  0.572 $\pm$ 0.054 &  0.0708 $\pm$ 0.0056 &  0.700 $\pm$ 0.084 &  0.0049 $\pm$ 0.0035 \\
                   Random Forest &  0.754 $\pm$ 0.038 &  0.0020 $\pm$ 0.0003 &  0.768 $\pm$ 0.038 &  0.0011 $\pm$ 0.0000 \\
          Support Vector Machine &  0.792 $\pm$ 0.042 &  0.0536 $\pm$ 0.0047 &  0.796 $\pm$ 0.033 &  0.0020 $\pm$ 0.0009 \\
\\\hline  \hline\multicolumn{5}{c}{ Normalized current \& Fourier transform}  \\
 \hline 
                   Decision Tree &  0.867 $\pm$ 0.035 &  0.0008 $\pm$ 0.0001 &  0.851 $\pm$ 0.043 &  0.0001 $\pm$ 0.0002 \\
                Gaussiam Process &  0.746 $\pm$ 0.032 &  0.2306 $\pm$ 0.0017 &  0.787 $\pm$ 0.042 &  0.0111 $\pm$ 0.0010 \\
           $k$-Nearest Neighbors &  0.853 $\pm$ 0.049 &  0.2399 $\pm$ 0.0471 &  0.859 $\pm$ 0.031 &  0.0093 $\pm$ 0.0010 \\
             Logistic Regression &  0.868 $\pm$ 0.036 &  0.0004 $\pm$ 0.0001 &  0.875 $\pm$ 0.036 &  0.0013 $\pm$ 0.0015 \\
           Multilayer Perceptron &  0.901 $\pm$ 0.032 &  0.0033 $\pm$ 0.0003 &  0.880 $\pm$ 0.028 &  0.0041 $\pm$ 0.0038 \\
 Quadratic Discriminant Analysis &  0.568 $\pm$ 0.066 &  0.1342 $\pm$ 0.0103 &  0.795 $\pm$ 0.048 &  0.0007 $\pm$ 0.0003 \\
                   Random Forest &  0.881 $\pm$ 0.033 &  0.0022 $\pm$ 0.0001 &  0.776 $\pm$ 0.043 &  0.0011 $\pm$ 0.0001 \\
          Support Vector Machine &  0.852 $\pm$ 0.045 &  0.0830 $\pm$ 0.0070 &  0.862 $\pm$ 0.032 &  0.0011 $\pm$ 0.0001 \\
\hline \hline
\end{tabular}

\caption{Classifier performances on double dot charge stability diagrams performed (from bottom to top)  on normalized current, Fourier transform and both and with and without PCA. Accuracies were computed over $n = 10$  train and test splits of the available data. }
\label{tab:dd_clf_metrics}
\end{table*}

\begin{table*}[!h]
\setlength{\tabcolsep}{8pt}
\renewcommand{\arraystretch}{1.2}
\small
\begin{tabular}{@{\extracolsep{6pt}}lcccc}
 \\ \hline  \hline & \multicolumn{2}{c}{PCA} & \multicolumn{2}{c}{no PCA} \\ \cline{2-3} \cline{4-5} 
classifier & accuracy & evaluation time [s] & accuracy & evaluation time [s] \\ 
\hline \multicolumn{5}{c}{ Normalized current}  \\
 \hline 
                   Decision Tree &  0.629 $\pm$ 0.054 &  0.0005 $\pm$ 0.0001 &  0.583 $\pm$ 0.052 &  0.0001 $\pm$ 0.0000 \\
                Gaussiam Process &  0.625 $\pm$ 0.047 &  0.1144 $\pm$ 0.0008 &  0.666 $\pm$ 0.042 &  0.0111 $\pm$ 0.0016 \\
           $k$-Nearest Neighbors &  0.627 $\pm$ 0.059 &  0.1753 $\pm$ 0.0183 &  0.653 $\pm$ 0.040 &  0.0102 $\pm$ 0.0012 \\
             Logistic Regression &  0.626 $\pm$ 0.037 &  0.0003 $\pm$ 0.0001 &  0.623 $\pm$ 0.049 &  0.0014 $\pm$ 0.0013 \\
           Multilayer Perceptron &  0.658 $\pm$ 0.047 &  0.0018 $\pm$ 0.0002 &  0.665 $\pm$ 0.046 &  0.0047 $\pm$ 0.0026 \\
 Quadratic Discriminant Analysis &  0.516 $\pm$ 0.056 &  0.0698 $\pm$ 0.0069 &  0.768 $\pm$ 0.082 &  0.0007 $\pm$ 0.0003 \\
                   Random Forest &  0.601 $\pm$ 0.048 &  0.0019 $\pm$ 0.0002 &  0.563 $\pm$ 0.050 &  0.0011 $\pm$ 0.0001 \\
          Support Vector Machine &  0.580 $\pm$ 0.042 &  0.0700 $\pm$ 0.0033 &  0.627 $\pm$ 0.051 &  0.0027 $\pm$ 0.0006 \\
\\\hline  \hline\multicolumn{5}{c}{ Fourier transform}  \\
 \hline 
                   Decision Tree &  0.724 $\pm$ 0.059 &  0.0004 $\pm$ 0.0001 &  0.736 $\pm$ 0.047 &  0.0001 $\pm$ 0.0000 \\
                Gaussiam Process &  0.782 $\pm$ 0.031 &  0.1142 $\pm$ 0.0012 &  0.776 $\pm$ 0.041 &  0.0147 $\pm$ 0.0016 \\
           $k$-Nearest Neighbors &  0.825 $\pm$ 0.037 &  0.1717 $\pm$ 0.0128 &  0.828 $\pm$ 0.038 &  0.0157 $\pm$ 0.0011 \\
             Logistic Regression &  0.816 $\pm$ 0.044 &  0.0003 $\pm$ 0.0001 &  0.798 $\pm$ 0.051 &  0.0009 $\pm$ 0.0006 \\
           Multilayer Perceptron &  0.838 $\pm$ 0.030 &  0.0018 $\pm$ 0.0001 &  0.837 $\pm$ 0.040 &  0.0066 $\pm$ 0.0043 \\
 Quadratic Discriminant Analysis &  0.523 $\pm$ 0.046 &  0.0713 $\pm$ 0.0085 &  0.592 $\pm$ 0.107 &  0.0010 $\pm$ 0.0002 \\
                   Random Forest &  0.695 $\pm$ 0.051 &  0.0020 $\pm$ 0.0002 &  0.637 $\pm$ 0.043 &  0.0011 $\pm$ 0.0001 \\
          Support Vector Machine &  0.790 $\pm$ 0.046 &  0.0469 $\pm$ 0.0034 &  0.778 $\pm$ 0.031 &  0.0024 $\pm$ 0.0003 \\
\\\hline  \hline\multicolumn{5}{c}{ Normalized current \& Fourier transform}  \\
 \hline 
                   Decision Tree &  0.698 $\pm$ 0.046 &  0.0008 $\pm$ 0.0000 &  0.684 $\pm$ 0.058 &  0.0001 $\pm$ 0.0000 \\
                Gaussiam Process &  0.669 $\pm$ 0.054 &  0.2313 $\pm$ 0.0020 &  0.699 $\pm$ 0.041 &  0.0152 $\pm$ 0.0020 \\
           $k$-Nearest Neighbors &  0.749 $\pm$ 0.045 &  0.3904 $\pm$ 0.0447 &  0.778 $\pm$ 0.041 &  0.0215 $\pm$ 0.0103 \\
             Logistic Regression &  0.798 $\pm$ 0.035 &  0.0004 $\pm$ 0.0000 &  0.789 $\pm$ 0.038 &  0.0027 $\pm$ 0.0027 \\
           Multilayer Perceptron &  0.836 $\pm$ 0.035 &  0.0033 $\pm$ 0.0001 &  0.820 $\pm$ 0.053 &  0.0050 $\pm$ 0.0036 \\
 Quadratic Discriminant Analysis &  0.510 $\pm$ 0.040 &  0.0894 $\pm$ 0.0522 &  0.647 $\pm$ 0.096 &  0.0010 $\pm$ 0.0003 \\
                   Random Forest &  0.684 $\pm$ 0.055 &  0.0022 $\pm$ 0.0001 &  0.607 $\pm$ 0.049 &  0.0011 $\pm$ 0.0001 \\
          Support Vector Machine &  0.769 $\pm$ 0.046 &  0.0753 $\pm$ 0.0062 &  0.769 $\pm$ 0.027 &  0.0026 $\pm$ 0.0005 \\
\hline \hline
\end{tabular}

\caption{Classifier performances on good single dot and good double dot charge stability diagrams performed (from bottom to top)  on normalized current, Fourier transform and both and with and without PCA. Accuracies were computed over $n = 10$  train and test splits of the available data. }
\label{tab:dr_clf_metrics}
\end{table*}

\end{document}